\documentclass[11pt]{article}

\usepackage[a4paper,margin=1in]{geometry}
\usepackage{amsmath,amssymb}
\usepackage{graphicx}
\usepackage{booktabs}
\usepackage{hyperref}
\usepackage{xcolor}
\usepackage{tabularx}
\usepackage{float}
\usepackage{longtable}
\usepackage{multirow}
\usepackage{algorithm}
\usepackage{algpseudocode}
\usepackage{array}
\usepackage{svg}
\usepackage{booktabs}

\newcounter{subsubsubsection}[subsubsection]
\renewcommand{\thesubsubsubsection}{\thesubsubsection.\arabic{subsubsubsection}}

\newcommand{\subsubsubsection}[1]{%
  \refstepcounter{subsubsubsection}%
  \vspace{0.5\baselineskip}%
  \noindent\textbf{\thesubsubsubsection\ #1}%
  \par\vspace{0.2\baselineskip}%
}

\title{VCT: A Verifiable Transcript System for LLM Conversations}

\author{
    Ruilin Xing\thanks{Email: ruilinxing8@gmail.com}, \ 
    Feihong Li, \ 
    Jiayue Liu, \ 
    Jiali Zheng\thanks{Corresponding author}, \ 
    Wei Liu\thanks{Co-corresponding author}, \ 
    Wanzhi Xie
    \\
    \small School of Computer Science and Electronic Information, Guangxi University
}

\date{}

\begin{document}

\maketitle

\begin{abstract}
Large language model (LLM) interaction records are increasingly vital in digital forensics and compliance auditing. However, traditional linear tamper-evident logs fail to capture the inherent non-linear evolution of LLM conversations, such as re-prompting based on historical queries, response regeneration, session deletion, multi-device concurrency, and selective sharing. To address this issue, this paper proposes \textbf{Verifiable Conversation Transcript (VCT)}, which abstracts complex non-linear LLM semantic operations into account-level authenticated state transitions. VCT constructs a three-tier cryptographic data structure: atomic Q\&A pairs form branch-level hash chains, branch tails aggregate into session-level Merkle roots, and all session roots are further aggregated into an account-level Merkle root anchored by joint signatures from both the user and the server. VCT introduces a serialized state transition protocol with deletion barriers to eliminate conflicts between deletion and modification, complemented by a deterministic state-merge protocol to preserve concurrent non-deletion incremental operations. Furthermore, incremental denial checks and a gossip protocol enable asynchronous user devices to autonomously detect view forks caused by malicious servers and generate non-repudiable forensic evidence. Security analysis demonstrates that, under standard cryptographic assumptions, VCT guarantees the integrity, consistency, verifiable shareability, and non-repudiation of account-level conversation records. Evaluation of a Python prototype shows that the cryptographic latency of core operations is within sub-millisecond to low-millisecond ranges. Under a realistic configuration with 21 KB of text, security metadata introduces a negligible storage overhead of only 0.9\%, validating the deployment feasibility of VCT for high-stakes forensic review on production-grade LLM platforms.
\end{abstract}

\noindent\textbf{Keywords:} 
large language models; verifiable conversation transcript; hash chain; Merkle tree; digital forensics; integrity; non-repudiation

\section{Introduction}
Large language models (LLMs) have become widely used tools in high-expertise workflows, including legal consultation, scientific writing, software engineering, and business analysis \cite{li2025fundamental,ye2025llms4all}. In these settings, human--AI conversations are no longer merely transient interaction traces. They increasingly serve as records of reasoning, sources of advice~\cite{huang2023survey}, and, in certain jurisdictions, potential legal evidence. Courts, auditors, and regulators have begun to rely on such records to assess intent, attribute responsibility, and review decision-making processes ~\cite{embry2026rule707,unodc2017digital,scanlon2023chatgpt}. This shift raises a fundamental question that current AI infrastructure has not adequately addressed: can the integrity and authenticity of LLM interaction records be independently verified?

Recent legal disputes make this question practically urgent. Conversation records between users and AI systems have already entered judicial proceedings and have been treated as evidentiary materials. In criminal investigations, the affidavit in \emph{United States v. Rinderknecht} ~\cite{rinderknecht2025} cited the defendant's queries to ChatGPT to infer subjective intent and post-event awareness. In civil lawsuits against AI service providers, the plaintiffs in \emph{Garcia v. Character Technologies} ~\cite{garcia2025character} and \emph{Raine v. OpenAI} ~\cite{raine2025openai} argued that model outputs themselves constituted central factual elements contributing to the alleged harm, and submitted complete conversation records as evidence of product defects and failures to warn. Yet in these cases, the authenticity and integrity of the conversation records ultimately depend on traditional evidentiary mechanisms, such as device seizure, assumptions that platform logs have not been tampered with, or post hoc admissions, rather than any cryptographic guarantee.

This fragility stems from a more basic limitation: existing LLM platforms do not provide a reliable mechanism for verifying what content a model actually generated, under which conversational context it was generated, and at what time it was produced.This temporal-anchoring problem has long been studied in the cryptographic timestamping
literature~\cite{haber1991timestamp}.  Unlike conventional digital evidence, LLM interaction records lack foundational authenticity guarantees. Screenshots can be easily forged, exported text can be silently edited, and platform-hosted sharing links often refer to mutable server-side state. More fundamentally, modern LLM interfaces allow users to regenerate responses, edit previous prompts, delete historical messages, and explore alternative conversation branches, making the notion of a ``true'' interaction history itself ambiguous. Even if these representation-level issues were resolved, a deeper systemic limitation would remain. In legal settings, the admissibility of electronic evidence depends on authenticity, integrity, and traceability ~\cite{unodc2017digital,anvar2014basheer,iso27037}. Existing LLM systems, however, operate on opaque centralized infrastructure, where service providers unilaterally control how interaction histories are stored and presented. This creates an evidentiary asymmetry: one party may present AI-generated records, while the other lacks a reliable means to determine whether those records are genuine, complete, or selectively constructed. This tension shows that traditional evidentiary mechanisms alone are insufficient for the verification challenges introduced by LLM interaction records.

Existing research does not fully address this problem. Tamper-evident secure logs ~\cite{schneier1999secure,bellare1997forward,holt2006logcrypt,ma2009new} typically provide integrity guarantees for a single linear append-only sequence through chained or aggregated authentication values, but they cannot express legitimate branching, historical edits, deletion operations, or the hierarchical state composed of branches, conversations, and accounts in LLM systems. Merkle trees and authenticated data structures ~\cite{merkle1987digital,papamanthou2011authenticated,miller2014authenticated} provide compact assertions and inclusion proofs; however, as general-purpose authentication mechanisms, they do not define the state objects or transition semantics specific to LLM conversations. Transparency logs ~\cite{laurie2013certificate,melara2015coniks} rely on a public auditing model that does not match the privacy boundary of LLM conversations. Their append-only consistency model cannot express leaf updates or replacement by deletion-state roots, and their fork semantics cannot distinguish legitimate user-created conversation branches from malicious server-induced forks. Secure messaging protocols ~\cite{cohn2017signal,barnes2023mls} target communication confidentiality, participant authentication, and key-update security. They do not provide account-level transcript proofs for third parties, non-repudiable sharing, or hierarchical conversation-state assertions over Q\&A atomic nodes. Verifiable reasoning systems ~\cite{lightman2023verify,ling2023deductive,jacovi2024weakest} assume that the underlying interaction records are authentic and complete, an assumption that does not hold in practice. Thus, the missing piece is not merely a lack of verification, but the lack of a verifiable assertion over evolving LLM conversation state itself.

The core issue is therefore one of problem formulation. The goal is not to build a better logging mechanism, but to construct cryptographic assertions over evolving conversational state, so that any disclosed interaction record can be independently verified against a tamper-evident, account-level global view. To address this challenge,\textbf{Verifiable Conversation Transcript (VCT)} is introduced as a system designed to provide integrity, consistency, verifiable sharing, and non-repudiation for LLM interaction records.

Unlike traditional linear logs or symmetric message streams, LLM conversation records exhibit multi-level and dynamically evolving structure. At the interaction level, each record is naturally organized as a Q\&A pair consisting of a user input, a model response, and associated metadata, requiring a clear atomicity boundary. Within a conversation, users may re-prompt based on historical questions or request regenerated responses, causing the conversation to branch; an integrity mechanism must therefore provide compact assertions over branch states without rewriting the entire history. At the account level, the verification target is no longer a single conversation but the cumulative state of multiple independent conversations, requiring conversation roots to be further aggregated into an account-level assertion. Multi-device concurrency further complicates account-state evolution. Devices may hold different local views before synchronization, and the protocol must distinguish benign concurrent divergence from server-induced account-level view forks while enabling consistency checking and accountability without assuming an honest server. Meanwhile, deletion operations in concurrent settings may conflict with non-deletion updates issued from other devices; the protocol must therefore define serialization rules between deletion and concurrent updates, and maintain a consistent user signing identity across devices. In addition, partial sharing introduces an independent verification requirement. Users often intend to selectively disclose only certain conversation fragments, but server-generated sharing links or exported artifacts may replace, remove, reorder, or omit the content that the user intended to share. Recipients therefore need a verification mechanism independent of the server-rendered view. These characteristics show that making LLM conversation records trustworthy is not merely a problem of single-message or single-conversation integrity, but an account-level state-authentication problem across branches, multiple conversations, concurrency, deletion, and sharing.

VCT addresses this account-level state-authentication problem by combining a layered authenticated structure with protocol-defined state transitions. To capture Q\&A atomicity, conversation branching, and multi-session aggregation, VCT models each Q\&A pair as a hash-chain node, using \texttt{parent\_hash} to preserve the sequential dependency along the same branch. The tail hashes of all branches within a conversation are aggregated into a conversation-level Merkle root, denoted \texttt{MTR\_con}. The roots of all conversations under an account are further aggregated into an account-level Merkle root, denoted \texttt{MTR\_acc}, which is jointly signed by the user and the server to produce a verifiable assertion over the complete account state. To handle multi-device concurrency and deletion semantics, VCT introduces a deterministic state-merge protocol that preserves non-deletion concurrent increments, models conversation deletion as an account-level serialized operation to avoid delete--modify conflicts, and uses a gossip protocol for synchronized devices to exchange signed account roots and detect server-induced account-level view forks. VCT also employs deterministic key derivation to maintain a consistent user signing identity across devices, without depending on the server to synchronize user private keys. For partial sharing, VCT reorganizes the shared nodes into an independent hash chain and requires both the user and the server to sign the chain tail, enabling the recipient to verify the shared content independently of the server. In this way, VCT binds branching, multi-session aggregation, concurrency, deletion, and sharing operations in LLM conversations to a unified account-level verifiable state assertion.

\noindent\textbf{Contributions.}
This paper makes the following contributions.

\begin{itemize}
\item \textbf{Verifiable Data Structure for Branching and Multi-Device Conversations.} 
A three-layer authenticated structure—comprising branch hash chains, conversation-level Merkle trees, and an account-level Merkle tree—is introduced to provide a unified representation of state transitions, including Q\&A atomic nodes, conversation branches, message appends, new conversation creation, and conversation deletion. This structure progressively binds branch-level interaction histories, conversation-level branch sets, and account-level multi-conversation states into a single account root, thereby enabling any disclosed node, branch, or conversation state to be traced back to an account-level verifiable state jointly confirmed by the user and the server.

\item \textbf{Multi-Device Concurrent Consistency Protocol with Accountability.} 
An account-level state-synchronization protocol is proposed to orchestrate concurrent multi-device interactions. The protocol preserves non-deletion concurrent updates within the same deletion epoch through deterministic state merging, and models conversation deletion as an account-level serialized operation to mitigate delete--modify conflicts. It further incorporates a device-side rejection mechanism for stale incremental replay, preventing a malicious server from returning an outdated state and later replaying omitted increments to evade detection. Finally, a gossip mechanism enables devices to exchange signed account roots over end-to-end encrypted channels, facilitating the detection of forked account views and producing accountable evidence backed by dual signatures.

\item \textbf{Verifiable Conversation Sharing Protocol.} 
A sharing mechanism is presented that couples an independent hash chain with dual signatures. A recipient only needs to verify the chain-tail hash and the corresponding signatures to confirm that the shared fragment remains untampered and jointly confirmed by the server and the data owner. Once generated, the sharing snapshot maintains its cryptographic verifiability independently of any subsequent appends, branches, or deletions within the original conversation.
\end{itemize}

\section{Related Work and Insights}

The core objective of Verifiable Conversation Transcript (VCT) is to provide account-level integrity, consistency, share verifiability, and non-repudiation for interaction records from human--LLM conversations. Unlike traditional system logs or ordinary message records, an LLM conversation is an evolving account state composed of Q\&A nodes, branching relations, deletion states, multi-session aggregation, and multi-device synchronization. VCT must therefore verify not only whether an individual record has been tampered with, but also whether the account state remains consistent across concurrent updates, deletion, and synchronization, and whether fragments derived from that account state have not been replaced, reordered, deleted, or selectively constructed during sharing.

This section compares VCT with six lines of related work: tamper-evident secure logs, Merkle trees and authenticated data structures, transparency logs, secure messaging protocols, fork consistency in untrusted storage, and verifiable reasoning.The key question is not whether these techniques can provide local integrity, but whether they can support the system-level semantics required by VCT, including branching LLM conversation structure, account-level cumulative assertions, consistency checking within account privacy boundaries, verifiable sharing, and non-repudiation in forensic settings. VCT draws on hash chains from secure logging, compact assertions and inclusion proofs from Merkle trees and authenticated data structures, signed roots and fork-detection ideas from transparency logs, transcript hashes from secure messaging protocols, and the broader concern for LLM output trustworthiness from verifiable reasoning. However, VCT does not aim to construct a public linear log, nor does it verify the correctness of model reasoning. Instead, it provides an account-level, branching, privacy-bounded verifiable state assertion for LLM interaction histories.

\subsection{Tamper-Evident Secure Logs}

Tamper-evident secure logs study how to maintain log records for post hoc auditing and forensics on partially untrusted hosts. Schneier and Kelsey~\cite{schneier1999secure} proposed a secure audit-log scheme based on hash chains and key evolution, so that even if an attacker compromises the logging host at some point, previously generated log entries cannot be modified or deleted without detection. The central goal of this line of work is to make tampering detectable during later audits, rather than to prevent all attacks. Bellare and Yee~\cite{bellare1997forward} further introduced the notion of forward integrity, formalizing the requirement that compromise of the current key should not invalidate the integrity of past log entries, and presented constructions satisfying this property. Holt's Logcrypt~\cite{holt2006logcrypt} combines forward security with public verification, reducing the dependence of log integrity checking on a private verifier and improving usability in third-party audit scenarios. Ma and Tsudik~\cite{ma2009new} improved the space efficiency and verifiability of secure logs through techniques such as forward-secure sequential aggregate authentication, focusing on authentication overhead and verification efficiency.

Secure logs provide a basic building block for the interaction layer of VCT. In an LLM conversation, a user request and the corresponding model response can be treated as a Q\&A atomic node. Each node cryptographically binds the user input, the model output, associated metadata, and the hash of its parent node, so that adjacent nodes along the same branch form a chained dependency. If an attacker modifies the input, output, or metadata of any round, the recomputed node hash will no longer match the parent hash recorded by its successor or the tail hash committed for that branch. Hash chains are therefore well suited to the interaction layer of VCT, where they provide sequential integrity within a single branch.

The difference between secure logs and VCT is twofold. At the data-model level, traditional secure logs assume a single linear append-only event sequence. LLM conversations, by contrast, allow users to edit historical prompts or regenerate responses, thereby creating legitimate branches; conversation deletion is also not an ordinary append operation, but a state transition that changes the account state. VCT therefore cannot simply place all interaction records into one linear log. It must instead represent parent--child and derivation relations among different conversation paths at the branch level. At the authentication-granularity level, traditional secure logs typically establish integrity over individual log entries, linear chain states, or aggregated authentication values, using mechanisms such as MACs, hash chains, forward-secure signatures, or sequential aggregate authentication. The authenticated object in VCT is not a single linear log, but a hierarchical state composed of branches, conversations, and accounts. VCT progressively accumulates these states into an account-level Merkle root and requires both the user and the server to sign that root, thereby covering all conversation states under the account with a single account root rather than signing each Q\&A node individually. Thus, secure logs offer a useful primitive for local sequential integrity, but they do not capture legitimate branching, deletion states, or account-level assertion semantics in LLM conversations.

\subsection{Merkle Trees and Authenticated Data Structures}

Merkle trees recursively aggregate hashes of data blocks into a root hash, thereby providing compact commitments and efficient inclusion proofs for large data sets. Merkle's classic work~\cite{merkle1987digital} shows that conventional cryptographic functions can be used to construct digital signature schemes and authentication structures, allowing a verifier to check the relationship between an element and a root commitment without reading the entire data set. Building on this idea, authenticated data structures (ADS) study how data can be stored and queried on untrusted servers, while requiring the server to provide efficiently verifiable proofs for query results. Papamanthou et al.~\cite{papamanthou2011authenticated} focus on correctness and integrity verification for outsourced databases, while Miller et al.~\cite{miller2014authenticated} propose general-purpose ADS from the perspective of programming languages and type systems, enabling a broader class of data structures to support authenticated operations and compact proofs.

Merkle trees and authenticated data structures (ADS) provide two core capabilities for VCT. First, a Merkle authentication structure can aggregate multiple objects into a single root hash and use a Merkle proof to show that a given object is committed by that root. Crosby and Wallach~\cite{crosby2009tamper} demonstrate the practical significance of this property for tamper-evident logging: replacing a flat hash-chain traversal with a tree-based aggregation structure reduces inclusion-proof size from linear to logarithmic in the number of log entries. For VCT, this means that a verifier can check whether a Q\&A node, a branch tail hash, or a conversation root is included in a signed account state with only logarithmic overhead, without reading all conversations under an account. Second, Merkle structures support local recomputation: when a leaf changes, only the authentication path from that leaf to the root needs to be recomputed. VCT uses this structural property to localize the effect of a state transition within the authenticated structure to the affected Q\&A node, the branch tail hash, the conversation-level Merkle path, and the account-level Merkle path.

However, ADS is a general authentication abstraction. It specifies how to authenticate data-structure states, query results, and update procedures, but it does not define how LLM-specific operations should modify the authenticated state. VCT must prove not only that ``a leaf belongs to a root,'' but also that a state transition conforms to the legal operation semantics of an LLM conversation system. Concretely, when a user appends a Q\&A pair to an existing branch, the branch tail hash is replaced by the new node hash. Since the conversation-level Merkle tree uses branch tail hashes as leaves, this operation updates an existing leaf in the conversation-level Merkle tree and further propagates to the account-level Merkle root. When the user regenerates a response or re-prompts from a historical node, the system does not overwrite the original history; instead, it creates a new branch at the historical node and inserts the new branch tail hash into the conversation-level Merkle tree. When a user creates a new conversation, the new conversation root is inserted into the account-level Merkle tree. When a user deletes a conversation, the system does not simply remove a leaf, but replaces the original conversation root with a deletion-state conversation root, thereby preserving a cryptographic assertion of the pre-deletion state.

Thus, Merkle trees and ADS provide the authentication mechanisms required by VCT, but not the complete state semantics of LLM conversations. VCT defines a three-layer state structure on top of these mechanisms: the interaction layer records Q\&A nodes and their parent--child relations, the conversation layer aggregates multiple branch tails within a single conversation, and the account layer aggregates all conversation states of a user. This structure represents branch extension, historical editing, response regeneration, new conversation creation, and conversation deletion in LLM conversations as verifiable account-state transitions.

\subsection{Transparency Logs}

Transparency-log systems combine Merkle trees, signed tree heads, inclusion proofs, and consistency proofs to detect unauthorized modifications to historical records by centralized services. Certificate Transparency (CT)~\cite{laurie2013certificate} requires TLS certificates to be recorded in a public, auditable, append-only log, allowing clients, monitors, and auditors to verify both certificate inclusion and append-only consistency between log roots observed at different points in time. CONIKS~\cite{melara2015coniks} applies similar transparency principles to user-key bindings in end-to-end encrypted communication, enabling end users to efficiently detect whether their key bindings are presented inconsistently by a provider, without relying on global third parties to monitor all entries.

Transparency logs provide two main insights for VCT. First, a server-signed root assertion makes it difficult for the server to later deny having published a particular state. Second, root comparison and consistency checking across different clients or devices can reveal equivocation or forked views. The gossip-based detection protocol in VCT adopts this idea in a multi-device setting for a single user. Devices exchange their locally stored account-level Merkle roots together with the corresponding dual signatures from the server and the user. If different devices receive incompatible account states, verifiable fork evidence can be produced based on the signed account roots.

However, transparency logs cannot be directly applied to VCT for three reasons. First, their auditing model does not match the privacy boundary of LLM conversations. Certificate entries in CT and key bindings in CONIKS are public or semi-public objects that can be observed and audited by external monitors. In contrast, LLM conversation records contain private information, including user inputs, model outputs, conversational context, and auxiliary file information. It is therefore impractical to require all users to collectively audit a public log. Consequently, VCT neither exposes complete conversation histories nor places all user records into a single global log. Instead, it constructs account-level Merkle roots within each user's account boundary and performs consistency checking primarily across devices belonging to the same user. Second, the consistency proofs used in transparency logs differ from the consistency semantics required by VCT. Append-only consistency proofs characterize a prefix-preserving relation in which existing leaves remain unchanged and only new leaves are appended\cite{tomescu2019transparency}. VCT, however, does not operate on a purely append-only log. To represent branching conversation states, VCT uses the current cumulative tail hash of each branch as a leaf in the conversation-level Merkle tree. As a result, legitimate interactions in an LLM conversation may not only append new records but also modify authenticated state leaves. For example, when a user appends a new Q\&A pair to an existing branch, the current branch-tail hash is replaced by the hash of the newly created node, causing both the conversation-level and account-level Merkle roots to change. Similarly, response regeneration or re-prompting from a historical question introduces new branch leaves, creating a new conversation introduces a new account-level leaf, and deleting a conversation replaces the original conversation root with a deletion-state root. VCT therefore must verify a set of legitimate state transitions defined by LLM conversation semantics, rather than a single notion of append-only consistency. Third, the fork semantics of transparency logs do not directly carry over to LLM conversations. In a linear transparency log, presenting incompatible histories to different observers typically indicates equivocation or a malicious fork. In contrast, a ``branch'' in an LLM conversation may be the result of legitimate user actions, such as editing a historical prompt or regenerating a response. VCT must therefore distinguish between two fundamentally different phenomena: legitimate branches created by user operations within a conversation and malicious forks caused by a server presenting incompatible account views to different devices of the same user. The former is explicitly represented as a normal state within the three-layer authenticated structure, whereas the latter is detected and attributed through account-level roots, dual signatures, and inter-device gossip verification.

Therefore, VCT borrows the notions of signed roots and fork detection from transparency logs, but does not inherit their globally visible, linear append-only logging model.

\subsection{Secure Messaging Protocols}

Secure messaging protocols primarily focus on authentication, confidentiality, forward secrecy, and key update between communicating parties. The Signal protocol provides end-to-end encryption, message authentication, forward secrecy, and post-compromise security for instant messaging through X3DH key agreement and the Double Ratchet mechanism. Cohn-Gordon et al.~\cite{cohn2017signal} conducted a formal security analysis of Signal's core key agreement and ratcheting construction. Messaging Layer Security (MLS)~\cite{barnes2023mls} targets asynchronous group communication, managing member updates using tree-based group state and maintaining a transcript hash within the protocol. It is worth noting that the transcript hash in MLS does not establish a complete log over all application-layer chat messages; rather, it computes a running hash over Proposal and Commit messages that affect group state, thereby binding the group key state to the history of prior protocol updates.

The insight these protocols offer for VCT is that system state can be bound to its historical trajectory through running hashes and key-state evolution. In particular, the transcript hash in MLS demonstrates that in dynamic, multi-party, and asynchronous update settings, a protocol can do more than protect individual messages---it can incorporate state transitions themselves into a verifiable history. What VCT borrows from this line of work is precisely this idea of ``binding state evolution to a historical assertion.''

Nevertheless, the goals and data models of secure messaging protocols differ fundamentally from those of VCT. Signal and MLS are primarily designed for communication confidentiality, participant authentication, and key-update security; they do not provide publicly verifiable transcript proofs for courts, auditors, or sharing recipients. Some protocols also emphasize deniability, whereas VCT requires non-repudiation in sharing and forensic settings. Even the MLS transcript hash, which comes closest to binding state to history, binds the evolutionary history of group protocol state rather than long-term verifiable LLM conversation content. Furthermore, in secure messaging protocols, messages are typically sent directly by communicating parties; the basic record unit in VCT, by contrast, is a Q\&A atomic node jointly constituted by user input and model output---only after the server generates model output A in response to user input Q does the tuple (Q, A, model\_config, file\_aux\_info, timestamp) enter the record as a complete node. This distinction makes it difficult for secure messaging protocols to directly express the state-transition semantics in LLM platforms that arise from response regeneration, historical editing, conversation deletion, account-level aggregation, and share export.

Accordingly, VCT does not compress the entire interaction history into a single running digest analogous to a transcript hash. Instead, it models LLM conversations as a hierarchical account state, with both the user and the server jointly signing the account root. For sharing scenarios, VCT further constructs an independent share chain and signs the share-chain tail. In this way, VCT extends the history-binding ideas of secure messaging protocols into a verifiable integrity, share verifiability, and multi-device accountability mechanism tailored to cloud-based LLM conversations.

\subsection{Fork Consistency in Untrusted Storage}
The notion of detecting inconsistent views served by an untrusted storage
provider has been studied extensively in the distributed systems security
literature under the name of \emph{fork consistency}~\cite{mazieres2002byzantine}.
SUNDR~\cite{li2004sundr} shows that a network file system can guarantee fork
consistency---clients can detect any integrity or consistency violation as long as they eventually exchange their locally observed, signed version
structures---whether or not the server is honest. Depot~\cite{mahajan2011depot} generalizes this guarantee to a Byzantine cloud-storage setting via
Fork-Join-Causal consistency, tolerating faulty clients in addition to a
faulty server.

VCT's gossip-based fork detection follows the same high-level principle as SUNDR/Depot---divergent views are made detectable
through the cross-exchange of signed, comparable digests. However, VCT
differs in object granularity and trust topology: SUNDR/Depot authenticate a
single shared file-system or key-value namespace across multiple mutually
distrusting \emph{users}, whereas VCT authenticates a hierarchical,
branching account state across multiple \emph{devices of the same user}, and
must additionally distinguish legitimate conversation branching from
server-induced forks, a distinction that has no counterpart in the
file-system fork-consistency model.

\subsection{Verifiable Reasoning vs. Verifiable Interaction}

An important line of recent research on LLM trustworthiness focuses on verifying the model's reasoning process. Lightman et al.~\cite{lightman2023verify} compare outcome supervision with process supervision, training process-reward models using step-level human feedback to improve the reliability of complex mathematical reasoning. Ling et al.~\cite{ling2023deductive} propose a framework for deductive verification of chain-of-thought reasoning, improving the rigor and inspectability of intermediate steps through a more structured natural-language reasoning format. Jacovi et al.~\cite{jacovi2024weakest} introduce the REVEAL benchmark for evaluating the fine-grained judgment capabilities of automatic verifiers on relevance, evidence attribution, and logical correctness within reasoning chains, and observe that existing verifiers still face substantial difficulties in detecting errors in such chains.

This body of work addresses the question of whether the reasoning produced by a model is correct, or whether intermediate reasoning steps are verifiable. While these contributions are important for improving the reliability of LLM outputs, they rest on an implicit assumption: that the prompts, responses, and reasoning traces being verified are authentic, complete, and untampered. In the context of legal forensics, auditing, and accountability tracing, this assumption does not always hold. Users may edit exported text, selectively excerpt favorable fragments, modify context, or forge screenshots; servers may return stale states, omit certain branches, present different account views to different devices, or replace, reorder, or delete content in sharing links and exported files. Even if a reasoning chain is logically verified, reasoning verification alone cannot support evidentiary authenticity and completeness if the underlying interaction record does not faithfully reflect what actually occurred. Put differently, the validity of reasoning verification depends on the authenticity of the interaction record---a precondition that existing work on verifiable reasoning does not guarantee.

A distinction is therefore drawn between verifiable reasoning and verifiable interaction, one that echoes broader work in the field of verifiable computation~\cite{walfish2015verifying}, where execution correctness and input authenticity are treated as two independent concerns. Specifically, verifiable reasoning is concerned with whether a model reaches sound conclusions given a specified input and reasoning trace—that is, it verifies the content of reasoning. Verifiable interaction is concerned with whether a given input, model output, context, and conversational state have indeed been incorporated into a signed, untampered historical assertion—that is, it verifies the interaction trace itself. VCT operates at the level of verifiable interaction: it does not directly assess whether model answers are correct, nor does it certify the faithfulness of the model's internal reasoning. Instead, it provides a more foundational evidentiary basis, enabling third parties to verify that a disclosed LLM conversation originates from a confirmed account state and has not been tampered with, reordered, omitted, or substituted in the course of disclosure.

Verifiable interaction can serve as a prerequisite for verifiable reasoning. If prompts, responses, or contextual records can be arbitrarily edited or selectively disclosed, then even a reasoning verifier that judges a given chain to be formally correct cannot confirm that the chain corresponds to a human--AI interaction that genuinely took place. Through node-level hashing, conversation-level branch assertions, account-level dual signatures, and a verifiable sharing mechanism, VCT binds the objects subject to reasoning verification to an inspectable interaction history. The two are not alternatives but operate at different layers: VCT guarantees the authenticity and completeness of interaction records, while verifiable reasoning builds upon this foundation to further evaluate the correctness of model outputs and reasoning processes.

\noindent\textbf{Related Work Summary}

Existing work collectively provides several technical building blocks required by VCT. Tamper-evident secure logs contribute ideas of sequential integrity, forward integrity, and post hoc auditing, but their linear append-only model and linear chain-based authentication objects cannot express the legitimate branching, historical editing, deletion-state operations, and account-level hierarchical state found in LLM conversations. Merkle trees and authenticated data structures provide compact assertions, inclusion proofs, and local update capabilities; however, as general-purpose authentication mechanisms, they do not define the state objects or state-transition semantics specific to LLM conversations. Transparency logs contribute ideas of signed roots, consistency proofs, and fork detection, but their public, linear, append-only log model is incompatible with the privacy boundaries of LLM conversations and cannot directly express the state consistency required by VCT, which encompasses leaf updates, new-leaf appends, and deletion-state replacements; moreover, their fork semantics cannot distinguish user-created legitimate branches from server-induced malicious forks. Secure messaging protocols contribute transcript hash techniques and experience with asynchronous state evolution, but their primary goals are communication confidentiality, participant authentication, and key-update security, rather than account-level transcript proofs, non-repudiable sharing, or hierarchical conversational-state assertions over Q\&A atomic nodes.Fork-consistency protocols for untrusted
storage~\cite{mazieres2002byzantine, li2004sundr, mahajan2011depot}
contribute the technique of detecting divergent views through the
cross-exchange of signed, comparable digests, but they authenticate a flat
file-system or key-value namespace shared by mutually distrusting clients,
rather than a hierarchical, branching account state belonging to a single
user, and therefore do not need to distinguish legitimate intra-account
branching from server-induced forks, a distinction central to VCT's gossip
protocol. Research on verifiable reasoning addresses whether model reasoning steps are correct, but does not certify whether the underlying interaction records are authentic, complete, and untampered.

VCT systematically reconstructs these technical foundations for the LLM conversation setting. At the data-structure level, VCT employs branch hash chains, a conversation-level Merkle tree, and an account-level Merkle tree, rather than a single linear log or a flat global log. At the signing granularity level, VCT applies dual signatures to the account-level Merkle root and the share-chain tail, rather than signing each Q\&A node individually. At the audit-model level, VCT does not rely on all users jointly auditing a public log; instead, it performs multi-device consistency checking within account privacy boundaries. At the forensic-semantics level, VCT explicitly distinguishes user-operation-induced legitimate branches from server-induced malicious forks---where the server presents incompatible states to different devices---and supports independent third-party verification of the authenticity, completeness, and non-repudiation of disclosed fragments through its verifiable sharing mechanism.

{\small
\begin{longtable}
{p{3.0cm}>{\raggedright}p{3.0cm}>{\raggedright}p{2.5cm}>{\raggedright\arraybackslash}p{6.0cm}}
\caption{Related Techniques and Their Limitations for VCT}
\label{tab:related_work_comparison}\\
\toprule
\textbf{Related Direction} & \textbf{Primary Problem Addressed} & \textbf{Applicable Mechanism} & \textbf{Why Cannot Be Directly Applied to VCT} \\
\midrule
\endfirsthead

\caption*{Table \ref{tab:related_work_comparison}: Related Techniques and Their Limitations for VCT (continued)}\\
\toprule
\textbf{Related Direction} & \textbf{Primary Problem Addressed} & \textbf{Applicable Mechanism} & \textbf{Why Cannot Be Directly Applied to VCT} \\
\midrule
\endhead

\midrule
\multicolumn{4}{r}{\textit{Continued on next page}} \\
\endfoot

\bottomrule
\endlastfoot

Secure Logs~\cite{schneier1999secure, bellare1997forward, holt2006logcrypt, ma2009new} & Tamper-evidence, forward integrity, post-hoc auditing for linear logs & Hash chains, public verification, aggregate authentication & Assumes a single linear append-only event sequence; cannot express legitimate branching, historical editing, or deletion in LLM conversations. The authenticated object is typically a single log entry, linear chain state, or aggregated authentication value, rather than the hierarchical state of branches, conversations, and accounts. \\
\midrule
Merkle / ADS~\cite{merkle1987digital, papamanthou2011authenticated, miller2014authenticated} & Compact assertions, inclusion proofs, query authenticity for untrusted storage & Merkle root, Merkle proof & Provides a general-purpose authentication abstraction, but its interface does not include LLM-specific state semantics, such as when a branch tail hash is updated, how a new branch is added, how a conversation root propagates to the account root, or how a pre-deletion conversation root is replaced by a deletion-state root. \\
\midrule
Transparency Logs~\cite{laurie2013certificate, melara2015coniks} & Append-only records, inclusion/consistency proofs, fork detection for public objects & Signed tree head, consistency proof, root gossip & Public or semi-public auditing does not match the privacy boundary of LLM conversations. Append-only consistency proofs cannot express leaf updates or deletion-state replacements. Fork semantics cannot distinguish user-created legitimate branches from server-induced malicious forks. \\
\midrule
Secure Messaging Protocols~\cite{cohn2017signal, barnes2023mls} & End-to-end communication security, key updates, group state consistency & Transcript hash & Does not provide third-party-verifiable account-level transcript proofs or non-repudiable sharing. The transcript hash primarily binds the evolution of protocol state rather than the hierarchical account state of LLM conversation content. Does not define LLM-specific state-transition semantics such as regeneration, historical editing, conversation deletion, multi-session aggregation, or share snapshots. \\
\midrule
Fork Consistency~\cite{mazieres2002byzantine, li2004sundr, mahajan2011depot} & Detecting divergent views and equivocation served by an untrusted storage provider to mutually distrusting clients & Signed version structures, cross-client digest exchange, fork-consistency guarantee & Authenticates a flat, shared file-system or key-value namespace across multiple distinct users, rather than a hierarchical, branching conversation state belonging to a single user across devices. Defines no notion of legitimate intra-account branching (e.g., response regeneration, historical editing) that must be distinguished from server-induced forks, and provides no deletion-state semantics or account-level aggregation across multiple independent conversations. \\
\midrule
Verifiable Reasoning~\cite{lightman2023verify, ling2023deductive, jacovi2024weakest} & Verifying reasoning steps, process supervision, CoT verifier evaluation & \multicolumn{1}{c}{\textbf{---}} & Assumes that underlying interaction records are authentic and complete, an assumption that does not hold in practice. \\

\end{longtable}
}
\section{Preliminaries and Problem Definition}

This section establishes the formal foundations of VCT. Notation and participants are first defined, followed by the core storage abstraction model underlying the protocol. The underlying cryptographic primitives and security assumptions are then detailed, and the threat model and security objectives are subsequently formalized. Collectively, these elements provide a unified formal basis and a precise problem boundary for the subsequent protocol design and security analysis.

\subsection{Notation and Tools}

The following notation is adopted throughout the remainder of this paper.

{\small
\begin{longtable}{p{3.5cm} >{\raggedright\arraybackslash}p{12.0cm}}
\caption{Notation and Description}
\label{tab:notation}\\
\toprule
\textbf{Notation} & \textbf{Description} \\
\midrule
\endfirsthead

\caption*{Table \ref{tab:notation}: Notation and Description (continued)}\\
\toprule
\textbf{Notation} & \textbf{Description} \\
\midrule
\endhead

\midrule
\multicolumn{2}{r}{\textit{Continued on next page}} \\
\endfoot

\bottomrule
\endlastfoot

$\text{Hash}(X)$ & Cryptographic hash value of $X$. \\

$\text{hash}_{\text{tail}}$ & Final cumulative hash value of each branch hash chain within the same conversation. \\

$\text{hash}_{\text{parent}}$ & Hash value of the parent node in a hash chain. \\

$\text{MTR}_{\text{con}}$ & Conversation-level Merkle root; constructed from the $\text{hash}_{\text{tail}}$ of all branches in the same conversation as leaves. \\

$\text{MTR}_{\text{con}}^{\text{del}}$ & Deletion-state conversation root, defined as $\text{MTR}_{\text{con}}^{\text{del}} = \text{Hash}(\texttt{"DEL\_SESSION"} \| \text{MTR}_{\text{con}}^{\text{old}} \| \text{timestamp}_{\text{del}})$. \\

$\text{MTR}_{\text{acc}}$ & Account-level Merkle root; constructed from $\text{MTR}_{\text{con}}$ or $\text{MTR}_{\text{con}}^{\text{del}}$ of all conversations belonging to the same user as leaves. \\

$\text{SIGN}_{\text{sk}_x}(Z)$ & Digital signature generated on message $Z$ using the private key $\text{sk}_x$ of entity $x$. \\

$\text{STR}_{\text{acc}}^{\text{u}}$ & User's signature on the account-level state root, typically $\text{STR}_{\text{acc}}^{\text{u}} \leftarrow \text{SIGN}_{\text{sk}_u}(\texttt{"ACCOUNT\_STATE"} \| \text{MTR}_{\text{acc}} \| \text{timestamp})$. \\

$\text{STR}_{\text{acc}}^{\text{s}}$ & Server's signature on the account-level state root, typically $\text{STR}_{\text{acc}}^{\text{s}} \leftarrow \text{SIGN}_{\text{sk}_s}(\texttt{"ACCOUNT\_STATE"} \| \text{MTR}_{\text{acc}} \| \text{timestamp})$. \\

$\text{model}_{\text{config}}$ & LLM model configuration parameters, including at least $\text{model}_{\text{id}}$, $\text{nonce}$, $\text{temperature}$, and may include other parameters that ensure deterministic output. \\

$\text{file\_aux\_info}$ & Composite data object containing uploaded files and their auxiliary information. \\

$Q$ & User input prompt. \\

$A$ & Response generated by the LLM model based on the user prompt. \\

$K_u$ & User root key, used to derive user signing private key and multi-device communication keys. \\

$\text{sk}_u, \text{sk}_s$ & Asymmetric signing private keys of user $u$ and server $s$, respectively. \\

$\text{pk}_u, \text{pk}_s$ & Asymmetric signing public keys corresponding to $\text{sk}_u$ and $\text{sk}_s$, respectively. \\

$\text{timestamp}$ & Ordinary timestamp used for temporal anchoring of node creation, state signing, and protocol events. \\

$\text{timestamp}_{\text{del}}$ & Timestamp of conversation deletion operation. \\

$\text{timestamp}_{\text{sig}}$ & Signature timestamp bound to $\text{STR}_{\text{acc}}^{\text{u}}$ returned by the server during "update conversation record". \\

$\text{timestamp}_{\text{update}}$ & Timestamp when a device initiates an "update conversation record" operation. \\

$\text{timestamp}_{\text{node}}$ & Timestamp of an incremental node, used for incremental freshness checking. \\

$\text{timestamp}_{\text{acc}}$ & Timestamp bound to the account state signature, used to verify the account state anchored by $\text{STR}_{\text{acc}}^{\text{s}}$ and $\text{STR}_{\text{acc}}^{\text{u}}$. \\

$\text{timestamp}_{\text{share}}$ & Timestamp of share snapshot generation, used to generate and verify $\text{share}_{\text{sig}}^{\text{s}}$ and $\text{share}_{\text{sig}}^{\text{u}}$. \\

$\text{share}_{\text{tail}}$ & Tail of the independent hash chain used for sharing. \\

$\text{share}_{\text{sig}}^{x}$ & Signature generated by entity $x$ on $\text{share}_{\text{tail}}$ during sharing, where $x \in \{u, s\}$. \\

$\text{DeriveKeyPair}(\cdot)$ & A deterministic key-derivation algorithm that maps a high-entropy seed $\mathit{seed}_u$ to a user public/private signing key pair $(sk_u, pk_u)$. \\

$\text{pwd}_u$ & User login password. \\

$\text{seed}_u$ & User key seed, derived from $K_u$ and $\text{pwd}_u$ via KDF, used to deterministically generate the user signing key pair. \\

$K_{\text{gossip}}$ & Gossip communication key derived from $K_u$ and $\text{pwd}_u$ using a KDF with domain separation label $\texttt{"gossip\_encryption"}$. \\

$\text{MerkleProofs}$ & Set of Merkle authentication paths used in the protocol to prove that a relevant node, branch tail, or conversation root is anchored to the corresponding signed Merkle root during conversation update or sharing verification. \\

$\text{MTR}_{\text{acc}}^{\text{base}}$ & The locally persisted account‑level Merkle root that the user’s device submits to the server as the state‑transition baseline for an update or sync request. \\

$\text{MTR}_{\text{acc}}^{\text{current}}$ & Server's current account-level Merkle root at the time of request processing. \\

$\text{MTR}_{\text{con}}^{\text{old}}$ & Conversation-level Merkle root before conversation deletion. \\

\end{longtable}
}

The following participants are involved in this paper.

\begin{table}[H]
\centering
\small
\caption{Participants and Description}
\label{tab:participants}
\begin{tabular}{p{3.5cm} >{\raggedright\arraybackslash}p{12.0cm}}
\toprule
\textbf{Participant} & \textbf{Description} \\
\midrule
User & The same user, who may own multiple devices and concurrently initiate conversations with the LLM. \\
LLM Model Server & Generates responses to user prompts and participates in conversation state storage, signing, and synchronization. Logically, it can be separated into a model inference service and a conversation storage service. \\
$\mathcal{TP}$ & Third-party entity used to verify the integrity of LLM conversation records and associated proofs. \\
$\mathcal{R}$ & Recipient, i.e., the party who views the shared conversation records. \\
\bottomrule
\end{tabular}
\end{table}

Based on the above notation and participant definitions, the system state and storage model of VCT is further described below.

\subsection{System State and Storage Model}

VCT targets LLM conversation systems built on a typical client-server (B/S) architecture, placing the persistent maintenance of complete conversation state on the server side while requiring user devices to retain only the minimal local state necessary for state verification and signature confirmation. The server maintains the committed state for each account, comprising Q\&A nodes, branch structures, the conversation-level Merkle root $\text{MTR}_{\text{con}}$, deletion-state records ($\text{MTR}_{\text{con}}^{\text{old}}$, $\text{MTR}_{\text{con}}^{\text{del}}$, $\text{timestamp}_{\text{del}}$), and the current account anchor ($\text{MTR}_{\text{acc}}$, $\text{timestamp}_{\text{acc}}$, $\text{STR}_{\text{acc}}^{\text{s}}$, $\text{STR}_{\text{acc}}^{\text{u}}$). To support device synchronization, deterministic merging, and state traceability, the server additionally maintains a complete $\text{MTR}_{\text{acc}}$ state-transition chain or an equivalent incremental log, and records the last confirmed account anchor and synchronization boundary ($\text{timestamp}_{\text{sig}}$, $\text{timestamp}_{\text{update}}$) for each device. Merkle intermediate nodes do not constitute semantic state that the protocol requires to be persistently stored; they may be retained as a server-side performance cache to support incremental Merkle updates and authentication path generation. In the absence of such a cache, the server can still reconstruct the relevant paths or root values from the underlying conversation state.

User devices are not required to retain complete conversation records, branch trees, or Merkle intermediate nodes over the long term. Instead, each device retains only the most recently verified account anchor, namely ($\text{MTR}_{\text{acc}}$, $\text{timestamp}_{\text{acc}}$, $\text{STR}_{\text{acc}}^{\text{s}}$, $\text{STR}_{\text{acc}}^{\text{u}}$), together with the server public key $\text{pk}_s$, the synchronization boundary ($\text{timestamp}_{\text{sig}}$, $\text{timestamp}_{\text{update}}$), and local key material $K_u$, $\text{salt}$, and $\text{iterations}$. The user signing private key $\text{sk}_u$ and the gossip communication key $K_{\text{gossip}}$ may be derived on demand or stored securely. Accordingly, during a conversation update, a user device does not recompute $\text{MTR}_{\text{acc}}$ from a complete local account state, nor does it rely on a full account view pushed by the server. Instead, it treats the locally stored $\text{MTR}_{\text{acc}}$ and its dual signatures as the account anchor, and verifies the legitimacy of the state transition against the $\text{update\_payload}$ and $\text{MerkleProofs}$ returned by the server.

\begin{longtable}{p{0.18\textwidth} >{\raggedright\arraybackslash}p{0.55\textwidth} p{0.22\textwidth}}
\caption{Storage Parties and Their Stored Content}
\label{tab:storage_model}\\
\toprule
\textbf{Storage Party} & \textbf{Stored Content} & \textbf{Persistence} \\
\midrule
\endfirsthead

\caption*{Table \ref{tab:storage_model}: Storage Parties and Their Stored Content (continued)}\\
\toprule
\textbf{Storage Party} & \textbf{Stored Content} & \textbf{Requirement} \\
\midrule
\endhead

\midrule
\multicolumn{3}{r}{\textit{Continued on next page}} \\
\endfoot

\bottomrule
\endlastfoot

Server & Q\&A nodes, branch structures, and conversation-level Merkle root $\text{MTR}_{\text{con}}$ & Persistent \\
Server & Deletion-state records ($\text{MTR}_{\text{con}}^{\text{old}}$, $\text{MTR}_{\text{con}}^{\text{del}}$, $\text{timestamp}_{\text{del}}$) & Persistent\\
Server & Current account-level Merkle root $\text{MTR}_{\text{acc}}$ and dual signatures ($\text{STR}_{\text{acc}}^{\text{s}}$, $\text{STR}_{\text{acc}}^{\text{u}}$) & Persistent \\
Server & User public key $\text{pk}_u$ and server signing private key $\text{sk}_s$ & Persistent \\
Server & Complete $\text{MTR}_{\text{acc}}$ state-transition chain or incremental log & Persistent \\
Server & Last confirmed account root and synchronization boundary ($\text{timestamp}_{\text{sig}}$, $\text{timestamp}_{\text{update}}$) for each device & Persistent \\
Server & Merkle intermediate nodes & Ephemeral Cache \\
User device & Most recently verified $\text{MTR}_{\text{acc}}$ and dual signatures ($\text{STR}_{\text{acc}}^{\text{s}}$, $\text{STR}_{\text{acc}}^{\text{u}}$) on the device & Persistent \\
User device & Device synchronization boundary ($\text{timestamp}_{\text{sig}}$, $\text{timestamp}_{\text{update}}$) & Persistent \\
User device & Server public key $\text{pk}_s$ & Persistent \\
User device & Local key material $K_u$, $\text{salt}$, $\text{iterations}$ & Persistent \\
User device & User signing private key $\text{sk}_u$ and gossip communication key $K_{\text{gossip}}$ & Derivable on demand or stored securely \\
\end{longtable}

\subsection{Cryptographic Preliminaries and Assumptions}

VCT is constructed from standard cryptographic primitives. Public keys used for verification are assumed to be authentically bound to their claimed identities. Unless stated otherwise, adversaries are modeled as probabilistic polynomial-time (PPT) algorithms.

\textbf{Hash Function.} Let $\mathsf{Hash}(\cdot)$ denote a cryptographic hash function. It is assumed that $\mathsf{Hash}(\cdot)$ satisfies collision resistance and second-preimage resistance. VCT uses hash values to bind protocol objects and their dependencies for integrity. Under these assumptions, any modification to authenticated content results in a change to the corresponding hash value, except with negligible probability. To prevent semantic ambiguity across distinct protocol objects, explicit domain-separation tags are incorporated into the relevant hash inputs.

\textbf{Merkle Authentication Structure.} VCT employs Merkle trees as an authenticated set structure. Assuming the underlying hash function satisfies collision resistance, a Merkle root serves as a compact authenticated commitment to a set state, and a Merkle inclusion proof reliably establishes that a given object is committed under the corresponding root. Merkle trees provide only set authentication and membership proofs; the legitimacy of state transitions is governed by the specific protocol definitions given in subsequent sections.

\textbf{Digital Signature.} Let $\sigma \leftarrow \mathsf{Sign}(sk_x,m)$ denote the digital signature produced by principal $x$ over message $m$ using private key $sk_x$, and let $\mathsf{Verify}(pk_x,m,\sigma)$ denote the corresponding verification algorithm. The signature scheme is assumed to satisfy existential unforgeability under chosen-message attacks (EUF-CMA)\cite{goldwasser1988digital}. 
VCT uses signatures to confirm protocol outputs such as account states and sharing snapshots. Under this assumption, an adversary cannot forge a valid signature on any message that has not been signed by the corresponding party, and any modification to the content of a signed message causes verification to fail.

\textbf{Key Derivation.} VCT assumes the availability of a secure key derivation function $\mathsf{KDF}(\cdot)$.

\textbf{Authenticated Inter-Device Channel.} Devices belonging to the same user are assumed to be able to establish authenticated and encrypted channels. Such channels provide confidentiality, integrity, and communicating-party authentication for account states and associated metadata exchanged across devices.

The primitives and security assumptions above constitute the cryptographic foundation of VCT. Subsequent sections present the protocol construction on this basis and formally analyze its security properties, including account-state integrity, multi-device consistency detection, share verifiability, and non-repudiation.

\subsection{Threat Model and Security Goals}

LLM conversation records differ fundamentally from conventional streaming session logs. Interactions between a user and an LLM exhibit a strict Q\&A pairing structure: each prompt $Q$ corresponds to exactly one response $A$. A $(Q, A)$ pair is treated as a complete node and committed to the data structure only after the server successfully returns the corresponding response $A$; if the server fails to return a valid response, the prompt $Q$ does not enter the persistent record and no verifiable state transition is formed.

\subsubsection{Threat Model}

Both the user and the server are assumed to be capable of deviating from the protocol, while the underlying cryptographic primitives are assumed to satisfy standard security assumptions. The threat sources considered in this work fall into the following four categories.

\begin{enumerate}
\item \textbf{User Framing the Server.} A malicious user may selectively disclose conversation content or fabricate contextual dependencies in order to attribute erroneous outputs to the server, or may attempt to repudiate previously confirmed conversation states and sharing snapshots.

\item \textbf{Server Tampering with Conversation Records.} A malicious server may forge, delete, or modify user inputs, model outputs, associated metadata, or historical account states.

\item \textbf{Inconsistent Multi-Device Views.} In a multi-device setting under the same account, a malicious server may present incompatible account states to different devices belonging to the same user, or may return a stale but previously valid $\text{MTR}_{\text{acc}}$ when a device initiates a synchronization update, and subsequently deliver omitted increments to evade consistency detection.

\item \textbf{Tampering with Sharing Links or Exported Content.} Since sharing links, PDFs, and other export packages may be generated by the server, a malicious server may substitute, reorder, delete, or omit Q\&A content that the user originally intended to share during the generation process, leaving the recipient $R$ unable to determine the completeness and authenticity of the disclosed content solely from the exported artifact.
\end{enumerate}

This work does not address model-parameter-level unlearning, training-data deletion, hardware-backed key storage, key revocation and rotation, or cross-device key-compromise recovery. These concerns are orthogonal and may be addressed in conjunction with the mechanisms proposed here, but fall outside the scope of the security goals of this work.

\subsubsection{Security Goals}

In response to the threat model above, the following four categories of security goals are defined and targeted.

\begin{enumerate}
\item \textbf{Integrity.}

\begin{itemize}
\item \textit{Node-level integrity:} The content of an individual Q\&A node cannot be tampered with without detection. Any modification breaks the $\text{parent}_{\text{hash}}$ hash-chain invariant or alters the branch $\text{hash}_{\text{tail}}$, propagating the change to $\text{MTR}_{\text{acc}}$.

\item \textit{Conversation-level integrity:} The set of all branches within a conversation cannot be added to, deleted from, or modified without detection. Any change to a branch alters $\text{MTR}_{\text{con}}$, which propagates to $\text{MTR}_{\text{acc}}$.

\item \textit{Account-level integrity:} The set of conversations within an account cannot be added to, deleted from, or modified without detection. Any change to a conversation alters $\text{MTR}_{\text{acc}}$, which is anchored by both $\text{STR}_{\text{acc}}^{\text{s}}$ and $\text{STR}_{\text{acc}}^{\text{u}}$ through dual signatures.
\end{itemize}

\item \textbf{Consistency.}

\begin{itemize}
\item \textit{Update authenticity:} Account-state updates resulting from user interactions on any device can be verified through the $\text{update\_payload}$ and $\text{MerkleProofs}$ returned by the server, confirming that the post-update account state $\text{MTR}_{\text{acc}}^{t+1}$ claimed by the server was derived from the local anchor $\text{MTR}_{\text{acc}}^{t}$ via a legitimate state transition.

\item \textit{Deletion serialization and deletion-epoch consistency:} $\textsc{DeleteSession}$ is modeled as an account-level serialized state transition that must be submitted against the server's current latest account root; non-deletion updates must be submitted against an account state that has already witnessed the latest deletion. This prevents undecidable merge semantics between deletion states and concurrent non-deletion updates.

\item \textit{Merge determinism:} The merge mechanism for non-deletion updates across multiple devices satisfies inclusiveness, determinism, and conflict-free preservation; concurrent modifications within the same conversation are retained as branches. Merge results are confirmed by dual signatures from both parties.

\item \textit{Incremental-replay rejection and gossip accountability:} Before accepting and signing the server-returned concurrent merge result, each device performs an incremental-freshness check to prevent the server from first returning a stale account state during an update and subsequently delivering omitted increments during synchronization to evade view-consistency detection. After all devices complete synchronization and gossip, they should converge to the same account state; if the server provides inconsistent account views, VCT enables the user side to produce verifiable fork evidence for accountability.
\end{itemize}

\item \textbf{Share Verifiability.}

\begin{itemize}
\item \textit{Sharing package authenticity:} Shared Q\&A nodes must first be proven, via $\text{MerkleProofs}$, to be anchored to a signed $\text{MTR}_{\text{acc}}$.

\item \textit{Sharing snapshot integrity:} Shared content is reorganized into an independent hash chain and confirmed by dual signatures from both the sharer and the server. The recipient $R$ can detect any substitution, reordering, deletion, or omission of shared content by recomputing the hash chain and verifying the dual signatures. Subsequent updates or deletions to the original conversation do not affect the verifiability of an already-generated sharing package.
\end{itemize}

\item \textbf{Non-repudiation.}

\begin{itemize}
\item \textit{Server non-repudiation:} The server's signature $\text{STR}_{\text{acc}}^{\text{s}}$ over $\text{MTR}_{\text{acc}}$ and its signature $\text{share}_{\text{sig}}^{\text{s}}$ over the independent share-chain tail $\text{share}_{\text{tail}}$ are non-repudiable.

\item \textit{User non-repudiation:} The user's signature $\text{STR}_{\text{acc}}^{\text{u}}$ over $\text{MTR}_{\text{acc}}$ and $\text{share}_{\text{sig}}^{\text{u}}$ over the independent share-chain tail $\text{share}_{\text{tail}}$ are non-repudiable.

\item \textit{Operation-content non-repudiation:} All signatures are directly bound to operation outcomes ($\text{MTR}_{\text{acc}}$ or $\text{share}_{\text{tail}}$); any malicious modification causes signature verification to fail.
\end{itemize}
\end{enumerate}

\section{Proposed Verifiable Conversation Transcript System}

\subsection{Verifiable Three-Layer Data Structure}

LLM conversations exhibit a strict Q\&A pairing structure; accordingly, a single user request together with its corresponding model response is treated as an atomic node for storage purposes. Each node contains the following fields.

\begin{longtable}{p{3.5cm} >{\raggedright\arraybackslash}p{12.0cm}}
\caption{Node Fields and Description}
\label{tab:node_fields}\\
\toprule
\textbf{Node Field} & \textbf{Description} \\
\midrule
\endfirsthead

\caption*{Table \ref{tab:node_fields}: Node Fields and Description (continued)}\\
\toprule
\textbf{Node Field} & \textbf{Description} \\
\midrule
\endhead

\midrule
\multicolumn{2}{r}{\textit{Continued on next page}} \\
\endfoot

\bottomrule
\endlastfoot

$\text{hash}_{\text{parent}}$ & Hash value of the parent node in a hash chain, used to form the hash chain. \\
$Q$ & The prompt submitted by the user. \\
$A$ & The response returned by the LLM. \\
$\text{model}_{\text{config}}$ & Model determinism parameters, including $\text{model}_{\text{id}}$, $\text{nonce}$, $\text{temperature}$, and related fields. \\
$\text{file\_aux\_info}$ & A composite data object comprising uploaded files and their associated auxiliary information. \\
$\text{timestamp}$ & Node creation timestamp. \\
\end{longtable}

\begin{figure}[htbp]
\centering
\includegraphics[width=0.8\textwidth]{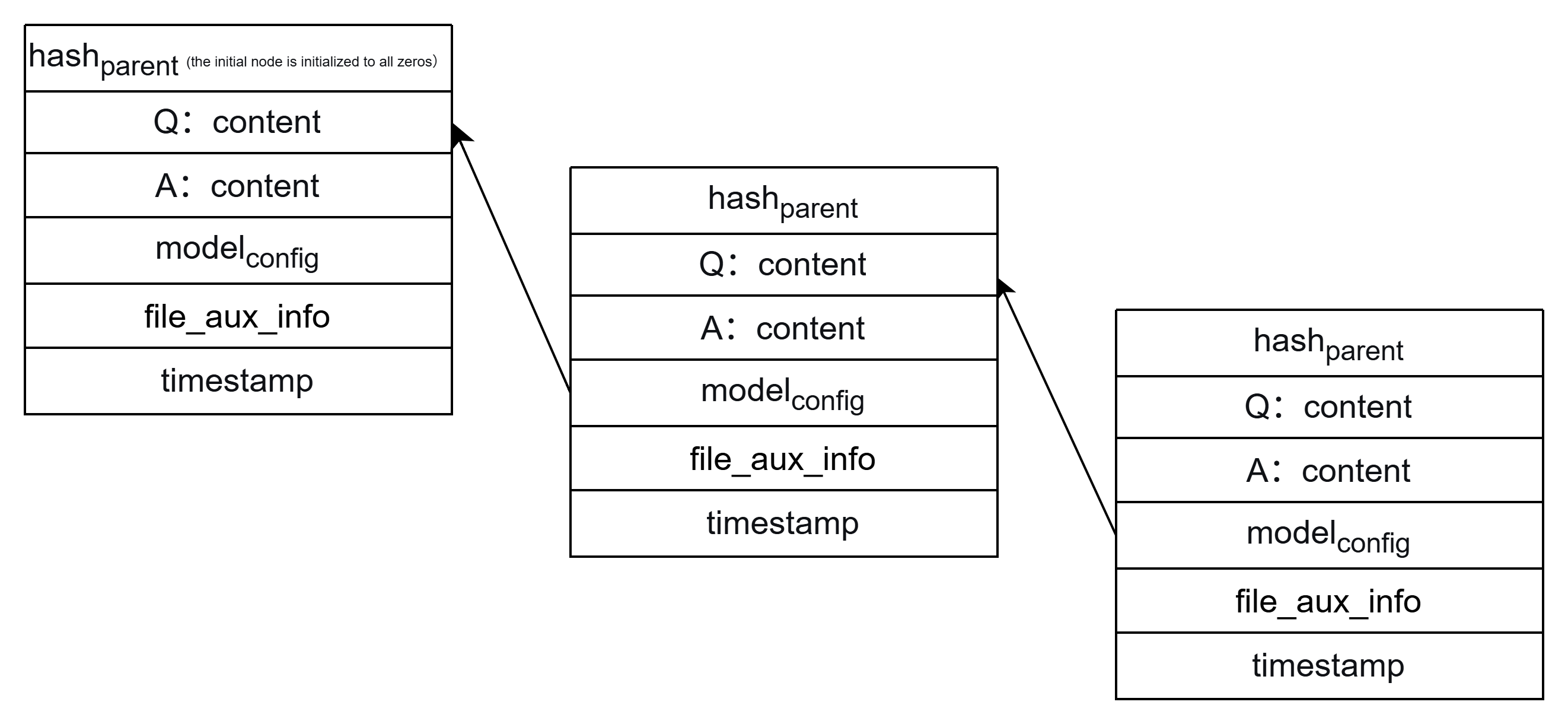}
\caption{Hash chain structure of Q\&A nodes in the interaction layer}
\label{fig:Q&A nodes}
\end{figure}

The node hash is defined as:

\[
\text{node}_{\text{hash}} = \text{Hash}(\text{hash}_{\text{parent}} \| Q \| A \| \text{model}_{\text{config}} \| \text{file\_aux\_info} \| \text{timestamp}).
\]

Each node explicitly stores only its parent hash $\text{hash}_{\text{parent}}$ and does not additionally store its own $\text{node}_{\text{hash}}$; the hash of a node is carried by the $\text{hash}_{\text{parent}}$ field of its child. The $\text{hash}_{\text{parent}}$ of the initial node in a conversation is set to the all-zero value, and the tail hash $\text{hash}_{\text{tail}}$ of a branch node is recomputed on demand as needed. This design trades a single hash computation for a reduction in global storage overhead without affecting the trustworthiness of $\text{hash}_{\text{tail}}$, since $\text{hash}_{\text{tail}}$ participates as a leaf in the next-layer Merkle tree assertion at every state update.

Intermediate hash nodes in the Merkle tree store only intermediate values of the form $\text{Hash}(h_i \| h_j)$. The conversation-level root $\text{MTR}_{\text{con}}$ is constructed from the $\text{hash}_{\text{tail}}$ values of all branches within the same conversation, and the account-level root $\text{MTR}_{\text{acc}}$ is constructed from the $\text{MTR}_{\text{con}}$ values of all conversations under the same account.

Because users may regenerate a response $A$ or re-edit a prompt $Q$, conversation records naturally give rise to branching. A linear hash chain cannot fully represent this structure; the system therefore adopts a three-layer verifiable storage structure.

\textbf{Layer 1: Hash-Chain Layer.} Each Q\&A pair constitutes a node, and nodes are linked via $\text{hash}_{\text{parent}}$ pointers. When a $\textsc{Regenerate}$ or $\textsc{EditQ}$ operation is performed, the original hash chain forks into multiple independent branches at the point of divergence. All branches share the assertion established before the fork and extend independently thereafter. The branch-tail assertion is defined as:

\[
\text{hash}_{\text{tail}} = \text{node}_{\text{hash}_n} = \text{Hash}(\text{hash}_{\text{parent}_{n-1}} \| Q \| A \| \text{model}_{\text{config}} \| \text{file\_aux\_info} \| \text{timestamp}).
\]

\textbf{Layer 2: Conversation Layer.} The $\text{hash}_{\text{tail}}$ values of all branches within the same conversation are used as leaves to construct a conversation-level Merkle tree whose root is $\text{MTR}_{\text{con}}$.

\textbf{Layer 3: Account Layer.} The $\text{MTR}_{\text{con}}$ values of all conversations under the same account are used as leaves to construct an account-level Merkle tree whose root is $\text{MTR}_{\text{acc}}$. The account root is signed separately by the user and the server:

\[
\text{STR}_{\text{acc}}^{\text{s}} \leftarrow \text{SIGN}_{\text{sk}_s}(\texttt{"ACCOUNT\_STATE"} \| \text{MTR}_{\text{acc}} \| \text{timestamp}),
\]
\[
\text{STR}_{\text{acc}}^{\text{u}} \leftarrow \text{SIGN}_{\text{sk}_u}(\texttt{"ACCOUNT\_STATE"} \| \text{MTR}_{\text{acc}} \| \text{timestamp}).
\]

\begin{figure}[htbp]
\centering
\includegraphics[width=0.8\textwidth]{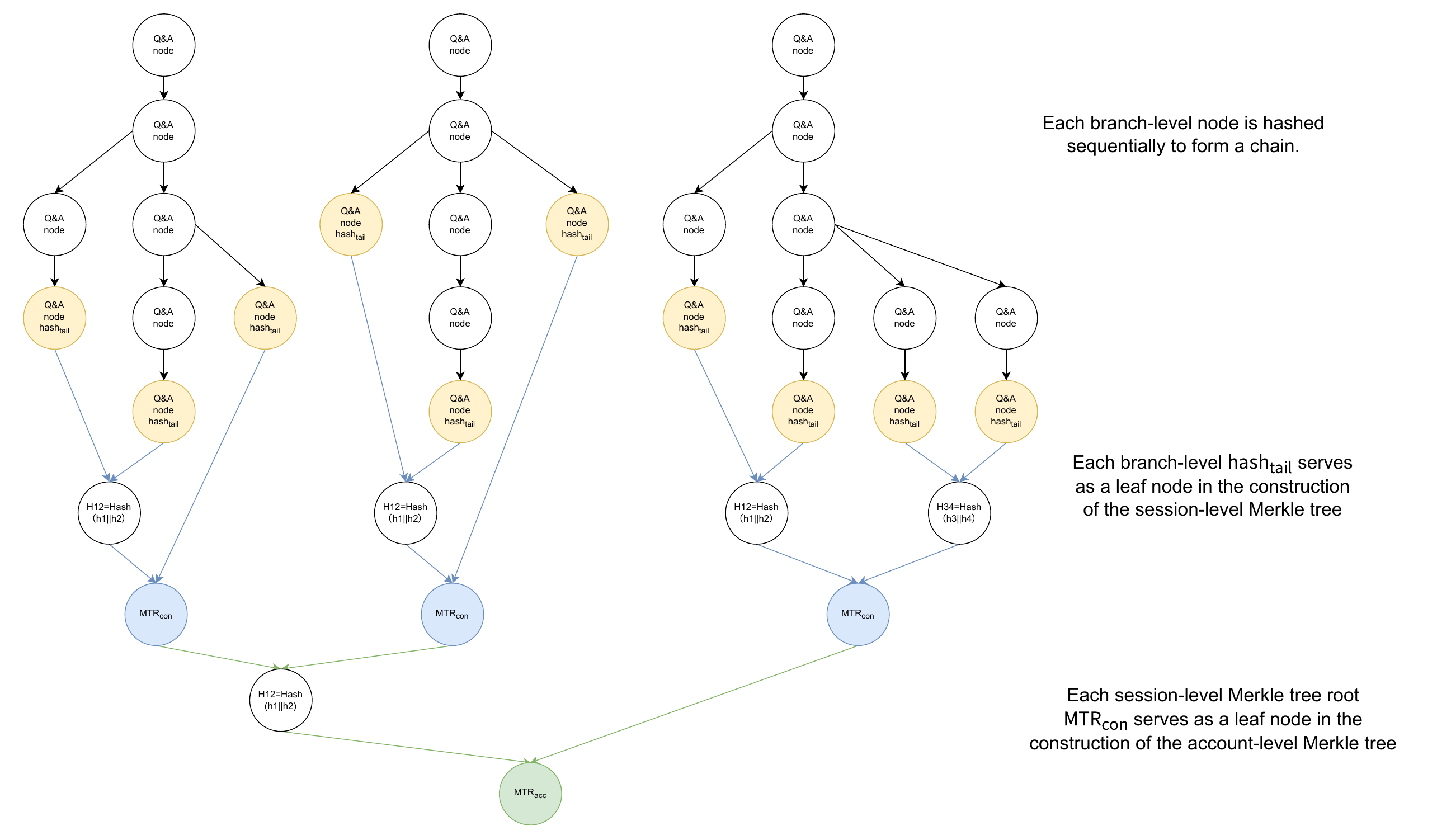}
\caption{Three-Layer Authenticated Data Structure of VCT}
\label{fig:account_root}
\end{figure}

The account-level Merkle root $\text{MTR}_{\text{acc}}$ and the corresponding user and server signatures $\text{STR}_{\text{acc}}^{\text{u}}$ and $\text{STR}_{\text{acc}}^{\text{s}}$ are publicly anchored.

\subsection{VCT System Modules}

\subsubsection{Key Generation and Device Registration}

To support authenticatable multi-device interaction and conversation-record signing, the system adopts a deterministic key derivation scheme.

\textbf{1. User Key and Multi-Device Consistency.}
Each user is associated with a high-entropy root key $K_u$, which is generated by the user from a random source at account initialization and stored securely on a trusted user device. The user additionally holds a login password $\text{pwd}_u$. The user key seed is derived jointly from $K_u$ and $\text{pwd}_u$ via a key derivation function~\cite{kaliski2000pkcs5}:

\[
\text{seed}_u \leftarrow \text{KDF}(K_u,\ \text{pwd}_u,\ \text{salt},\ \text{iterations},\ \texttt{"user\_signing"},\ \text{key\_len}),
\]

where $K_u$ is the root key generated and securely retained by the user; $\text{pwd}_u$ is the login password known only to the user; $\text{salt}$ is stored on the user side together with $K_u$; $\text{iterations}$ denotes the number of slow-stretching iterations; $\texttt{"user\_signing"}$ serves as a domain-separation tag; and $\text{key\_len}$ specifies the byte length of the KDF output seed.

The signing key pair is subsequently derived from this seed:

\[
(\text{sk}_u,\ \text{pk}_u) \leftarrow \text{DeriveKeyPair}(\text{seed}_u).
\]

This stateless derivation mechanism allows all devices holding the same $K_u$ and $\text{pwd}_u$ to derive identical $\text{sk}_u$ and $\text{pk}_u$, thereby eliminating the need for device-specific key registration and explicit key synchronization. Signatures produced by any device can be verified under the same user public key, and a verifier need only maintain a single public key per user.

The user employs $\text{sk}_u$ to produce account-level and share-level signed assertions:

\[
\text{STR}_{\text{acc}}^{\text{u}} \leftarrow \text{SIGN}_{\text{sk}_u}(\texttt{"ACCOUNT\_STATE"} \| \text{MTR}_{\text{acc}} \| \text{timestamp}),
\]
\[
\text{share}_{\text{sig}}^{\text{u}} \leftarrow \text{SIGN}_{\text{sk}_u}(\texttt{"SHARE\_SNAPSHOT"} \| \text{share}_{\text{tail}} \| \text{timestamp}).
\]

\textbf{2. Server Key Generation.}
A long-term key pair $(\text{sk}_s,\ \text{pk}_s)$ is generated at system initialization. The server uses $\text{sk}_s$ to sign the account states and sharing data it publishes:

\[
\text{STR}_{\text{acc}}^{\text{s}} \leftarrow \text{SIGN}_{\text{sk}_s}(\texttt{"ACCOUNT\_STATE"} \| \text{MTR}_{\text{acc}} \| \text{timestamp}),
\]
\[
\text{share}_{\text{sig}}^{\text{s}} \leftarrow \text{SIGN}_{\text{sk}_s}(\texttt{"SHARE\_SNAPSHOT"} \| \text{share}_{\text{tail}} \| \text{timestamp}).
\]

These signatures provide publicly verifiable and non-repudiable assertions over the data published by the server.

\textbf{3. New Device Registration.}
When a new device joins, it obtains the user root key $K_u$ and the key derivation parameters from an existing trusted device via a secure out-of-band channel such as a QR code, while the login password $\text{pwd}_u$ is entered locally on the new device by the user. The new device then computes:

\[
\text{seed}_u \leftarrow \text{KDF}(K_u,\ \text{pwd}_u,\ \text{salt},\ \text{iterations},\ \texttt{"user\_signing"},\ \text{key\_len}),
\]
\[
(\text{sk}_u,\ \text{pk}_u) \leftarrow \text{DeriveKeyPair}(\text{seed}_u).
\]

In this process, $\text{pwd}_u$ is never transmitted between devices; consequently, intercepting $K_u$ alone or compromising $\text{pwd}_u$ alone is insufficient to derive $\text{sk}_u$.

\subsubsection{Conversation State Update}

Conversation updates are modeled as deterministic state transitions. Each time a user initiates an update, the server computes the next state $\text{state}_{t+1}$ from the current account state $\text{state}_t$ and recomputes the account-level Merkle root $\text{MTR}_{\text{acc}}$. The system supports four classes of operations:

\begin{itemize}
\item \text{NewSession}: A new independent hash chain is initialized, and the new conversation root is inserted into the account-level Merkle tree.

\item \text{Append}: A new Q\&A node is appended to the tail of an existing branch within a conversation, and the corresponding $\text{hash}_{\text{tail}}$ is updated.

\item \text{Branch}: A new branch is created from an already-authenticated historical node, for instance when regenerating response $A$ or re-editing prompt $Q$.

\item \text{DeleteSession}: The target conversation is replaced by a deletion-state conversation root; its plaintext content is hidden or removed in accordance with applicable legal requirements, a behavior that constitutes a storage-layer policy and does not alter the definition of the authenticated state transition.
\end{itemize}

All of the above operations are completed through a unified confirmation protocol. Algorithm 1 integrates server-side state transition, proof generation, user-side proof verification, dual-party signing, and final storage into a single procedure. The server first checks whether the baseline account root submitted by the device constitutes a legitimate starting point for the current state transition; upon a successful check, it executes the corresponding state transition. The server then generates $\text{update\_payload}$ and $\text{MerkleProofs}$ from the pre- and post-transition account states and signs the new account root. The user device, using the local account anchor as its baseline, verifies the state-transition proof against the $\text{update\_payload}$ and $\text{MerkleProofs}$, and performs a receive-side freshness check where required. Upon successful verification, the user produces a signature over the same account root, and the server finally verifies the user signature and commits the new account state. The specific conditions checked are determined by the operation type and synchronization scenario.

\begin{algorithm}
\caption{\text{VerifiedConversationUpdate}}
\label{alg:update}
\begin{algorithmic}[1]
\State \textbf{Input:} $op \in \{\text{NewSession},\text{Append},\text{Branch},\text{DeleteSession},\text{Merge}\}$;   $\text{anchor}_t = (\text{MTR}_{\text{acc}_t},\ \text{STR}_{\text{acc}_t}^{\text{s}},\ \text{STR}_{\text{acc}_t}^{\text{u}})$;
$\text{sync\_boundary} = (\text{timestamp}_{\text{sig}},\ \text{timestamp}_{\text{update}})$; 
\State \textbf{Output:} $\text{state}_{t+1},\text{MTR}_{\text{acc}_{t+1}},\text{STR}_{\text{acc}_{t+1}}^{\text{s}},\text{STR}_{\text{acc}_{t+1}}^{\text{u}}$

\ =====Phase 1: State Transition and Proof Generation (Server-side)=====

\State server checks $\text{CheckUpdateAdmissibility}(op,\text{MTR}_{\text{acc}}^{\text{base}},\text{state}_t)$
\State server checks $\textsc{CheckUpdateAdmissibility}(op,\text{MTR}_{\text{acc}}^{\text{base}},\text{state}_t)$
\If{check fails} reject and return required info \EndIf
\State \textbf{switch} $op$
\State \hspace{\algorithmicindent} \textbf{case} $\text{NewSession}$: $\text{state}_{t+1} \leftarrow \text{NewSession}(\text{state}_t,\text{input})$
\State \hspace{\algorithmicindent} \textbf{case} $\textsc{Append}$: $\text{state}_{t+1} \leftarrow \text{AppendMessage}(\text{state}_t,\text{input})$
\State \hspace{\algorithmicindent} \textbf{case} $\textsc{Branch}$: $\text{state}_{t+1} \leftarrow \text{BranchMessage}(\text{state}_t,\text{input})$
\State \hspace{\algorithmicindent} \textbf{case} $\text{DeleteSession}$: $\text{state}_{t+1} \leftarrow \textsc{DeleteSession}(\text{state}_t,\text{input})$
\State \hspace{\algorithmicindent} \textbf{case} $\textsc{Merge}$: $\text{state}_{t+1} \leftarrow \text{MergeStates}(\text{state}_t,\text{input}.\mathcal{U})$

\State $\text{MTR}_{\text{acc}_{t+1}}\leftarrow\text{ComputeAccountRoot}(\text{state}_{t+1})$
\State $\text{update\_payload}\leftarrow\text{ExtractUpdatePayload}(op,\text{state}_t,\text{state}_{t+1},\text{input})$
\State $\text{MerkleProofs}\leftarrow\text{GenerateTransitionProofs}(op,\text{MTR}_{\text{acc}}^{\text{base}},\text{MTR}_{\text{acc}_{t+1}},\text{update\_payload})$
\State $\text{STR}_{\text{acc}_{t+1}}^{\text{s}}\leftarrow\mathsf{Sign}_{sk_s}(\texttt{"ACCOUNT\_STATE"}\|\text{MTR}_{\text{acc}_{t+1}}\|\text{timestamp})$
\State \textbf server sends $(\text{update\_payload},\text{MTR}_{\text{acc}}^{\text{base}},\text{MTR}_{\text{acc}_{t+1}},\text{timestamp},\text{MerkleProofs},\text{STR}_{\text{acc}_{t+1}}^{\text{s}})$

\ =====Phase 2: Proof Verification and Freshness Check (User-side)=====

\State user checks $\text{MTR}_{\text{acc}}^{\text{base}}=\text{locally stored }\text{MTR}_{\text{acc}_t}$
\If{check fails} reject and abort \EndIf
\State $\tau_{\text{nodes}}\leftarrow\text{ExtractRemoteNodeTimestamps}(\text{update\_payload})$
\State user performs $\text{IncrementalReplayFreshnessCheck}(\tau_{\text{nodes}},\text{timestamp}_{\text{sig}},\text{timestamp}_{\text{update}})$
\If{check fails} reject and abort \EndIf
\State user performs $\text{VerifyTransitionProof}(op,\text{MTR}_{\text{acc}}^{\text{base}},\text{MTR}_{\text{acc}_{t+1}},\text{update\_payload},\text{MerkleProofs})$
\If{verification fails} reject and abort \EndIf
\State verify $\text{STR}_{\text{acc}_{t+1}}^{\text{s}}$ using $pk_s$ \If{fails} reject and abort \EndIf

\ ===============Phase 3: User Commitment==============
\State$\text{STR}_{\text{acc}_{t+1}}^{\text{u}}\leftarrow\mathsf{Sign}_{sk_u}(\texttt{"ACCOUNT\_STATE"}\|\text{MTR}_{\text{acc}_{t+1}}\|\text{timestamp})$
\State user sends $(\text{MTR}_{\text{acc}_{t+1}},\text{timestamp},\text{STR}_{\text{acc}_{t+1}}^{\text{u}})$ to server

\ ==============Phase 4: Server Confirmation==============
\State$\text{STR}_{\text{acc}_{t+1}}^{\text{u}}$ using $pk_u$ \If{fails} reject and abort \EndIf
\State server stores $(\text{MTR}_{\text{acc}_{t+1}},\text{STR}_{\text{acc}_{t+1}}^{\text{s}},\text{STR}_{\text{acc}_{t+1}}^{\text{u}})$; send \textit{success}

\ =================Phase 5: Finalization=================
\State$(\text{MTR}_{\text{acc}_{t+1}},\text{STR}_{\text{acc}_{t+1}}^{\text{s}},\text{STR}_{\text{acc}_{t+1}}^{\text{u}})$
\State \Return $\text{state}_{t+1},\text{MTR}_{\text{acc}_{t+1}},\text{STR}_{\text{acc}_{t+1}}^{\text{s}},\text{STR}_{\text{acc}_{t+1}}^{\text{u}}$
\end{algorithmic}
\end{algorithm}

The following presents the four basic state transitions: \text{NewSession}, \text{Append}, \text{Branch}, and \text{DeleteSession}.

\paragraph{Case 1: New Session Creation.}
Upon creating a new session, the system generates an initial Q\&A node whose $\text{parent\_hash}$ is set to the all-zero value; the node hash serves as the $\text{hash}_{\text{tail}}$ of the new branch. This $\text{hash}_{\text{tail}}$ is used to construct a new conversation-level root $\text{MTR}_{\text{con}}$, which is then inserted as a new leaf into the account-level Merkle tree. This process corresponds to a leaf-append operation on the account-level Merkle tree.

\begin{algorithm}
\caption{\text{NewSession}$(\text{state},Q\&A))$}
\label{alg:newsession}
\begin{algorithmic}[1]
\State \textbf{Input:} current account state and initial Q\&A data
\State \textbf{Output:} updated state

\State create node $N$ with $\text{parent\_hash} \leftarrow \mathbf{0}$
\State compute $\text{node\_hash} \leftarrow \mathsf{Hash}(\text{parent\_hash} \| Q \| A \| \text{model\_config} \| \text{file\_aux\_info} \| \text{timestamp})$
\State create a new session containing a single branch whose tail is $\text{node\_hash}$
\State update the session root and account state
\State \Return updated state
\end{algorithmic}
\end{algorithm}

\begin{figure}[htbp]
\centering
\includegraphics[width=0.8\textwidth]{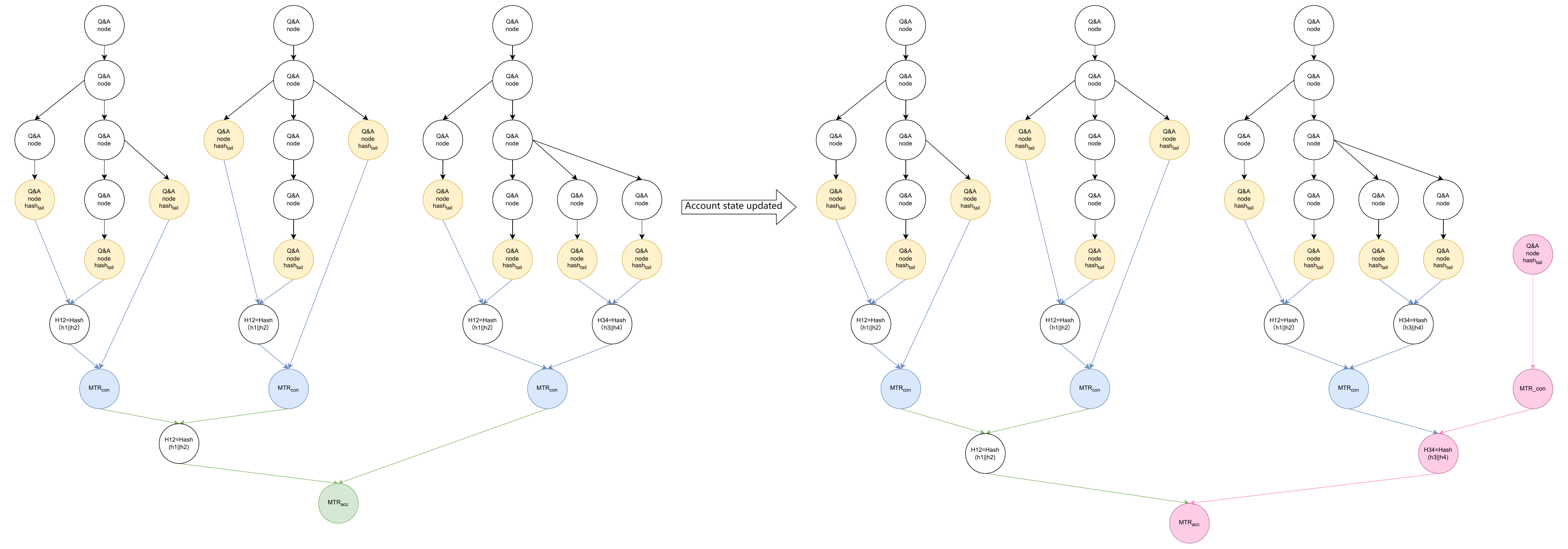}
\caption{ Account-State Transition for New Session Creation}
\label{fig: New Session Creation}
\end{figure}

\textbf{User-side proof verification.} 
Let $\text{MTR}_{\text{acc}_t}$ and $\text{MTR}_{\text{acc}_{t+1}}$ denote the account roots before and after the update, respectively. The server returns the newly created Q\&A node, the updated account root $\text{MTR}_{\text{acc}_{t+1}}$, and the corresponding $\text{MerkleProofs}$. The initial node hash is recomputed from $\text{parent\_hash} = \mathbf{0}$, $Q$, $A$, $\text{model\_config}$, $\text{file\_aux\_info}$, and $\text{timestamp}$, and the result is taken as the $\text{hash}_{\text{tail}}$ of the sole branch in the new conversation:

\[
\text{hash}_{\text{tail}} = \mathsf{Hash}(\text{parent\_hash} \| Q \| A \| \text{model\_config} \| \text{file\_aux\_info} \| \text{timestamp}).
\]

Since the conversation contains exactly one branch at this point, the new conversation root $\text{MTR}_{\text{con\_new}}$ is derived directly from $\text{hash}_{\text{tail}}$. The $\text{MerkleProofs}$ are then used to verify the append-style account-state transition from $\text{MTR}_{\text{acc}_t}$ to $\text{MTR}_{\text{acc}_{t+1}}$: all pre-existing conversation roots must remain consistent with the locally anchored account state, and $\text{MTR}_{\text{con\_new}}$ must be correctly incorporated as a new leaf in the account-level Merkle tree. If the account root recomputed along the proof path matches the server-returned $\text{MTR}_{\text{acc}_{t+1}}$, the $\textsc{NewSession}$ operation passes user-side verification. The proof size is $O(\log n)$, where $n$ is the number of conversations in the account prior to the update.

\paragraph{Case 2: Message Append.}
A message append modifies only the chain tail of the target branch. The current $\text{hash}_{\text{tail}}$ is used as the $\text{parent\_hash}$ of the new node; the new node hash is computed, and the corresponding leaf in the conversation-level Merkle tree is replaced with the updated $\text{hash}_{\text{tail}}$. The conversation root $\text{MTR}_{\text{con}}$ is then updated, and the change is propagated to the account-level root $\text{MTR}_{\text{acc}}$.

\begin{algorithm}
\caption{\text{AppendMessage}$(\text{state},\ (\text{session\_id},\ \text{branch\_id},Q\&A ))$}
\label{alg:append}
\begin{algorithmic}[1]
\State \textbf{Input:} current account state, target branch, and new Q\&A data
\State \textbf{Output:} updated state

\State locate the target branch and retrieve its current tail
\State create node $N$ with $\text{parent\_hash}$ equal to the current branch tail
\State compute $\text{node\_hash} \leftarrow \mathsf{Hash}(\text{parent\_hash} \| Q \| A \| \text{model\_config} \| \text{file\_aux\_info} \| \text{timestamp})$
\State append $N$ to the branch and replace the branch tail with $\text{node\_hash}$
\State update the session root and account state
\State \Return updated state
\end{algorithmic}
\end{algorithm}

\begin{figure}[htbp]
\centering
\includegraphics[width=0.8\textwidth]{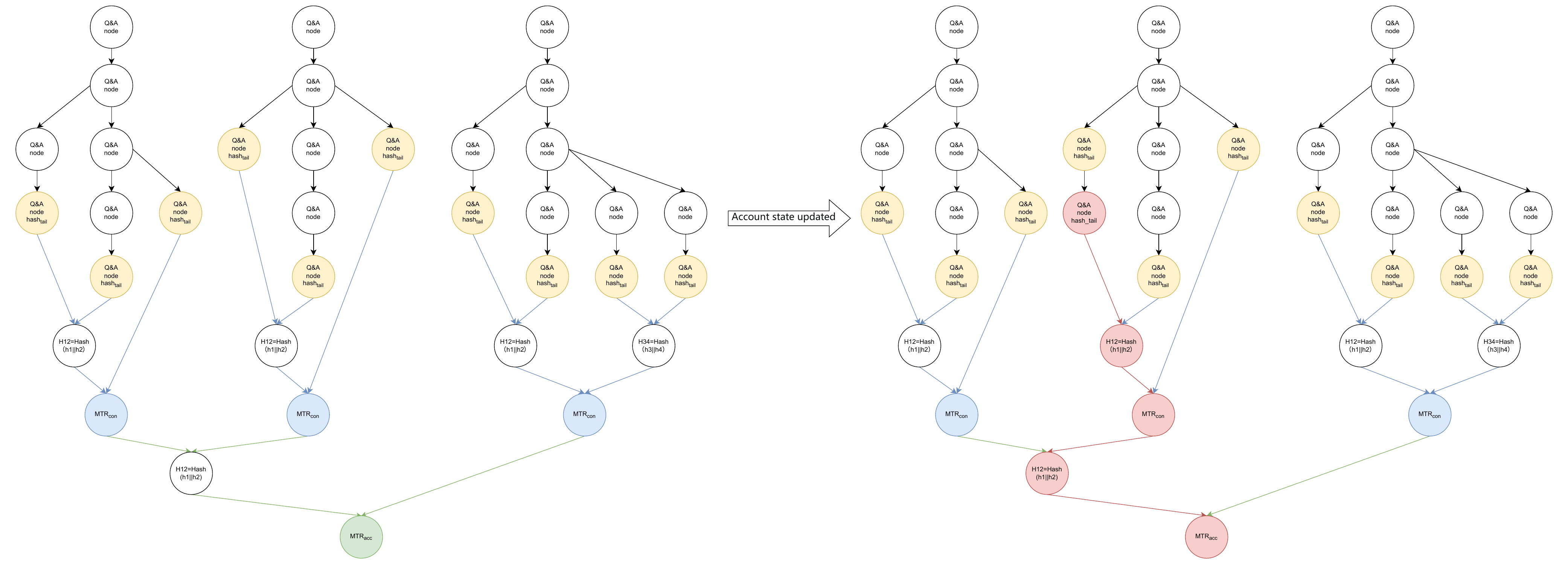}
\caption{Account-State Transition for Message Appending}
\label{fig: Message Appending}
\end{figure}

\textbf{User-side proof verification.} 
Let $\text{MTR}_{\text{con}_t}$ and $\text{MTR}_{\text{con}_{t+1}}$ denote the conversation roots before and after the update, and let $\text{MTR}_{\text{acc}_t}$ and $\text{MTR}_{\text{acc}_{t+1}}$ denote the corresponding account roots. A message append modifies only the tail of the target branch; no new conversation or branch is created. The server returns the new Q\&A node, the previous branch tail $\text{hash}_{\text{tail}}^{\text{old}}$, the updated account root $\text{MTR}_{\text{acc}_{t+1}}$, and the corresponding $\text{MerkleProofs}$.

Two checks are performed first: $\text{hash}_{\text{tail}}^{\text{old}}$ must be a valid branch leaf under $\text{MTR}_{\text{con}_t}$, and $\text{MTR}_{\text{con}_t}$ must be anchored to the locally stored account root $\text{MTR}_{\text{acc}_t}$. With $\text{parent\_hash}$ set to $\text{hash}_{\text{tail}}^{\text{old}}$, the new branch tail is then recomputed as:

\[
\text{hash}_{\text{tail}}^{\text{new}} = \mathsf{Hash}(\text{parent\_hash} \| Q \| A \| \text{model\_config} \| \text{file\_aux\_info} \| \text{timestamp}).
\]

$\text{hash}_{\text{tail}}^{\text{old}}$ is replaced by $\text{hash}_{\text{tail}}^{\text{new}}$, and $\text{MTR}_{\text{con}_{t+1}}$ is recomputed along the conversation-level proof path. $\text{MTR}_{\text{con}_t}$ is then replaced by $\text{MTR}_{\text{con}_{t+1}}$ in the account-level tree, and $\text{MTR}_{\text{acc}_{t+1}}$ is recomputed along the account-level proof path. If the recomputed account root matches the server-returned value, the append operation passes user-side verification. The proof size is $O(\log m + \log n)$, where $m$ is the number of branches in the target conversation and $n$ is the number of conversations in the account.

\paragraph{Case 3: Branch Creation.}
Adding a new branch to an existing conversation is equivalent to introducing a new hash chain. The hash of the branching-point node is used as the $\text{parent\_hash}$ of the new node; the $\text{hash}_{\text{tail}}$ of the new branch is computed and inserted as a new leaf into the conversation-level Merkle tree. This operation modifies $\text{MTR}_{\text{con}}$ and propagates the change to the account-level root $\text{MTR}_{\text{acc}}$.

\begin{algorithm}
\caption{\text{BranchMessage}$(\text{state},\ (\text{session\_id},\ \text{parent\_node},\ Q,\ A,\ \text{params}))$}
\label{alg:branch}
\begin{algorithmic}[1]
\State \textbf{Input:} current account state, branching point, and new Q\&A data
\State \textbf{Output:} updated state

\State locate and verify $\text{parent\_node}$ in the target session
\State create node $N$ with $\text{parent\_hash} \leftarrow \mathsf{Hash}(\text{parent\_node})$
\State compute $\text{node\_hash} \leftarrow \mathsf{Hash}(\text{parent\_hash} \| Q \| A \| \text{model\_config} \| \text{file\_aux\_info} \| \text{timestamp})$
\State create a new branch ending at $\text{node\_hash}$
\State update the session root and account state
\State \Return updated state
\end{algorithmic}
\end{algorithm}

\begin{figure}[htbp]
\centering
\includegraphics[width=0.8\textwidth]{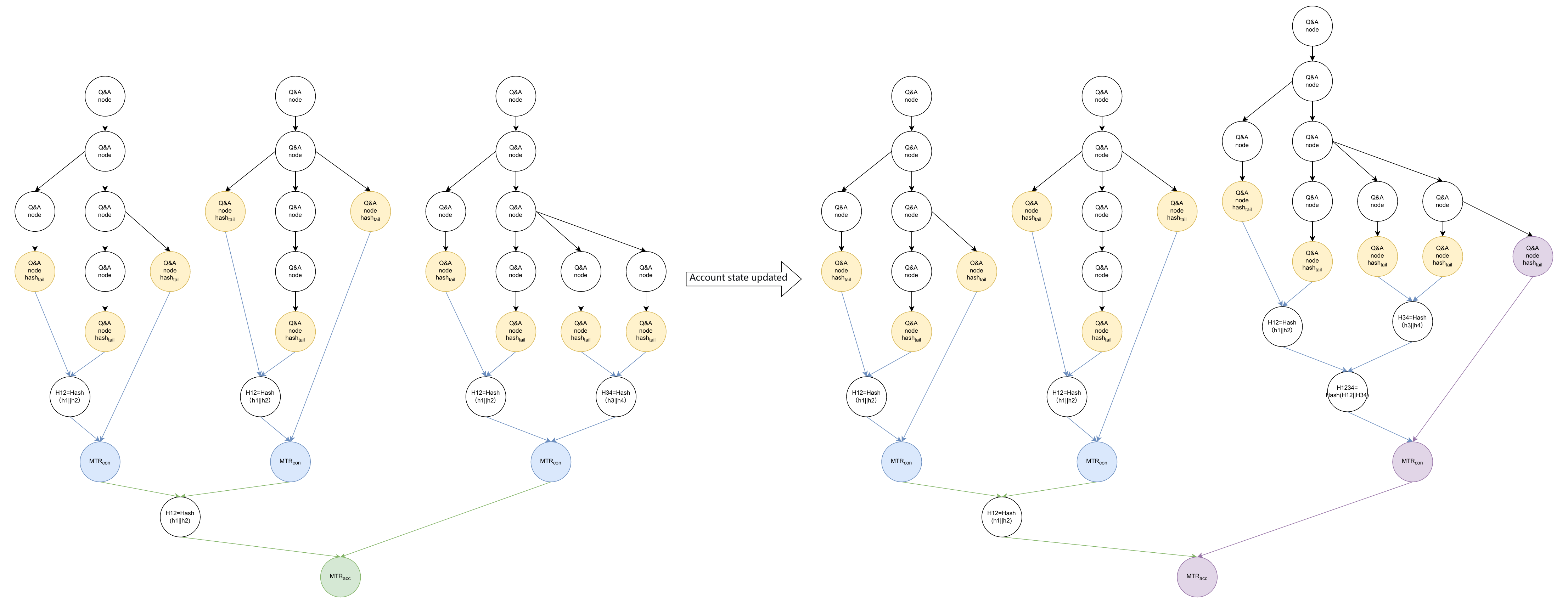}
\caption{Account-State Transition for Branch Creation}
\label{fig: Branch Creation}
\end{figure}

\textbf{User-side proof verification.} 
Branch creation does not overwrite the existing history; instead, a new branch chain is created beyond an already-authenticated historical node. The server returns the branching-point node, the new Q\&A node, the updated account root $\text{MTR}_{\text{acc}_{t+1}}$, and the corresponding $\text{MerkleProofs}$. Since the user device retains only the account-level anchor and neither the full intra-conversation hash chain nor intermediate authentication paths, the server must supply a successor-chain proof from the branching-point node to some prior branch tail $\text{hash}_{\text{tail}}^{\text{old}}$, together with the authentication paths from $\text{hash}_{\text{tail}}^{\text{old}}$ to $\text{MTR}_{\text{con}_t}$ and from $\text{MTR}_{\text{con}_t}$ to $\text{MTR}_{\text{acc}_t}$. These proofs are used to verify that the branching-point node is anchored to the pre-update conversation root $\text{MTR}_{\text{con}_t}$, which is in turn anchored to the locally stored account root $\text{MTR}_{\text{acc}_t}$.

Upon successful verification, the hash of the branching-point node is set as the $\text{parent\_hash}$ of the new node, and the new branch tail is recomputed from $Q$, $A$, $\text{model\_config}$, $\text{file\_aux\_info}$, and $\text{timestamp}$:

\[
\text{hash}_{\text{tail}}^{\text{new}} = \mathsf{Hash}(\text{parent\_hash} \| Q \| A \| \text{model\_config} \| \text{file\_aux\_info} \| \text{timestamp}).
\]

$\text{hash}_{\text{tail}}^{\text{new}}$ is then incorporated as a new branch leaf into the conversation-level Merkle tree, and $\text{MTR}_{\text{con}_{t+1}}$ is recomputed. $\text{MTR}_{\text{con}_t}$ is subsequently replaced by $\text{MTR}_{\text{con}_{t+1}}$ in the account-level tree, and $\text{MTR}_{\text{acc}_{t+1}}$ is recomputed. If the recomputed account root matches the server-returned value, the branch operation passes user-side verification. If the successor chain from the branching point to the prior branch tail has length $l$, the total verification material is $O(l + \log m + \log n)$, where $m$ denotes the number of branches in the target conversation and $n$ denotes the number of conversations in the account.

\paragraph{Case 4: Session Deletion.}
Session deletion does not remove a leaf from the account-level Merkle tree; instead, the original conversation root is replaced by a deletion-state conversation root. All other nodes in the target conversation are deleted or hidden in accordance with applicable legal requirements. Let $\text{MTR}_{\text{con}}^{\text{old}}$ denote the root of the target conversation prior to deletion. The deletion-state conversation root is defined as:

\[
\text{MTR}_{\text{con}}^{\text{del}} = \mathsf{Hash}(\texttt{"DEL\_SESSION"} \| \text{MTR}_{\text{con}}^{\text{old}} \| \text{timestamp}_{\text{del}}),
\]

where $\text{MTR}_{\text{con}}^{\text{old}}$ preserves the historical commitment of the deleted conversation, $\text{timestamp}_{\text{del}}$ records the deletion time, and $\texttt{"DEL\_SESSION"}$ serves as a domain-separation tag. The leaf corresponding to the target conversation in the account-level Merkle tree is replaced from $\text{MTR}_{\text{con}}^{\text{old}}$ to $\text{MTR}_{\text{con}}^{\text{del}}$, after which $\text{MTR}_{\text{acc}}$ is recomputed.

\begin{algorithm}
\caption{\text{DeleteSession}$(\text{state},\ \text{input})$}
\label{alg:deletesession}
\begin{algorithmic}[1]
\State \textbf{Input:} current account state and target session information
\State \textbf{Output:} updated state or rejection

\If{$\text{MTR}_{\text{acc}}^{\text{base}} \neq \text{current account root}$}
    \State reject
\EndIf
\State locate $\text{target\_session}$; if already deleted, reject
\State compute $\text{MTR}_{\text{con}}^{\text{del}} \leftarrow \mathsf{Hash}(\texttt{"DEL\_SESSION"} \| \text{MTR}_{\text{con}}^{\text{old}} \| \text{timestamp}_{\text{del}})$
\State mark the target session as deleted and record the deletion witness
\State replace $\text{MTR}_{\text{con}}^{\text{old}}$ with $\text{MTR}_{\text{con}}^{\text{del}}$ in the account state
\State \Return updated state
\end{algorithmic}
\end{algorithm}

\begin{figure}[htbp]
\centering
\includegraphics[width=0.8\textwidth]{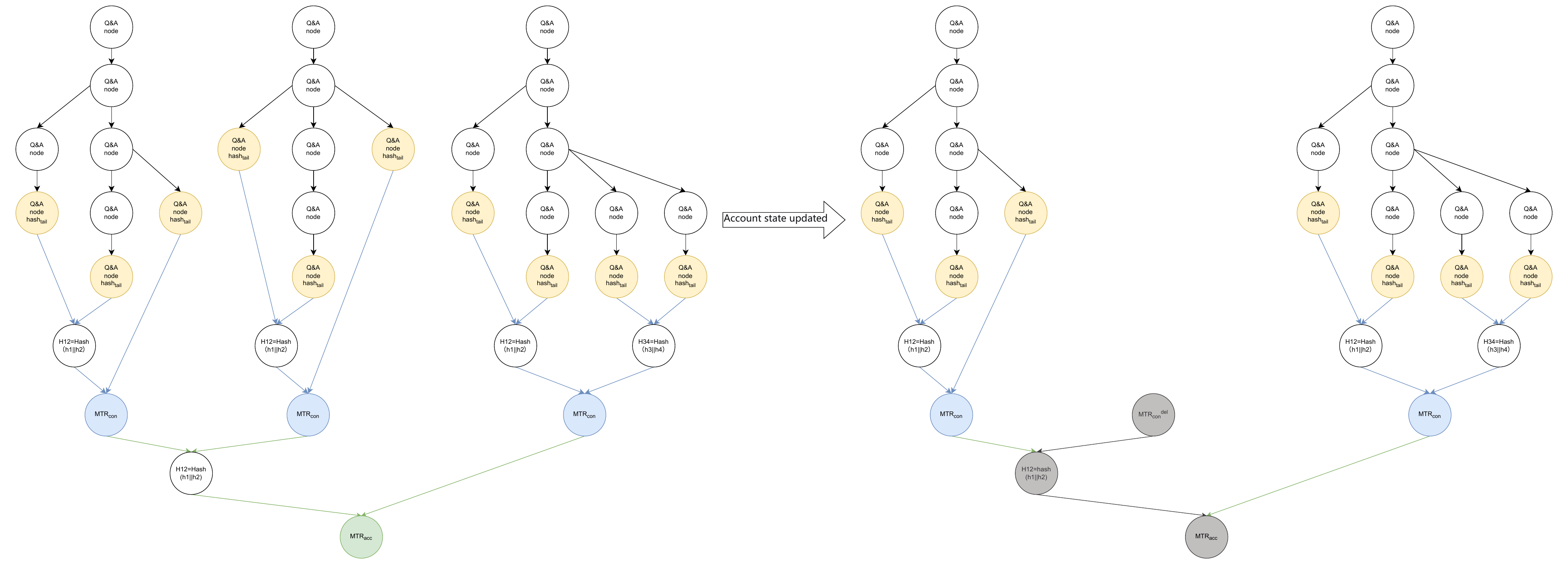}
\caption{Account-State Transition for Session Deletion}
\label{fig:Session Deletion}
\end{figure}

\textbf{User-side proof verification.} 
Session deletion is modeled in the authentication structure as an account-level leaf replacement rather than a direct removal of a leaf from the account-level Merkle tree. The server returns the pre-deletion conversation root $\text{MTR}_{\text{con}}^{\text{old}}$, the deletion timestamp $\text{timestamp}_{\text{del}}$, the deletion-state conversation root $\text{MTR}_{\text{con}}^{\text{del}}$, the updated account root $\text{MTR}_{\text{acc}_{t+1}}$, and the corresponding $\text{MerkleProofs}$.

Two checks are performed. First, $\text{MerkleProofs}$ are used to verify that $\text{MTR}_{\text{con}}^{\text{old}}$ is the current leaf for the target conversation under the locally anchored account root $\text{MTR}_{\text{acc}_t}$. Second, the deletion-state root is recomputed and verified against its definition:

\[
\text{MTR}_{\text{con}}^{\text{del}} = \mathsf{Hash}(\texttt{"DEL\_SESSION"} \| \text{MTR}_{\text{con}}^{\text{old}} \| \text{timestamp}_{\text{del}}).
\]

The account-level leaf $\text{MTR}_{\text{con}}^{\text{old}}$ is then replaced by $\text{MTR}_{\text{con}}^{\text{del}}$, and $\text{MTR}_{\text{acc}_{t+1}}$ is recomputed along the account-level authentication path. If the recomputed account root matches the server-returned value, the deletion operation passes user-side verification. This process ensures that a session deletion does not erase the cryptographic commitment to the pre-deletion conversation state; rather, it converts that state into a verifiable deletion record. The proof size is $O(\log n)$ hash values, where $n$ denotes the number of conversations in the account. Serialization constraints between deletion and concurrent updates are handled by the multi-device consistency protocol described in the following section.

\paragraph{Complexity Analysis.}
The cryptographic update complexity for server-side account-state reconstruction and user-side account-state transition verification across the four operation types is summarized as follows.

\begin{table}[H]
\centering
\small
\caption{Cryptographic complexity of core operations}
\label{tab:complexity}
\begin{tabular}{lcc}
\toprule
\textbf{Operation} & \textbf{Server (Update)} & \textbf{User (Verify)} \\
\midrule
\text{NewSession} & $O(\log n)$ & $O(\log n)$ \\
\text{AppendMessage} & $O(\log m + \log n)$ & $O(\log m + \log n)$ \\
\text{BranchMessage} & $O(\log m + \log n)$ & $O(l + \log m + \log n)$ \\
\text{DeleteSession} & $O(\log n)$ & $O(\log n)$ \\
\bottomrule
\end{tabular}
\end{table}

Here $n$ denotes the number of conversations in the account, $m$ denotes the number of branches in the target conversation, and $l$ denotes the length of the successor chain from the branching point to the prior branch tail. Session deletion additionally involves hiding or removing plaintext Q\&A content at the database layer, incurring a storage overhead of $O(S)$, where $S$ is the number of Q\&A nodes in the target conversation; this overhead does not affect the cryptographic state-update complexity. 

\subsubsection{Multi‑Device Concurrent Merge and Accountable Synchronization}

This section presents the view-merging and eventual-consistency accountability mechanism for multi-device settings. The system permits multiple devices belonging to the same user to interact with the LLM concurrently; however, during concurrent operation each device uses only its own locally anchored account state as the model context, and does not automatically incorporate state updates produced by other devices that have not yet been merged locally. A device triggers account-state merging through an \textsc{UpdateConversationRecord}(denoted $\texttt{UCR}$ for brevity) operation to obtain concurrent updates from other devices during that period; upon completion of the merge, the new account-level Merkle root $\text{MTR}_{\text{acc}}$ is jointly confirmed by dual signatures from the server and the user.

To prevent undecidable semantic conflicts between deletion states and concurrently modified states, $\textsc{DeleteSession}$ is defined in VCT as an account-level serialized state transition rather than an ordinary commutative update. Specifically, a deletion operation must be submitted against the server's current account root and, once committed, advances the account into a new deletion epoch. The multi-device merge algorithm processes only non-deletion updates—namely $\textsc{NewSession}$, $\textsc{Append}$, and $\textsc{Branch}$—within the same deletion epoch.

Although a committed deletion state does not participate in ordinary concurrent merges, it must be witnessed as a synchronization barrier by all subsequent device states. If a device's local account anchor does not yet reflect the latest deletion state, the server must, before accepting any further updates from that device, return a state-transition proof demonstrating that the deletion transition has occurred. The device uses this proof to verify that the target conversation root has been replaced from $\text{MTR}_{\text{con}}^{\text{old}}$ to $\text{MTR}_{\text{con}}^{\text{del}}$, and advances its local account anchor to the account root that incorporates the deletion state. Only after crossing this deletion barrier are non-deletion updates produced by the device considered to belong to the current deletion epoch and eligible for the subsequent deterministic merge procedure.

\subsubsubsection{Concurrent Interaction Model}

When multiple devices interact with the LLM simultaneously, each device maintains an independent local conversation view. Concurrent increments across devices remain mutually invisible and are not incorporated into each other's LLM context until the user explicitly invokes an \textsc{UCR} operation. This concurrent interaction model is captured in Algorithm 6.

\begin{algorithm}
\caption{\text{ConcurrentChat}}
\label{alg:concurrentchat}
\begin{algorithmic}[1]
\State \textit{// Each device independently interacts with the LLM}
\State $\textit{device}_i.\textit{local} \leftarrow \textit{device}_i.\textit{local} \cup \mathsf{LLM}(\textit{device}_i.\textit{local} \| \textit{input}_i)$
\State \textit{// The device invokes \text{UCR} to trigger merging}
\Procedure{SYNC}{$\textit{device}_k$}
    \State $(\textit{state\_new},\ \textit{MTR\_acc\_new},\ \textit{STR\_acc\_s},\ \textit{STR\_acc\_u})$
    \State \hspace{2.5em} $\leftarrow \text{MergeStates}(\textit{device}_k,\ \textit{global},\ \textit{device}_k.\textit{id},\ \textit{version})$
    \State $\textit{global} \leftarrow \textit{state\_new}$
    \State $\textit{device}_k.\textit{local} \leftarrow \textsc{Sessions}(\textit{MTR\_acc\_new})$
    \State \Return $\textit{device}_k.\textit{local}$
\EndProcedure
\end{algorithmic}
\end{algorithm}

The notion of "independent interaction" in Algorithm 6 refers solely to the independence of model context; it does not imply that the system is unaware of the operation target. Each state update submitted to the server still carries protocol metadata—including the operation type, the target conversation, and the corresponding baseline state—which the server uses to execute the merge, verification, and signing procedures.

\subsubsubsection{Account-Level Serialization of Session Deletion}

Session deletion is modeled as an account-level serialized state transition. Before issuing a deletion request, the device must first invoke $\textsc{UCR}$ to obtain the current global account state as declared by the server. The deletion request carries the account root $\text{MTR}_{\text{acc}}^{\text{base}}$ observed by the device, as well as the conversation-level root $\text{MTR}_{\text{con}}^{\text{old}}$ of the target conversation under that account state. The server accepts the deletion request only when $\text{MTR}_{\text{acc}}^{\text{base}}$ exactly matches the current global account root $\text{MTR}_{\text{acc}}^{\text{current}}$. Since the account-level Merkle root already authenticates the target conversation root $\text{MTR}_{\text{con}}^{\text{old}}$, the account-root consistency check guarantees that the deletion operation targets the latest conversation state observed by the device.

These rules endow the deletion operation with strict account-level serial semantics. Because a $\textsc{DeleteSession}$ operation must be submitted against the server's current latest account root, it cannot be submitted concurrently with other operations. If the account root on which a deletion request is based lags behind $\text{MTR}_{\text{acc}}^{\text{current}}$, committed updates exist in the account state that the device has not yet witnessed; the server rejects the deletion request and requires the device to synchronize first. Once a deletion is committed serially, the resulting $\text{MTR}_{\text{con}}^{\text{del}}$ is incorporated into the new account state and advances the deletion epoch.

In addition to the strict freshness check on deletion operations, the server enforces a deletion-epoch check on subsequent non-deletion updates. If a device submits a $\textsc{NewSession}$, $\textsc{Append}$, or $\textsc{Branch}$ request based on an account root that does not yet reflect the latest deletion state, the server does not accept the update directly; instead, it returns the missing state transition, the deletion-state witness, and the corresponding $\text{MerkleProofs}$, enabling the device to advance its local account anchor to the account root that incorporates the deletion state and thereby enter the current deletion epoch. Once the account state has been updated, if the target conversation remains active, the non-deletion update may be resubmitted against the current deletion epoch; if the target conversation is already in a deleted state, the server rejects any subsequent $\textsc{Append}$ or $\textsc{Branch}$ directed at that conversation.

These two checks together constitute the deletion serialization principle: a $\textsc{DeleteSession}$ operation must be submitted against the current latest account root, and non-deletion updates must be submitted against an account state that has already witnessed the latest deletion. Consequently, $\textsc{DeleteSession}$ does not enter the concurrent merge procedure but instead forms an account-level synchronization barrier. This mechanism ensures that all devices have observed the latest session deletion result before producing further updates, thereby preventing undecidable merge semantics between deletion states and concurrent append or branch operations.

\begin{table}[H]
\centering
\small
\caption{Operation types and their handling in concurrent settings}
\label{tab:deletion_serialization}
\begin{tabular}{c p{0.15\textwidth}
p{0.25\textwidth} p{0.35\textwidth}}
\toprule
\textbf{Operation} & \textbf{Concurrent Merge} & \textbf{Additional Acceptance Condition} & \textbf{Outcome} \\
\midrule
$\texttt{NewSession}$ & Yes & --- & Incorporated as a new conversation leaf at the account level. \\
$\texttt{Append}$ & Yes & Target conversation must not be in a deleted state. & The branch $\text{hash}_{\text{tail}}$ is updated; changes propagate to $\text{MTR}_{\text{con}}$ and $\text{MTR}_{\text{acc}}$. \\
$\texttt{Branch}$ & Yes & Target conversation must not be in a deleted state. & A new branch leaf is inserted; changes propagate to $\text{MTR}_{\text{con}}$ and $\text{MTR}_{\text{acc}}$. \\
$\texttt{DeleteSession}$ & No & $\text{MTR}_{\text{acc}}^{\text{base}} = \text{MTR}_{\text{acc}}^{\text{current}}$ & Committed serially; $\text{MTR}_{\text{con}}^{\text{del}}$ generated and $\text{MTR}_{\text{acc}}$ updated; deletion state serves as a synchronization barrier. \\
\bottomrule
\end{tabular}
\end{table}

\subsubsubsection{Merge Protocol}
A device obtains the current global account state declared by the server by invoking the $\texttt{UCR}$ operation. This operation calls the view-merge algorithm, which consolidates non-deletion increments concurrently produced by multiple devices into a unified account state. The deterministic merge function is defined as follows.

\noindent \textbf{Definition 1 (Merge Function).} 
The merge function

\[
\texttt{Merge}(M_1, M_2, \ldots, M_n) \rightarrow M_{\text{merge}}
\]

satisfies the following three properties:

\begin{itemize}
\item \textbf{Inclusion:} $M_1 \preceq M_{\text{merge}},\ M_2 \preceq M_{\text{merge}},\ \ldots,\ M_n \preceq M_{\text{merge}}$. A valid state-transition path exists from every input state $M_i$ to $M_{\text{merge}}$.

\item \textbf{Determinism:} Identical input sets always yield a unique output.

\item \textbf{Conflict-free preservation:} Conflicting non-deletion updates are materialized as distinct branches; no valid update produced by any device is discarded~\cite{shapiro2011conflict,kleppmann2017json}.
\end{itemize}

Based on this definition, the state-merge procedure in VCT partitions merge scenarios into two cases depending on whether concurrent updates target the same conversation.

\paragraph{Merging of Concurrent Updates on Distinct Conversations}
If two devices each create a new conversation, the resulting conversations are ordered by the timestamp of their respective initial nodes and inserted into the account-level conversation set; this is equivalent to appending leaves to the account-level Merkle tree. If one device creates a new conversation while another modifies an existing one, a leaf insertion and a leaf update are applied simultaneously to the account-level Merkle tree, after which $\text{MTR}_{\text{acc}}$ is recomputed. If two devices modify different existing conversations, two distinct leaves are updated; the account-level authentication structure update alone costs $O(\log n)$, where $n$ is the number of conversations in the account. When intra-conversation state updates are also considered, the total cost equals the sum of the respective per-conversation update costs and the account-level path update cost.

\begin{figure}[htbp]
\centering
\includegraphics[width=\linewidth]{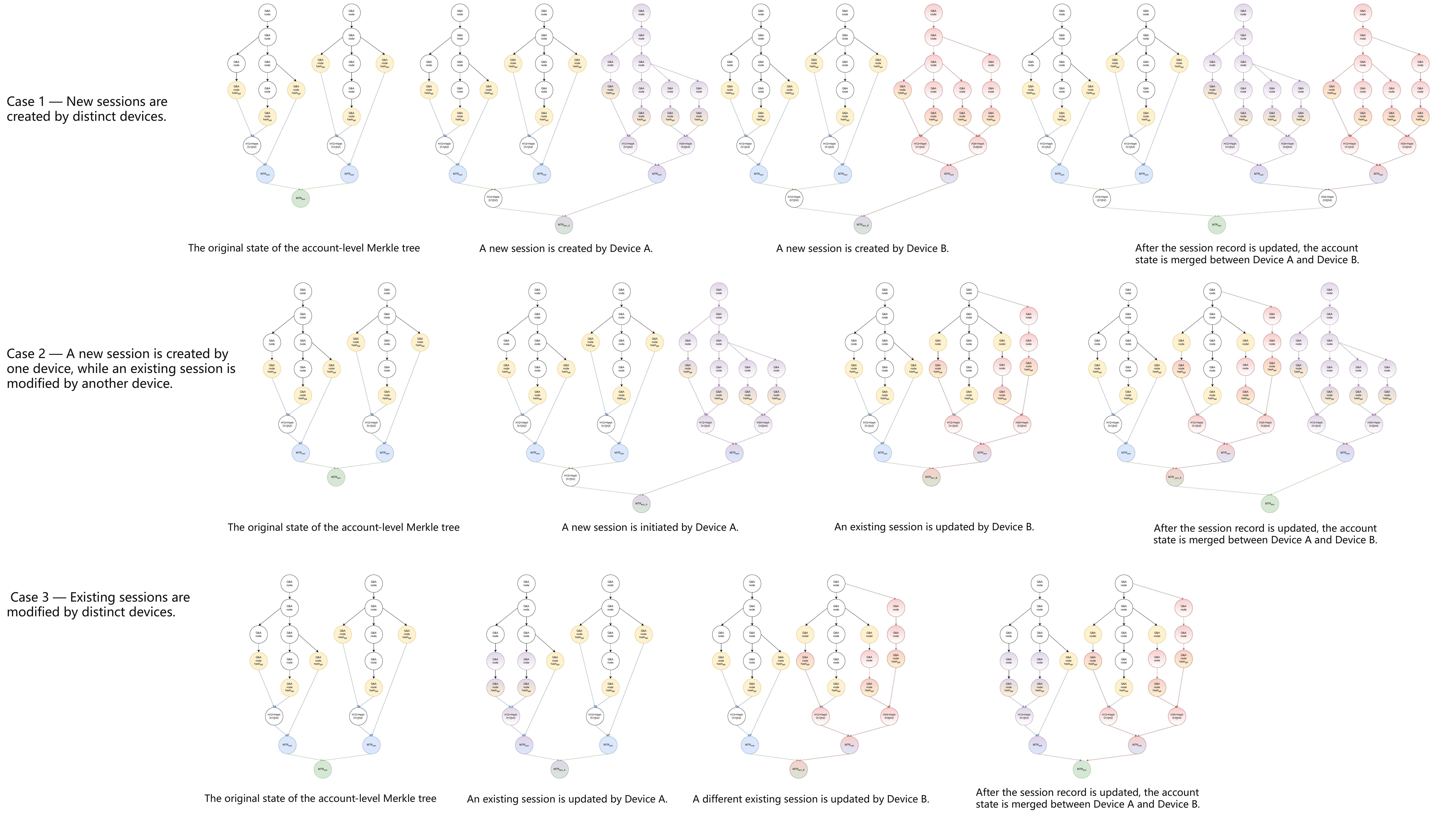}
\caption{Merge scenarios on the same conversation.}
\label{fig:Merge distinct conversations}
\end{figure}

\paragraph{Merging of Concurrent Updates on the Same Conversation.}
If multiple devices concurrently modify the same existing conversation, concurrent updates are materialized as distinct branches. Regardless of whether a device appends a message to the tail of an existing branch or creates a new branch from a historical node, each such update is treated at the merge layer as the insertion of one or more branches into the conversation tree. When multiple devices create branches from the same parent node, branches are ordered by their initial-node timestamps with respect to existing sibling timestamps; when devices target different parent nodes, ordering is applied independently under each respective parent.

This results in an update to the conversation-level Merkle root $\text{MTR}_{\text{con}}$, which is then propagated to the account-level root $\text{MTR}_{\text{acc}}$. The complexity is $O(\log m + \log n + \sum_i \log L_i)$, where $m$ is the number of branches in the target conversation after the merge, $n$ is the number of conversations in the account, and $L_i$ is the number of children of the $i$-th affected parent node after the merge. These three terms correspond, respectively, to the conversation-level Merkle path update, the account-level Merkle path update, and the cost of inserting new branches into an ordered child set supporting $O(\log L_i)$-time insertion by timestamp.

\begin{figure}[htbp]
\centering
\includegraphics[width=\linewidth]{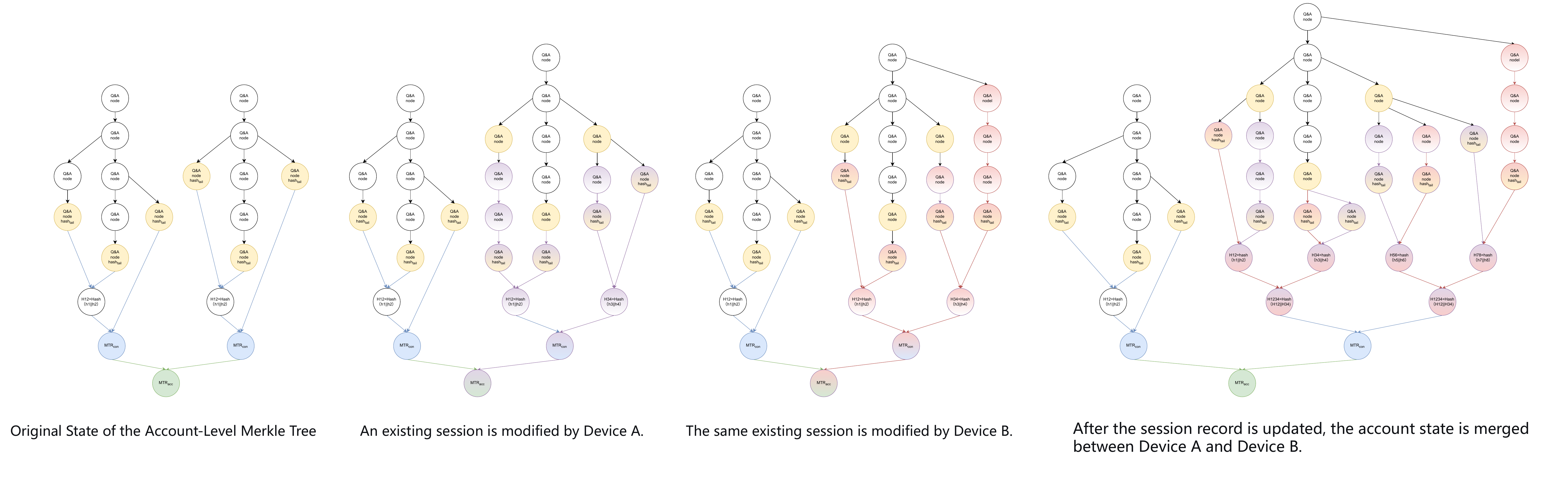}
\caption{Merge scenarios on the same conversation.}
\label{fig:Merge same conversations}
\end{figure}

\paragraph{Merge Protocol}Unlike single-device basic update operations, the increments being merged have already been witnessed by the devices that produced them and are anchored to user-signed account states. The correctness of the merge can therefore be further verified by the user side by leveraging these pre-existing signatures.

As described in Section 3.2, a complete account-state history with associated state-transition records, or an equivalent delta log, is maintained on the server. Upon invoking $\texttt{UCR}$, a device submits only its locally held account root $\text{MTR}_{\text{acc}}^{\text{base}}$. The account state anchored by $\text{MTR}_{\text{acc}}^{\text{base}}$ is taken as the synchronization baseline; non-deletion updates belonging to the same deletion epoch that have been committed by other devices but not yet witnessed by the synchronizing device are extracted from the signed account-state history and deterministically merged into the device's local view via Algorithm 7.

\begin{algorithm}[H]
\caption{$\texttt{MergeStates}(\text{MTR}_{\text{acc}}^{\text{base}}, H_{\text{acc}})$}
\label{alg:mergestates}
\begin{algorithmic}[1]
\State \textbf{Input:}
\State \hspace{\algorithmicindent} $\text{MTR}_{\text{acc}}^{\text{base}}$: account root submitted by the synchronizing device
\State \hspace{\algorithmicindent} $H_{\text{acc}}$: signed account-state history with associated transition records, or an equivalent delta log
\State \textbf{Output:}
\State \hspace{\algorithmicindent} $S_{\text{mer}}$: merged account state
\State \textit{// Precondition:}
\State \hspace{\algorithmicindent} $\text{MTR}_{\text{acc}}^{\text{base}}$ has passed the deletion-epoch check.
\State \hspace{\algorithmicindent} $\texttt{DeleteSession}$ is serialized at the account level and does not enter $\texttt{MergeStates}$.
\State \hspace{\algorithmicindent} The server can recover the account state $S_{\text{base}}$ anchored by $\text{MTR}_{\text{acc}}^{\text{base}}$, or an equivalent delta representation.

\State $S_{\text{base}} \leftarrow \text{ResolveAccountState}(\text{MTR}_{\text{acc}}^{\text{base}}, H_{\text{acc}})$
\State $\mathcal{U} \leftarrow \text{CollectUnseenMergeableUpdates}(H_{\text{acc}}, \text{MTR}_{\text{acc}}^{\text{base}})$
\State $S_{\text{mer}} \leftarrow \textsc{Copy}(S_{\text{base}})$

\For{each $u \in \texttt{NewSessionUpdates}(\mathcal{U})$ in deterministic order}
    \State $S_{\text{mer}} \leftarrow \text{InsertNewSession}(S_{\text{mer}}, u)$
\EndFor

\For{each $\textit{session\_id} \in \texttt{ExistingSessionsAffectedBy}(\mathcal{U})$ in deterministic order}
    \State $\mathcal{G} \leftarrow \texttt{UpdatesForSession}(\mathcal{U}, \textit{session\_id})$
    \State merge $\texttt{Append}$ updates in $\mathcal{G}$ by extending the corresponding branch tails
    \State merge $\texttt{Branch}$ updates in $\mathcal{G}$ by inserting new branch tails at their authenticated branching points
    \State update the session root of $\textit{session\_id}$ in $S_{\text{mer}}$
\EndFor

\State $\texttt{SortSessions}(S_{\text{mer}})$ by $(\textit{created\_at}, \textit{session\_id})$
\State \Return $S_{\text{mer}}$
\end{algorithmic}
\end{algorithm}

Here $H_{\text{acc}}$ denotes the dual-signed account-state history maintained by the server for each device, together with associated state-transition records, or an equivalent delta log. The account state $S_{\text{base}}$ anchored by $\text{MTR}_{\text{acc}}^{\text{base}}$ is recovered via $\textsc{ResolveAccountState}$, and the set $\mathcal{U}$ of same-epoch non-deletion updates that have been committed and confirmed by other devices but are not yet reflected in $\text{MTR}_{\text{acc}}^{\text{base}}$ is extracted via $\textsc{CollectUnseenMergeableUpdates}$.

Starting from $S_{\text{base}}$, Algorithm 7 deterministically incorporates the non-deletion increments in $\mathcal{U}$ into the synchronizing device's account view. Algorithm 7 produces only the merged account state $S_{\text{mer}}$; this result is then submitted as an $op = \texttt{Merge}$ state transition to Algorithm 1 for account-root computation and dual-signature confirmation. Specifically, the new account root $\text{MTR}_{\text{acc}_{t+1}}$ is computed from $S_{\text{mer}}$; the associated $\text{update\_payload}$ and $\text{MerkleProofs}$ are generated and $\text{STR}_{\text{acc}_{t+1}}^{\text{s}}$ is issued by the server. On the user side, the account root is recomputed from $\text{update\_payload}$ and $\text{MerkleProofs}$, and each incorporated update is verified to have been anchored to an account state previously signed by the same user private key on another device. Upon successful state-transition proof and signature verification, $\text{STR}_{\text{acc}_{t+1}}^{\text{u}}$ is issued over the confirmed $\text{MTR}_{\text{acc}_{t+1}}$, and the account state is finalized upon server-side verification of the user signature. The merge operation thus shares the same proof-verification and dual-signature confirmation procedure as all other account updates.

Devices are not required to invoke $\texttt{UCR}$ simultaneously; verification and signature confirmation of the merge result are performed by the device that triggers synchronization. The set $\mathcal{U}$ of unseen non-deletion updates may vary across devices depending on each device's local account root and the timing of synchronization, but the merge rules remain invariant.

\subsubsubsection{Incremental-Replay Rejection and Gossip Pre-check}

To prevent the server from returning a stale but previously valid account state during synchronization and subsequently delivering deliberately omitted increments in follow-up updates, an incremental-replay freshness check is enforced on the device side when accepting merged increments. Let $\text{timestamp}_{\text{sig}}$ denote the maximum timestamp bound to the user signatures $\text{STR}_{\text{acc}}^{\text{u}}$ over the confirmed account states returned in the transition proofs of the last $\texttt{UCR}$ invocation, and let $\text{timestamp}_{\text{update}}$ denote the time at which that invocation was initiated. Since $\text{STR}_{\text{acc}}^{\text{u}}$ is produced under the user private key and bound to the account-level Merkle root $\text{MTR}_{\text{acc}}$, increments already reflected in that signed state prior to $\text{timestamp}_{\text{sig}}$ cannot be omitted without invalidating signature verification. The check therefore focuses on whether increments that existed between $\text{timestamp}_{\text{sig}}$ and $\text{timestamp}_{\text{update}}$ were selectively withheld by the server.

Specifically, if an incremental node received during a subsequent synchronization or merge carries a timestamp $\text{timestamp}_{\text{node}}$ satisfying

\[
\text{timestamp}_{\text{sig}} < \text{timestamp}_{\text{node}} \leq \text{timestamp}_{\text{update}},
\]

the node is determined to have existed at the time of the device's last $\texttt{UCR}$ invocation but was not incorporated into the latest account state claimed by the server. The increment is accordingly rejected and treated as evidence that the server failed to return the current latest account state at that time. Conversely, if $\text{timestamp}_{\text{node}} > \text{timestamp}_{\text{update}}$, the node may be a legitimate new update produced after the device's last update request and is not rejected under this rule.

This check is complementary to the gossip accountability mechanism described below: incremental-replay rejection prevents the server from patching a single device's view through a strategy of first returning a stale state and subsequently delivering omitted increments.

\subsubsubsection{Gossip-Based Fork Detection and Accountability}

The gossip protocol is employed to detect whether inconsistent account views have been served to different devices~\cite{ryan2014enhanced,li2004sundr, mahajan2011depot}. Gossip is executed opportunistically whenever communication is available; when view inconsistency is suspected, all devices may be required to come online simultaneously and execute a full gossip round. Prior to gossip execution, all participating devices must complete $\texttt{UCR}$, and LLM interactions are suspended during the gossip round to prevent account-state changes from occurring mid-comparison.

Inter-device gossip communication is protected by an end-to-end encrypted channel. The communication key is derived jointly from the user root key $K_u$ and the login password $\text{pwd}_u$, following the same derivation structure as the user signing key but under a distinct domain-separation tag:

\[
K_{\text{gossip}} \leftarrow \mathsf{KDF}(K_u,\ \text{pwd}_u,\ \text{salt},\ \text{iterations},\ \texttt{"gossip\_encryption"},\ \text{key\_len}),
\]

where $K_u$ is the high-entropy user root key shared across all devices; $\text{pwd}_u$ is the user login password; $\text{salt}$ and $\text{iterations}$ are shared with the signing-key derivation; $\texttt{"gossip\_encryption"}$ serves as a domain-separation tag ensuring that $K_{\text{gossip}}$ is cryptographically independent of the signing key $\text{sk}_u$; and $\text{key\_len}$ specifies the output length in bytes. The key is derived and securely stored upon device initialization and reused in subsequent gossip rounds.

The core gossip procedure is as follows. A secure channel is established under $K_{\text{gossip}}$, and each device exchanges its latest signed state $\mathsf{Enc}_{K_{\text{gossip}}}(\text{MTR}_{\text{acc}} \| \text{STR}_{\text{acc}}^{\text{u}} \| \text{STR}_{\text{acc}}^{\text{s}})$. Upon decryption, the user and server signatures are verified and the account-level Merkle roots are compared. Matching roots indicate a consistent view; diverging roots indicate a fork. Fork evidence consists of two distinct valid signed states; since the server cannot forge user signatures and cannot repudiate its own, the evidence is cryptographically non-repudiable. Algorithm 8 performs online consistency detection between devices.

\begin{algorithm}[H]
\caption{$\texttt{GossipProtocol}(D_A, D_B)$}
\label{alg:gossip}
\begin{algorithmic}[1]
\State \textbf{Input:}
\State \hspace{\algorithmicindent} $D_A$: $(\text{MTR}_{\text{acc}_A},\ \text{timestamp}_A,\ \text{STR}_{\text{acc}_A}^{\text{u}},\ \text{STR}_{\text{acc}_A}^{\text{s}})$
\State \hspace{\algorithmicindent} $D_B$: $(\text{MTR}_{\text{acc}_B},\ \text{timestamp}_B,\ \text{STR}_{\text{acc}_B}^{\text{u}},\ \text{STR}_{\text{acc}_B}^{\text{s}})$
\State \textbf{Output:}
\State \hspace{\algorithmicindent} $\texttt{Consistent}$ or $\texttt{ForkDetected}$ with evidence
\State \textbf{Precondition:}
\State \hspace{\algorithmicindent} Both $D_A$ and $D_B$ have completed $\texttt{UCR}$; no new LLM interaction occurs during gossip.

\If{$K_{\text{gossip}}$ is not initialized}
    \State $K_{\text{gossip}} \leftarrow \mathsf{PBKDF2}(K_u,\ \text{pwd}_u,\ \text{salt},\ \text{iterations},\ \texttt{"gossip\_encryption"},\ \text{key\_len}=32)$
    \State $\texttt{SecureStore}(K_{\text{gossip}})$
\EndIf

\State $\textit{channel} \leftarrow \texttt{EncChannel}(K_{\text{gossip}})$
\State $\textit{msg}_A \leftarrow \mathsf{Enc}_{K_{\text{gossip}}}(\text{MTR}_{\text{acc}_A} \| \text{timestamp}_A \| \text{STR}_{\text{acc}_A}^{\text{u}} \| \text{STR}_{\text{acc}_A}^{\text{s}})$
\State send $\textit{msg}_A$ to $D_B$; receive $\textit{msg}_B$ from $D_B$
\State $(\text{MTR}_{\text{acc}_B},\ \text{timestamp}_B,\ \text{STR}_{\text{acc}_B}^{\text{u}},\ \text{STR}_{\text{acc}_B}^{\text{s}}) \leftarrow \mathsf{Dec}_{K_{\text{gossip}}}(\textit{msg}_B)$

\If{$\mathsf{Verify}(pk_u,\ \texttt{"ACCOUNT\_STATE"} \| \text{MTR}_{\text{acc}_A} \| \text{timestamp}_A,\ \text{STR}_{\text{acc}_A}^{\text{u}}) = \bot$} \Return $\bot$ \EndIf
\If{$\mathsf{Verify}(pk_s,\ \texttt{"ACCOUNT\_STATE"} \| \text{MTR}_{\text{acc}_A} \| \text{timestamp}_A,\ \text{STR}_{\text{acc}_A}^{\text{s}}) = \bot$}  \Return $\bot$ \EndIf
\If{$\mathsf{Verify}(pk_u,\ \texttt{"ACCOUNT\_STATE"} \| \text{MTR}_{\text{acc}_B} \| \text{timestamp}_B,\ \text{STR}_{\text{acc}_B}^{\text{u}}) = \bot$} \Return $\bot$ \EndIf
\If{$\mathsf{Verify}(pk_s,\ \texttt{"ACCOUNT\_STATE"} \| \text{MTR}_{\text{acc}_B} \| \text{timestamp}_B,\ \text{STR}_{\text{acc}_B}^{\text{s}}) = \bot$} \Return $\bot$ \EndIf

\If{$\text{MTR}_{\text{acc}_A} = \text{MTR}_{\text{acc}_B}$}
    \State \Return $\texttt{Consistent}$
\Else
    \State \Return $\texttt{ForkDetected}$ with evidence:
    \State \hspace{2em} $(\text{MTR}_{\text{acc}_A},\ \text{timestamp}_A,\ \text{STR}_{\text{acc}_A}^{\text{u}},\ \text{STR}_{\text{acc}_A}^{\text{s}})$,
    \State \hspace{2em} $(\text{MTR}_{\text{acc}_B},\ \text{timestamp}_B,\ \text{STR}_{\text{acc}_B}^{\text{u}},\ \text{STR}_{\text{acc}_B}^{\text{s}})$
\EndIf
\end{algorithmic}
\end{algorithm}

Algorithm 9 is not a gossip participant; it is an offline verification procedure for use by a third party $\mathcal{TP}$ or auditor to validate fork evidence produced by Algorithm 8. If both evidence records carry valid user and server signatures and bind distinct values of $\text{MTR}_{\text{acc}}$, the evidence constitutes a verifiable proof that inconsistent account views were served to different devices.

\begin{algorithm}[H]
\caption{$\texttt{VerifyForkEvidence}(\textit{evidence})$}
\label{alg:verifyfork}
\begin{algorithmic}[1]
\State \textbf{Input:}
\State \hspace{\algorithmicindent} $\textit{evidence} = (E_A, E_B)$, where $E_i = (\text{MTR}_{\text{acc}_i},\ \text{STR}_{\text{acc}_i}^{\text{u}},\ \text{STR}_{\text{acc}_i}^{\text{s}})$
\State \textbf{Output:}
\State \hspace{\algorithmicindent} $\texttt{ValidFork}$ or $\texttt{Invalid}$

\If{$\texttt{VerifySignedState}(E_A) = \bot$} \Return $\texttt{Invalid}$ \EndIf
\If{$\texttt{VerifySignedState}(E_B) = \bot$}  \Return $\texttt{Invalid}$ \EndIf
\If{$\text{MTR}_{\text{acc}_A} = \text{MTR}_{\text{acc}_B}$} \Return $\texttt{Invalid}$ \EndIf

\State \Return $\texttt{ValidFork}$
\end{algorithmic}
\end{algorithm}

Once a full gossip round is eventually executed across all devices, any attempt by the server to serve incompatible states to different devices is detected. If a stale but valid account root was returned to a device during synchronization and the server subsequently attempts to deliver omitted increments, those increments are rejected by the device based on the synchronization boundary and the signed timestamps. Should the server continue to withhold the omitted state, the divergence between the $\text{STR}_{\text{acc}}^{\text{u}}$ and $\text{STR}_{\text{acc}}^{\text{s}}$ values held by different devices is exposed during gossip.

\subsubsubsection{Device Online Synchronization Protocol}

Upon coming online, a device initiates an $\texttt{UCR}$ request to the server to synchronize to the current latest account state. Let $\text{MTR}_{\text{acc}_D}$ denote the device's locally held account-level Merkle root. A state-transition proof package is returned by the server, comprising the relevant $\text{update\_payload}$, $\text{MerkleProofs}$, any applicable deletion-state witnesses, and the target account root $\text{MTR}_{\text{acc}}^{\text{current}}$.

Using the locally verified $\text{MTR}_{\text{acc}_D}$ as the anchor, the returned $\text{update\_payload}$, $\text{MerkleProofs}$, and deletion-state witnesses are verified in deterministic order, and the target account root $\text{MTR}_{\text{acc}}'$ is reconstructed. If $\text{MTR}_{\text{acc}}' = \text{MTR}_{\text{acc}}^{\text{current}}$ and all state-transition verifications pass, the device issues a signature to confirm and updates its local account anchor. This synchronization procedure does not rely on server honesty: a stale but previously valid state may be returned by the server, but any attempt to serve inconsistent views to different devices is ultimately detected and attributed through subsequent synchronization and gossip rounds.

\subsubsubsection{New Device Enrollment Protocol}

New device enrollment begins with the key-provisioning procedure described in Section 4.2.1: $K_u$ is securely obtained from an existing trusted device, and $\text{pwd}_u$ is entered locally by the user on the new device, from which $\text{sk}_u$, $\text{pk}_u$, and $K_{\text{gossip}}$ are derived. This phase does not rely on the server to synchronize the user private key.

Upon completion of key provisioning, the device online synchronization protocol of Section 4.2.3.6 is reused to obtain and verify the current account anchor $(\text{MTR}_{\text{acc}},\ \text{timestamp}_{\text{acc}},\ \text{STR}_{\text{acc}}^{\text{s}},\\ \text{STR}_{\text{acc}}^{\text{u}})$ as declared by the server. If participation in subsequent updates, signature confirmation, and gossip is sufficient, only this account anchor need be retained on the new device. If recovery of historical conversation content is required, the corresponding Q\&A payload, $\text{update\_payload}$, and $\text{MerkleProofs}$ are returned by the server on demand, enabling the new device to verify the binding between the recovered content and the account anchor.

\subsubsection{Verifiable Conversation Sharing}

The goal of verifiable conversation sharing is to enable a recipient $R$ to independently verify the integrity and provenance of a shared Q\&A fragment without trusting the server's process of generating sharing links or export files. The selected Q\&A pairs are first verified to be anchored to the account-level state $\text{MTR}_{\text{acc}}$; they are then reorganized into an independent hash chain, over whose tail separate signatures are issued by the server and the user.

The protocol proceeds in three phases.

\paragraph{Phase 1: User Selection and Proof Package Verification.}
The user designates the Q\&A pairs to be shared. A proof package is generated by the server:

\[
\textit{proof} = \{\text{selected\_nodes},\ \text{MerkleProofs},\ \text{MTR}_{\text{acc}},\ \text{STR}_{\text{acc}}^{\text{s}},\ \text{STR}_{\text{acc}}^{\text{u}}\},
\]

where $\text{selected\_nodes}$ is the set of Q\&A nodes designated by the user, $\text{MerkleProofs}$ establishes that these nodes are anchored to the account-level Merkle root $\text{MTR}_{\text{acc}}$, and $\text{STR}_{\text{acc}}^{\text{s}}$ and $\text{STR}_{\text{acc}}^{\text{u}}$ are the server and user signatures over $\text{MTR}_{\text{acc}}$, respectively. Upon receipt of the proof package, the Merkle proof for each selected node is verified, along with $\text{STR}_{\text{acc}}^{\text{s}}$ and $\text{STR}_{\text{acc}}^{\text{u}}$. If any verification fails, the sharing procedure is aborted.

\paragraph{Phase 2: Independent Share Chain Construction and Dual Signing.}
Upon confirmation that the selected Q\&A nodes are anchored to a signed account state, the nodes are reorganized by the server into an independent hash chain. This chain carries no dependency on unshared nodes in the original conversation record and contains no external pointers into the original conversation tree. Let the tail of this independent chain be:

\[
\text{share}_{\text{tail}} = \texttt{BuildIndependentChain}(\text{selected\_nodes}).
\]

A signature over $\text{share}_{\text{tail}}$ and the timestamp is first issued by the server:

\[
\text{share}_{\text{sig}}^{\text{s}} \leftarrow \mathsf{Sign}_{sk_s}(\texttt{"SHARE\_SNAPSHOT"} \| \text{share}_{\text{tail}} \| \text{timestamp}_{\text{share}}).
\]

The tuple $(\text{share}_{\text{tail}},\ \text{timestamp}_{\text{share}},\ \text{share}_{\text{sig}}^{\text{s}})$ is transmitted to the user. The independent chain tail is recomputed from $\text{selected\_nodes}$ and verified to match the server-declared $\text{share}_{\text{tail}}$; the server signature is verified under $pk_s$. Upon successful verification, a signature over the same chain tail is issued by the user:

\[
\text{share}_{\text{sig}}^{\text{u}} \leftarrow \mathsf{Sign}_{sk_u}(\texttt{"SHARE\_SNAPSHOT"} \| \text{share}_{\text{tail}} \| \text{timestamp}_{\text{share}}).
\]

Upon receipt of $\text{share}_{\text{sig}}^{\text{u}}$, the user signature is verified by the server under $pk_u$. A successful verification confirms that both parties have reached cryptographic agreement over the same sharing snapshot.

\paragraph{Phase 3: Share Package Generation.}
The final share package is defined as:

\[
\{\text{selected\_nodes},\ \text{share}_{\text{tail}},\ \text{share}_{\text{sig}}^{\text{s}},\ \text{share}_{\text{sig}}^{\text{u}},\ \text{timestamp}_{\text{share}}\}.
\]

This package may be further rendered as a sharing link, PDF, or other export format. To confirm that the shared content is unmodified and has been jointly confirmed by both parties, the recipient recomputes the $\text{share}_{\text{tail}}$ of the independent chain and verifies the dual signatures. Since the share chain is independent of the original conversation tree, subsequent appends, edits, branches, or deletions to the original conversation do not affect the verifiability of an already-generated sharing snapshot.

\begin{algorithm}[H]
\caption{$\texttt{ShareTranscript}$}
\label{alg:sharetranscript}
\begin{algorithmic}[1]
\State \textbf{Input:} $\text{selected\_nodes}$, $\text{MTR}_{\text{acc}}$, $\text{STR}_{\text{acc}}^{\text{s}}$, $\text{STR}_{\text{acc}}^{\text{u}}$, $\text{MerkleProofs}$

\State \textbf{Output:} $\textit{share\_package} = \{\text{selected\_nodes},\ \text{share}_{\text{tail}},\ \text{timestamp}_{\text{share}},\ \text{share}_{\text{sig}}^{\text{s}},\ \text{share}_{\text{sig}}^{\text{u}}\}$

\ =========Phase 1: User verification of the proof package=========
\State $\textit{proof} \leftarrow \{\text{selected\_nodes},\ \text{MerkleProofs},\ \text{MTR}_{\text{acc}},\ \text{timestamp}_{\text{acc}},\ \text{STR}_{\text{acc}}^{\text{s}},\ \text{STR}_{\text{acc}}^{\text{u}}\}$
\State send $\textit{proof}$ to user

\State $\texttt{For}$ {each $\textit{node} \in \text{selected\_nodes}$} $\texttt{do}$
    $\texttt{VerifyMerkleProof}(\textit{node},\ \text{MerkleProofs},\ \text{MTR}_{\text{acc}})$

\State $\mathsf{Verify}(pk_s,\ \texttt{"ACCOUNT\_STATE"} \| \text{MTR}_{\text{acc}} \| \text{timestamp}_{\text{acc}},\ \text{STR}_{\text{acc}}^{\text{s}})$
\State $\mathsf{Verify}(pk_u,\ \texttt{"ACCOUNT\_STATE"} \| \text{MTR}_{\text{acc}} \| \text{timestamp}_{\text{acc}},\ \text{STR}_{\text{acc}}^{\text{u}})$

\ == Phase 2: Construction and dual signing of the independent share chain==
\State $\text{share}_{\text{tail}} \leftarrow \texttt{BuildIndependentChain}(\text{selected\_nodes})$
\State $\text{share}_{\text{sig}}^{\text{s}} \leftarrow \mathsf{Sign}_{sk_s}(\texttt{"SHARE\_SNAPSHOT"} \| \text{share}_{\text{tail}} \| \text{timestamp}_{\text{share}})$
\State send $(\text{share}_{\text{tail}},\ \text{timestamp}_{\text{share}},\ \text{share}_{\text{sig}}^{\text{s}})$ to user
\State $\texttt{ComputeChainTail}(\text{selected\_nodes}) = \text{share}_{\text{tail}}$
\State $\mathsf{Verify}(pk_s,\ \texttt{"SHARE\_SNAPSHOT"} \| \text{share}_{\text{tail}} \| \text{timestamp}_{\text{share}},\ \text{share}_{\text{sig}}^{\text{s}})$
\State $\text{share}_{\text{sig}}^{\text{u}} \leftarrow \mathsf{Sign}_{sk_u}(\texttt{"SHARE\_SNAPSHOT"} \| \text{share}_{\text{tail}} \| \text{timestamp}_{\text{share}})$
\State send $\text{share}_{\text{sig}}^{\text{u}}$ to server
\State $\mathsf{Verify}(pk_u,\ \texttt{"SHARE\_SNAPSHOT"} \| \text{share}_{\text{tail}} \| \text{timestamp}_{\text{share}},\ \text{share}_{\text{sig}}^{\text{u}})$

\ =============Phase 3: Output share package=============
\State \Return $\{\text{selected\_nodes},\ \text{share}_{\text{tail}},\ \text{timestamp}_{\text{share}},\ \text{share}_{\text{sig}}^{\text{s}},\ \text{share}_{\text{sig}}^{\text{u}}\}$
\end{algorithmic}
\end{algorithm}

The protocol enforces a separation between original-conversation-state verification and sharing-snapshot confirmation. The former establishes that the shared nodes originate from a signed account state; the latter fixes the shared content via an independent hash chain and dual signatures. Any substitution, deletion, or reordering of shared nodes during share-link generation or file export is detectable; equally, repudiation of the sharing snapshot by the user is precluded.

\subsection{Protocol Communication Overhead}

This section analyzes the protocol communication overhead introduced by VCT beyond a typical LLM request–response exchange. Application-layer payloads—including the original user prompt, model response, and uploaded file content—are excluded from the analysis. Only the root hashes, signatures, authentication paths, incremental metadata, and protocol control fields required to achieve verifiable state confirmation, multi-device synchronization, consistency detection, and verifiable sharing are accounted for. Let $|h|$ denote the hash output length and $|\sigma|$ the signature length. Let $n$ denote the number of conversations in the account, $m$ the number of branches in the target conversation, $k$ the number of increments involved in a synchronization or merge, $s$ the number of shared nodes, and $d$ the number of devices belonging to the same user that participate in gossip. For the $\texttt{Branch}$ operation, $l$ denotes the length of the successor chain from the branching point to a previously authenticated branch tail. Concrete instantiation parameters—including hash output length, signature length, and cryptographic parameters—are provided in Section 5.2.

\subsubsection{Per-Update Confirmation Overhead}

The communication overhead of an account-state update arises from the bidirectional confirmation procedure of Algorithm 1. The server transmits to the user the operation-specific $\text{update\_payload}$, the updated account root $\text{MTR}_{\text{acc}}$, $\text{MerkleProofs}$, and the server signature $\text{STR}_{\text{acc}}^{\text{s}}$; upon successful verification, $\text{STR}_{\text{acc}}^{\text{u}}$ over the same account root is returned by the user. Prompt, response, and uploaded file content in $\text{update\_payload}$ constitute application-layer data and are excluded from VCT-specific overhead; operation type, session identifier, branch identifier, timestamp, and necessary hash values contribute only constant-size control fields. Therefore, the additional communication overhead of a single account-state update consists primarily of the baseline account root, $\text{MerkleProofs}$, the updated account root, and two signatures.

The per-confirmation overhead for each of the four basic update operations is summarized below.

\noindent $\texttt{NewSession}$: Upon new session creation, $\text{MTR}_{\text{con\_new}}$ is inserted as a new leaf into the account-level Merkle tree. An account-level append proof and the authentication path for the new conversation root must be supplied; the incremental overhead is therefore
\[
O(\log n \cdot |h| + 2|h| + 2|\sigma|).
\]

\noindent $\texttt{Append}$: A message append replaces only the $\text{hash}_{\text{tail}}$ of the target branch and propagates the change to the conversation root and account root. The conversation-level authentication path from the old $\text{hash}_{\text{tail}}$ to $\text{MTR}_{\text{con}}$, and the account-level path from $\text{MTR}_{\text{con}}$ to $\text{MTR}_{\text{acc}}$, must be supplied; the incremental overhead is therefore
\[
O((\log m + \log n) \cdot |h| + 2|h| + 2|\sigma|).
\]

\noindent $\texttt{Branch}$: Branch creation requires proof that the branching point belongs to an authenticated history. Since the user device retains neither the full conversation chain nor intermediate authentication paths, a successor-chain witness from the branching point to a prior authenticated branch tail must be supplied, together with authentication paths from that branch tail to $\text{MTR}_{\text{con}}$ and from $\text{MTR}_{\text{con}}$ to $\text{MTR}_{\text{acc}}$; the incremental overhead is therefore
\[
O((l + \log m + \log n) \cdot |h| + 2|h| + 2|\sigma|).
\]

\noindent $\texttt{DeleteSession}$: Session deletion is modeled as an account-level leaf replacement, substituting $\text{MTR}_{\text{con}}^{\text{del}}$ for $\text{MTR}_{\text{con}}^{\text{old}}$. The authentication path of $\text{MTR}_{\text{con}}^{\text{old}}$ in the account-level Merkle tree must be supplied; $\text{MTR}_{\text{con}}^{\text{del}}$ can be recomputed and verified locally by the user from $\text{MTR}_{\text{con}}^{\text{old}}$ and $\text{timestamp}_{\text{del}}$. The incremental overhead is therefore
\[
O(\log n \cdot |h| + 2|h| + 2|\sigma|).
\]

\subsubsection{State Synchronization and Merge Overhead}

The communication overhead of account-state synchronization and merging depends on the number and type of increments to be synchronized. Let $\text{MTR}_{\text{acc}}^{\text{base}}$ denote the account root locally anchored by the lagging device. The synchronization target returned by the server varies by scenario: a full synchronization targets the server's current account root $\text{MTR}_{\text{acc}}^{\text{current}}$; submission of a non-deletion update requires synchronization only to the account root reflecting the latest deletion state; submission of $\texttt{DeleteSession}$ requires synchronization to $\text{MTR}_{\text{acc}}^{\text{current}}$, since a deletion must be submitted against the server's current latest account state.

Let the interval from $\text{MTR}_{\text{acc}}^{\text{base}}$ to the target account root contain $k$ non-deletion increments and $r$ deletion-state transitions that have been serially committed but not yet witnessed by the device. The server returns the relevant $\text{update\_payload}$, $\text{MerkleProofs}$, the target account root, and dual signatures attesting to the state transitions. Using $\text{MTR}_{\text{acc}}^{\text{base}}$ as the anchor, these state changes are verified in deterministic order and the target account root is reconstructed.

For non-deletion increments, let $\Pi_i$ denote the proof material size for the $i$-th increment. When only non-deletion increments are present, the synchronization and merge overhead is:

\[
O\!\left(\sum_{i=1}^{k} \Pi_i + |h| + 2|\sigma|\right),
\]

where $\Pi_i$ is determined by the operation type: $O(\log n_i \cdot |h|)$ for $\texttt{NewSession}$; $O((\log m_i + \log n_i) \cdot |h|)$ for $\texttt{Append}$; and $O((l_i + \log m_i + \log n_i) \cdot |h|)$ for $\texttt{Branch}$. Here $n_i$ denotes the number of conversations in the account at the time of the $i$-th update, $m_i$ the number of branches in the target conversation, and $l_i$ the length of the successor chain from the branching point to the previously authenticated branch tail. The latest local account roots $\text{MTR}_{\text{acc}}$ and $\text{STR}_{\text{acc}}^{\text{u}}$ carried in the proof material for concurrent increments constitute a constant number of hash values whose count depends only on the number of devices and does not affect the asymptotic overhead.

Deletion operations do not enter $\texttt{MergeStates}$ as ordinary concurrent increments; instead, each forms an account-level serialization boundary. When the synchronization interval contains deletion-state transitions, a deletion-state witness must also be returned by the server. This witness establishes that $\text{MTR}_{\text{con}}^{\text{old}}$ is the target conversation leaf under the pre-deletion account root, and is used to verify:

\[
\text{MTR}_{\text{con}}^{\text{del}} = \mathsf{Hash}(\texttt{"DEL\_SESSION"} \| \text{MTR}_{\text{con}}^{\text{old}} \| \text{timestamp}_{\text{del}}).
\]

The device then verifies along the account-level authentication path that $\text{MTR}_{\text{con}}^{\text{del}}$ is incorporated into the post-deletion account root. Let $\Delta_j$ denote the proof material size of the $j$-th deletion-state witness; in general:

\[
\Delta_j = O(\log n_j \cdot |h|).
\]

When dual signatures are attached to each deletion-state account root, the synchronization overhead for an interval containing $r$ deletion-state transitions is:

\[
O\!\left(\sum_{i=1}^{k} \Pi_i + \sum_{j=1}^{r}(\Delta_j + |h| + 2|\sigma|) + |h| + 2|\sigma|\right),
\]

where $\sum_{j=1}^{r}(\Delta_j + |h| + 2|\sigma|)$ corresponds to the deletion-state witnesses together with the dual signatures over their respective deletion-state account roots, and the final term $|h| + 2|\sigma|$ corresponds to the target account root and its dual signatures. When no deletion-state transitions are present in the synchronization interval, $r = 0$ and the expression reduces to the overhead of ordinary non-deletion increment synchronization.

\subsubsection{Gossip Overhead}

The communication overhead of the gossip protocol is digest-level. Rather than exchanging Q\&A content, conversation trees, or $\text{MerkleProofs}$, devices exchange only their respective latest signed account anchors $(\text{MTR}_{\text{acc}},\ \text{STR}_{\text{acc}}^{\text{u}},\ \text{STR}_{\text{acc}}^{\text{s}})$. The per-pair gossip communication cost is

\[
O(|h| + 2|\sigma|),
\]

independent of the number of conversations, Q\&A nodes, or conversation content in the account. When $d$ devices execute a full pairwise gossip round, the total communication cost is

\[
O(d^2(|h| + 2|\sigma|)).
\]

Since the number of devices per user is typically small and each gossip message contains only an account root hash and two signatures, gossip overhead remains low in practical deployments.

\subsubsection{New Device Onboarding Overhead}

New device onboarding comprises two phases: inter-device key transfer and account-state synchronization. The user root key $K_u$ and key derivation parameters ($\text{salt}$, $\text{iterations}$) are first obtained by the new device from an existing trusted device via a secure out-of-band channel such as a QR code, incurring a transfer cost of

\[
O(|K_u| + |\text{salt}| + |\text{iterations}|).
\]

Under fixed security parameters, this phase incurs constant communication overhead.

After $K_u$ is obtained and $\text{sk}_u$ is derived, the current account anchor—$\text{MTR}_{\text{acc}}$ and its dual signatures $(\text{STR}_{\text{acc}}^{\text{s}},\ \text{STR}_{\text{acc}}^{\text{u}})$—must be acquired. If the new device is only required to participate in subsequent signature confirmation, synchronization, and gossip without immediately recovering the full conversation history, the VCT-specific overhead of account onboarding is

\[
O(|h| + 2|\sigma|).
\]

If full historical conversation recovery is required, the historical Q\&A payload constitutes application-state synchronization and is excluded from VCT-specific overhead; the additional verification cost follows the same structure as that of a lagging-device synchronization, consisting primarily of $\text{MerkleProofs}$ for the increments to be recovered, the final account root, and dual signatures:

\[
O\!\left(\sum_{i=1}^{k} |\Pi_i| + |h| + 2|\sigma|\right).
\]

The minimum onboarding overhead is therefore constant; bandwidth for full history recovery is dominated by the volume of historical content the user chooses to restore, while the VCT-specific overhead remains governed by proof paths and the account anchor.

\subsubsection{Verifiable Sharing Overhead}

The communication overhead of the verifiable sharing protocol is computed following the three-phase procedure of Section 4.2.4. During share generation, the server returns to the user a provenance proof package comprising $\text{selected\_nodes}$, the corresponding $\text{MerkleProofs}$, the account root $\text{MTR}_{\text{acc}}$, and the dual signatures $(\text{STR}_{\text{acc}}^{\text{s}},\ \text{STR}_{\text{acc}}^{\text{u}})$ over the account root. The Q\&A content of $\text{selected\_nodes}$ constitutes user-initiated disclosure and is excluded from VCT-specific overhead; $\text{MerkleProofs}$, $\text{MTR}_{\text{acc}}$, and the dual signatures constitute the additional verification cost of the provenance proof. When $s$ nodes are shared and an independent authentication path is provided per node, the incremental provenance-proof overhead is

\[
O(s(\log m + \log n) \cdot |h| + |h| + 2|\sigma|).
\]

When multiple shared nodes reside in the same conversation or share Merkle path segments, the server may employ batch proofs or multi-proofs to reduce constant factors and redundant path overhead; the expression above gives an upper bound for the case of independent paths.

Upon user confirmation of the provenance proof, $\text{share}_{\text{tail}}$ and $\text{share}_{\text{sig}}^{\text{s}}$ are transmitted by the server, and $\text{share}_{\text{sig}}^{\text{u}}$ is returned by the user. This dual-signing phase introduces an additional overhead of

\[
O(|h| + 2|\sigma|).
\]

The total VCT-specific overhead for share generation is therefore

\[
O(s(\log m + \log n) \cdot |h| + 2|h| + 4|\sigma|),
\]

which, omitting constant factors, simplifies to

\[
O(s(\log m + \log n) \cdot |h| + |\sigma|).
\]

The final public share package contains $\text{selected\_nodes}$, $\text{share}_{\text{tail}}$, $\text{share}_{\text{sig}}^{\text{s}}$, $\text{share}_{\text{sig}}^{\text{u}}$, and $\text{timestamp}$. Verification by the recipient requires only reconstruction of the independent share chain and verification of the dual signatures over the chain tail, with no access to the full account state and no further server interaction required. Excluding the shared content itself, the VCT-specific overhead of the public share package is $O(|h| + 2|\sigma|)$.

\subsection{Security Analysis}

The security properties of VCT are analyzed in this section based on the threat model, security goals, and cryptographic assumptions established in Section 3. Unless stated otherwise, the following global assumptions are taken to hold throughout: $\mathsf{Hash}(\cdot)$ satisfies collision resistance and second-preimage resistance; the digital signature scheme satisfies existential unforgeability under chosen-message attacks; the KDF and the underlying authenticated encryption mechanism are secure; user signatures are generated by user devices only after protocol-specified verification is completed; and the gossip channel between devices belonging to the same user provides authenticity and integrity. For properties that depend on protocol liveness—such as multi-device eventual consistency—it is additionally assumed that honest devices eventually complete synchronization and execute gossip.

\subsubsection{Integrity Verification}

\noindent \textbf{Theorem 1 (Node Integrity).} If any field $Q$, $A$, $\text{model}_{\text{config}}$, $\text{file\_aux\_info}$, or $\text{timestamp}$ of a Q\&A node is modified by an adversary, the modification is detected during branch hash-chain verification or account-state verification.

\noindent \textit{Proof.} Consider a branch hash chain consisting of nodes $N_1, N_2, \ldots, N_m$, with branch tail $\text{hash}_{\text{tail}}$. For any $i < m$, the hash-chain invariant requires $N_{i+1}.\text{parent\_hash} = \mathsf{Hash}(N_i)$, where

\[
\mathsf{Hash}(N_i) = \mathsf{Hash}(\text{parent\_hash}_i \| Q_i \| A_i \| \text{model}_{\text{config}_i} \| \text{file\_aux\_info}_i \| \text{timestamp}_i).
\]

If an intermediate node $N_k$ with $k < m$ is modified by the adversary, the recomputed $\mathsf{Hash}(N_k)$ differs from its original value, while $N_{k+1}.\text{parent\_hash}$ retains the original value; the hash-chain invariant is thereby violated. If the tail node $N_m$ is modified, the recomputed $\text{hash}_{\text{tail}}$ changes, which propagates to the conversation-level root $\text{MTR}_{\text{con}}$ and subsequently to the account-level root $\text{MTR}_{\text{acc}}$. Since any accepted $\text{MTR}_{\text{acc}}$ must be jointly anchored by the server signature $\text{STR}_{\text{acc}}^{\text{s}}$ and the user signature $\text{STR}_{\text{acc}}^{\text{u}}$, the modified account root cannot pass verification without a valid forgery. Any tampering with node content is therefore detected. $\square$

\noindent \textbf{Theorem 2 (Conversation Integrity).} If any branch within a conversation is added, deleted, or modified by an adversary, the modification is detected during conversation-level or account-level verification.

\noindent \textit{Proof.} Let the branch set of conversation $C$ be $\{B_1, B_2, \ldots, B_b\}$ with corresponding branch tails $\{\text{hash}_{\text{tail}_1}, \text{hash}_{\text{tail}_2}, \ldots, \text{hash}_{\text{tail}_b}\}$. The conversation-level Merkle root is defined as

\[
\text{MTR}_{\text{con}} = \mathsf{MerkleRoot}(\{\text{hash}_{\text{tail}_1}, \text{hash}_{\text{tail}_2}, \ldots, \text{hash}_{\text{tail}_b}\}).
\]

Any modification to branch content alters the corresponding $\text{hash}_{\text{tail}}$; any addition or deletion of a branch alters the leaf set of the conversation-level Merkle tree. Unless a hash collision can be found, the recomputed $\text{MTR}_{\text{con}}$ must therefore differ from its original value. Since $\text{MTR}_{\text{con}}$ participates as a leaf in the computation of $\text{MTR}_{\text{acc}}$, the change propagates to the account-level root. A valid account state must pass dual-signature verification under $\text{STR}_{\text{acc}}^{\text{s}}$ and $\text{STR}_{\text{acc}}^{\text{u}}$, and no valid signature can be forged by the adversary over the modified account root. Any tampering with the branch set or branch content of a conversation is therefore detected. 

\noindent \textbf{Theorem 3 (Account Integrity).} If any conversation in a user account is added, deleted, or modified by an adversary, the modification is detected during account-level verification.

\noindent \textit{Proof.} Let the conversation set of the account be $\{C_1, C_2, \ldots, C_c\}$ with corresponding conversation roots $\{r_1, r_2, \ldots, r_c\}$, where each $r_j$ is either an ordinary conversation root $\text{MTR}_{\text{con}}$ or a deletion-state conversation root $\text{MTR}_{\text{con}}^{\text{del}}$. The account-level Merkle root is defined as

\[
\text{MTR}_{\text{acc}} = \mathsf{MerkleRoot}(\{r_1, r_2, \ldots, r_c\}).
\]

Any addition, deletion, or substitution of conversation leaves by the adversary alters the leaf set or leaf order of the account-level Merkle tree, thereby changing $\text{MTR}_{\text{acc}}$, unless a hash collision can be constructed. Since a valid account state must simultaneously pass verification under the server signature $\text{STR}_{\text{acc}}^{\text{s}}$ and the user signature $\text{STR}_{\text{acc}}^{\text{u}}$, the modified account state cannot be forged without access to the corresponding private keys. Any unauthorized modification to the account-level conversation state is therefore detected. 

\subsubsection{Multi-Device View Consistency Verification}

\noindent \textbf{Theorem 4 (State-Transition Authenticity).} 
Whether a device performs an independent single-interaction update or merges increments from other devices via $\texttt{UpdateConversationRecord}$ in a concurrent multi-device setting, a verified account-state root $\text{MTR}_{\text{acc}}$ and its associated signatures guarantee the authenticity of the corresponding state transition.

\noindent \textit{Proof.} 
During an update or synchronization, the server returns $\text{update\_payload}$, $\text{MerkleProofs}$, $\text{MTR}_{\text{acc}_{t+1}}$, and the server signature $\text{STR}_{\text{acc}_{t+1}}^{\text{s}}$. Using the locally anchored $\text{MTR}_{\text{acc}_t}$ as the baseline, proof-path verification is performed according to the operation type: for $\texttt{NewSession}$, it is verified that the new $\text{MTR}_{\text{con}}$ is incorporated into the account-level Merkle tree; for $\texttt{Append}$, the old $\text{hash}_{\text{tail}}$ is verified to be anchored to $\text{MTR}_{\text{con}_t}$, and the new $\text{hash}_{\text{tail}}$, $\text{MTR}_{\text{con}_{t+1}}$, and $\text{MTR}_{\text{acc}_{t+1}}$ are recomputed; for $\texttt{Branch}$, the branching point is verified to be anchored to $\text{MTR}_{\text{con}_t}$ via the prior branch tail, and the conversation root and account root after branch insertion are recomputed; for $\texttt{DeleteSession}$, $\text{MTR}_{\text{con}}^{\text{old}}$ is verified to be the current conversation leaf under $\text{MTR}_{\text{acc}_t}$, and $\text{MTR}_{\text{con}}^{\text{del}}$ and $\text{MTR}_{\text{acc}_{t+1}}$ are recomputed. In the merge scenario, concurrent increments being merged have typically already been verified by the corresponding devices and anchored to their then-confirmed account states via user signatures; the server therefore cannot introduce fabricated increments that were never confirmed by any user device. If the server omits, substitutes, or forges any operation-related object, the recomputed $\text{MTR}_{\text{acc}}'$ will differ from the server-returned $\text{MTR}_{\text{acc}_{t+1}}$, or $\text{STR}_{\text{acc}_{t+1}}^{\text{s}}$ will fail verification. Under the assumptions of collision-resistant hashing and unforgeable signatures, no incorrectly anchored update state can be accepted by a device. 

\noindent \textbf{Theorem 5 (Deletion Serialization and Deletion-Epoch Admission).} 
In a concurrent multi-device setting, VCT prevents deletion operations from producing undecidable merge conflicts with concurrent non-deletion updates.

\noindent \textit{Proof.} 
$\texttt{DeleteSession}$ is modeled in VCT as an account-level serialized operation. A deletion request must satisfy $\text{MTR}_{\text{acc}}^{\text{base}} = \text{MTR}_{\text{acc}}^{\text{current}}$, meaning a deletion may only be submitted against the server's current latest account root. If a deletion request is submitted against a stale account root, the server rejects it per protocol and requires the device to synchronize first. Upon commitment of a deletion, the target conversation root is replaced from $\text{MTR}_{\text{con}}^{\text{old}}$ to the deletion-state root $\text{MTR}_{\text{con}}^{\text{del}}$, and a new deletion epoch is established. Subsequently, before accepting any non-deletion update—$\texttt{NewSession}$, $\texttt{Append}$, or $\texttt{Branch}$—the server checks whether the $\text{MTR}_{\text{acc}}^{\text{base}}$ submitted by the device already reflects the latest deletion state. If the device remains in a prior deletion epoch, the server rejects the update and returns the deletion-state witness. Deletion operations are thereby handled as account-level serialized transitions, and non-deletion updates are admitted to the subsequent merge procedure only after the latest deletion state has been witnessed, preventing undecidable merge semantics between deletion states and concurrent non-deletion updates. 

\noindent \textbf{Theorem 6 (Merge Determinism).} 
For any set of concurrent non-deletion updates within the same deletion epoch, $\texttt{MergeStates}$ produces a uniquely determined output that is independent of merge ordering.

\noindent \textit{Proof.} 
Per Definition~1 in Section 4.2.3.3, $\texttt{Merge}$ satisfies inclusion, determinism, and conflict-free preservation. Concurrent updates targeting distinct conversations act on different leaves of the account-level Merkle tree; concurrent updates targeting the same conversation are preserved as distinct branches and ordered by deterministic rules—specifically, the initial-node timestamps of branches and the sibling timestamps under each parent node. The merge result depends solely on the input increment set, parent–child relationships, and timestamp ordering rules, and is independent of the order in which the server processes those increments. For any fixed input set, $\texttt{MergeStates}$ therefore produces a uniquely determined merged state. 

\noindent \textbf{Theorem 7 (Incremental-Replay Rejection).} 
The server cannot cause a device to accept omitted increments without detection by first returning a stale signed state and subsequently delivering those increments in a follow-up synchronization.

\noindent \textit{Proof.} 
Before accepting a synchronization or merge result and issuing a user signature, the device inspects the timestamp set of remote incremental nodes to be accepted in the current response. If any $\text{timestamp}_{\text{node}}$ satisfies $\text{timestamp}_{\text{sig}} < \text{timestamp}_{\text{node}} \leq \text{timestamp}_{\text{update}}$, the node is determined to have been produced before the last synchronization request was initiated but was not incorporated into the latest merged account state previously returned by the server and confirmed by the user; the merge result is accordingly rejected. Since the server cannot forge the user-signature-bound $\text{timestamp}_{\text{sig}}$ or $\text{timestamp}_{\text{node}}$, omitted increments cannot be accepted without detection via a strategy of returning a stale state followed by post-hoc delivery. 

\noindent \textbf{Theorem 8 (View Eventual Consistency and Gossip Accountability).} 
Upon execution of the gossip protocol, VCT satisfies the following dichotomy: either all devices belonging to the same user eventually converge to the same account state, or two devices exist that hold distinct account roots each bearing valid dual signatures, thereby constituting publicly verifiable fork evidence.

\noindent \textit{Proof.} 
All conversation states under the same account are aggregated into the account-level Merkle root $\text{MTR}_{\text{acc}}$, which is confirmed by dual signatures from the server and the user. During synchronization, the server merges non-deletion updates from the same deletion epoch that the device has not yet witnessed into the device's account view via $\texttt{MergeStates}$, and completes state-transition proof verification and dual-signature confirmation via Algorithm 1. If the server returns state transitions honestly, all devices eventually converge to the same $\text{MTR}_{\text{acc}}$. If incompatible account states are served to different devices, those devices exchange and verify their respective $(\text{MTR}_{\text{acc}},\ \text{STR}_{\text{acc}}^{\text{u}},\ \text{STR}_{\text{acc}}^{\text{s}})$ tuples during gossip. Whenever two devices hold distinct values of $\text{MTR}_{\text{acc}}$, the two valid but differing dual-signed account roots constitute fork evidence. Since the server cannot forge user signatures and cannot repudiate its own, the evidence is non-repudiable and suitable for accountability. 

\subsubsection{Shared Transcript Integrity Verification}

\noindent \textbf{Theorem 9 (Share Package Authenticity).} 
Every shared Q\&A node contained in a share package is verifiable as originating from an account state jointly confirmed by the user and the server; no unauthenticated content can be inserted into a share package by the server without detection.

\noindent \textit{Proof.} 
Per Section 4.2.4, the provenance proof package is defined as

\[
\textit{proof} = \{\text{selected\_nodes},\ \text{MerkleProofs},\ \text{MTR}_{\text{acc}},\ \text{STR}_{\text{acc}}^{\text{s}},\ \text{STR}_{\text{acc}}^{\text{u}}\}.
\]

Prior to generating the sharing snapshot, the server signature $\text{STR}_{\text{acc}}^{\text{s}}$ and the user signature $\text{STR}_{\text{acc}}^{\text{u}}$ are verified by the user device, confirming that the account root $\text{MTR}_{\text{acc}}$ has been mutually confirmed by both parties. The $\text{MerkleProofs}$ are then used to verify that each node to be shared belongs to the conversation structure authenticated by that account state and is ultimately anchored to $\text{MTR}_{\text{acc}}$. If the server attempts to insert a fabricated node $N'$ that does not appear in the account history, a valid Merkle inclusion proof for $N'$ must be produced. Since $N'$ did not participate in the computation of $\text{MTR}_{\text{acc}}$, verification fails unless the server can construct a forged Merkle proof or find a hash collision. Furthermore, if the server attempts to fabricate a new account root that incorporates $N'$, a valid user signature $\text{STR}_{\text{acc}}^{\text{u}}$ over that root must also be forged, which contradicts the existential unforgeability of the signature scheme. Under the assumptions of collision-resistant hashing and unforgeable signatures, every shared node in a share package is therefore verifiable as originating from an account state jointly confirmed by the user and the server, and no unauthenticated content can be inserted by the server. 

\noindent \textbf{Theorem 10 (Sharing Snapshot Integrity).} 
Once a sharing snapshot is generated, its verification is independent of the subsequent state of the original conversation tree; any modification, deletion, reordering, or omission of content in the share package is detected by the recipient $R$ upon verification.

\noindent \textit{Proof.} 
The sharing protocol reorganizes $\text{selected\_nodes}$ into an independent share chain and produces the chain tail $\text{share}_{\text{tail}}$. Subsequent appends, branches, or deletions in the original conversation tree affect only later values of $\text{MTR}_{\text{con}}$ or $\text{MTR}_{\text{acc}}$ and do not participate in the recomputation of an already-generated $\text{share}_{\text{tail}}$. Once a sharing snapshot is generated, its verification object is therefore decoupled from the original conversation tree. During verification, the recipient $R$ recomputes the share chain tail $\text{share}_{\text{tail}}'$ according to the share-chain construction rules, checks that $\text{share}_{\text{tail}}' = \text{share}_{\text{tail}}$, and verifies $\text{share}_{\text{sig}}^{\text{s}}$ and $\text{share}_{\text{sig}}^{\text{u}}$. Any content modification, node deletion, reordering, or omission from the signed share set alters $\text{share}_{\text{tail}}'$. If the adversary simultaneously replaces $\text{share}_{\text{tail}}$, valid signatures from both the server and the user over the new $\text{share}_{\text{tail}}$ must be forged, which contradicts the unforgeability of the signature scheme. Any tampering, reordering, deletion, or omission within a signed sharing snapshot is therefore detected. 

\subsubsection{Non-repudiation of Commitments}

\noindent \textbf{Theorem 11 (Operation Non-repudiation).} 
Any conversation update, state merge, session deletion, or share generation that has been confirmed by both parties is cryptographically non-repudiable.

\noindent \textit{Proof.} 
Account state-transition confirmation is accomplished through dual signatures: the server produces $\text{STR}_{\text{acc}}^{\text{s}}$ over the account root $\text{MTR}_{\text{acc}}$, and the user device, upon successful state-transition verification, produces $\text{STR}_{\text{acc}}^{\text{u}}$ over the same $\text{MTR}_{\text{acc}}$. State merges reuse the same account state-transition confirmation procedure. Session deletion produces the deletion-state conversation root $\text{MTR}_{\text{con}}^{\text{del}}$, which is jointly confirmed by both parties via signatures over the new account root $\text{MTR}_{\text{acc}}$. For sharing snapshots, $\text{share}_{\text{sig}}^{\text{s}}$ and $\text{share}_{\text{sig}}^{\text{u}}$ are produced by the server and the user, respectively, over the independent share chain tail $\text{share}_{\text{tail}}$. Since the signature scheme satisfies existential unforgeability~\cite{goldwasser1988digital,kremer2002formal,zhou1996fair}, $\text{STR}_{\text{acc}}^{\text{s}}$ and $\text{share}_{\text{sig}}^{\text{s}}$ can only be produced under the server private key, and $\text{STR}_{\text{acc}}^{\text{u}}$ and $\text{share}_{\text{sig}}^{\text{u}}$ can only be produced under the user private key. Each signature is bound directly to the account root $\text{MTR}_{\text{acc}}$ or the share chain tail $\text{share}_{\text{tail}}$; any modification to an operation outcome or shared content causes signature verification to fail. Neither the user nor the server can therefore repudiate any operation or its bound content that has been confirmed. 

\subsubsection{Summary of Security Guarantees}

\begin{longtable}{p{0.22\textwidth} p{0.52\textwidth} p{0.22\textwidth}}
\caption{Summary of Security Properties}
\label{tab:security_guarantees}\\
\toprule
\textbf{Security Property} & \textbf{Mechanism and Final Anchor} & \textbf{Cryptographic Assumptions} \\
\midrule
\endfirsthead

\caption*{Table \ref{tab:security_guarantees}: Summary of Security Properties (continued)}\\
\toprule
\textbf{Security Property} & \textbf{Mechanism and Final Anchor} & \textbf{Cryptographic Assumptions} \\
\midrule
\endhead

\midrule
\multicolumn{3}{r}{\textit{Continued on next page}} \\
\endfoot

\bottomrule
\endlastfoot

Node Integrity & $\text{parent\_hash}$ hash chain; recomputation of tail $\text{hash}_{\text{tail}}$ $\rightarrow$ anchors to $\text{STR}_{\text{acc}}$ ($\text{hash}_{\text{tail}} \rightarrow \text{MTR}_{\text{con}} \rightarrow \text{MTR}_{\text{acc}}$) & Collision-resistant hash + Unforgeable signatures \\
Conversation Integrity & Conversation-level Merkle tree over $\text{hash}_{\text{tail}}$ $\rightarrow$ anchors to $\text{STR}_{\text{acc}}$ ($\text{MTR}_{\text{con}} \rightarrow \text{MTR}_{\text{acc}}$) & Collision-resistant hash + Unforgeable signatures \\
Account Integrity & Account-level Merkle tree + $\text{STR}_{\text{acc}}^{\text{s}} / \text{STR}_{\text{acc}}^{\text{u}}$ $\rightarrow$ anchors to $\text{MTR}_{\text{acc}}$ & Collision-resistant hash + Unforgeable signatures \\
State-Transition Authenticity & $\text{MerkleProofs}$ + local recomputation + root reconstruction + dual-signature verification $\rightarrow$ anchors to $\text{MTR}_{\text{acc}}^{\text{new}} / \text{STR}_{\text{acc}}^{\text{s}} / \text{STR}_{\text{acc}}^{\text{u}}$ & Collision-resistant hash + Unforgeable signatures \\
Deletion Serialization & $\texttt{DeleteSession}$ requires $\text{MTR}_{\text{acc}}^{\text{base}} = \text{MTR}_{\text{acc}}^{\text{cur}}$; post-deletion new epoch formed with $\text{MTR}_{\text{con}}^{\text{del}}$ $\rightarrow$ anchors to $\text{MTR}_{\text{con}}^{\text{del}} / \text{MTR}_{\text{acc}} / \text{dual signatures}$ & Protocol check + Collision-resistant hash + Unforgeable signatures \\
Merge Determinism & $\texttt{MergeStates}$ + deterministic ordering + branch preservation $\rightarrow$ anchors to merged $\text{MTR}_{\text{acc}}$ + dual signatures & Deterministic ordering + Collision-resistant hash + Unforgeable signatures \\
Incremental-Replay Rejection & Rejection of remote increments satisfying $\text{timestamp}_{\text{sig}} < \text{timestamp}_{\text{node}} \leq \text{timestamp}_{\text{update}}$ $\rightarrow$ anchors to synchronization boundary and set of accepted increments & Unforgeable signatures \\
View Consistency and Gossip Accountability & Device synchronization + gossip exchange and verification of $(\text{MTR}_{\text{acc}}, \text{STR}_{\text{acc}}^{\text{u}}, \text{STR}_{\text{acc}}^{\text{s}})$; divergent roots produce fork evidence $\rightarrow$ anchors to consistent $\text{MTR}_{\text{acc}}$ or two distinct dual-signed roots & Unforgeable signatures + Collision-resistant hash + Gossip channel security + eventual synchronization \\
Share Authenticity & Dual-signed $\text{MTR}_{\text{acc}}$; $\text{MerkleProofs}$ prove nodes belong to that state $\rightarrow$ anchors to $\text{MTR}_{\text{acc}}$ + dual signatures & Collision-resistant hash + Unforgeable signatures \\
Share Integrity & Reorganize $\text{selected\_nodes}$ into independent hash chain; recompute $\text{share}_{\text{tail}}$ and verify dual signatures $\rightarrow$ anchors to $\text{share}_{\text{tail}} / \text{share}_{\text{sig}}^{\text{s}} / \text{share}_{\text{sig}}^{\text{u}}$ & Collision-resistant hash + Unforgeable signatures \\

\end{longtable}

\section{Experimental Evaluation}

\subsection{Prototype System and Evaluation Scope}

A prototype system is implemented in Python to evaluate the protocol overhead and security-goal achievability of VCT. The prototype employs standard cryptographic libraries—SHA-256, Ed25519, and PBKDF2-HMAC-SHA256—and realizes the core protocol paths defined in Section 4, including Q\&A node hash chains, conversation-level and account-level Merkle trees, the four state-update operations, multi-device synchronization, gossip-based fork detection, and the verifiable sharing protocol.

Account-state objects in the prototype are kept consistent with the formal model. Q\&A nodes form branch hash chains via $\text{parent\_hash}$; each conversation constructs $\text{MTR}_{\text{con}}$ from the $\text{hash}_{\text{tail}}$ values of its internal branches; the account layer constructs $\text{MTR}_{\text{acc}}$ from the conversation roots or deletion-state conversation roots, with dual signatures issued separately by the server and the user. Multi-device synchronization employs deterministic merging of non-deletion updates; deletion operations follow the account-level serialization rules; the sharing procedure reorganizes selected nodes into an independent share chain with appended dual signatures.

It should be noted that the account state-update protocol is designed such that the server recomputes only the Merkle path from the affected leaf to the root when updating $\text{MTR}_{\text{acc}}$. For implementation simplicity, the current prototype adopts linear recomputation or cache-flush updates on the server side. The experimental results therefore reflect end-to-end overhead under a conservative server-side implementation rather than the theoretical optimum achievable with incremental Merkle maintenance.

\subsection{Experimental Setup}

\noindent \textbf{Platform.} 
All experiments are conducted on an Intel Core i7-13700H (2.40 GHz, 14 cores / 20 threads) with 32 GB RAM, a 256 GB NVMe SSD, and Windows 11 (10.0.26200). Measurements cover only the cryptographic computation and in-memory processing of VCT, excluding LLM inference, network I/O, persistent storage I/O, and client-side rendering.

\noindent \textbf{Cryptographic and Merkle Configuration.} 
The prototype instantiates SHA-256~\cite{nist2015fips180} (32-byte digest, represented as 64 hexadecimal characters) and the Ed25519 digital signature algorithm~\cite{bernstein2012high,rfc8032} (32-byte public and private keys, 64-byte signatures). The KDF is instantiated as PBKDF2-HMAC-SHA256~\cite{kaliski2000pkcs5,nist800132}. The Merkle tree follows the recursive binary tree structure of RFC 6962, with leaves as pre-hashed commitments and internal nodes aggregated as $\mathsf{Hash}(\textit{left} \| \textit{right})$. Only the tree structure of RFC 6962 is adopted; the leaf and internal-node domain-separation prefixes (0x00 and 0x01) are not used, so the results do not guarantee interoperability with RFC 6962 implementations. The empty-tree root is defined as the fixed constant $\texttt{ZERO\_HASH} = \texttt{"0"} \times 64$, serving solely as an implementation placeholder.

\noindent \textbf{Measurement Methodology and Latency Definition.} 
Timing measurements are based on a monotonic clock with microsecond precision. Unless stated otherwise, each experimental configuration is repeated independently 50 times, and results are reported as the sample mean with sample standard deviation.

\noindent \textbf{State-Update Operation Latency (End-to-End).} 
For account state-update operations—$\texttt{NewSession}$, $\texttt{Append}$, $\texttt{Branch}$, $\texttt{DeleteSession}$, and $\texttt{Merge}$—latency is defined as the end-to-end time required to complete one logical operation, denoted $T_{e2e}$ and computed as:

\[
T_{e2e} = t_{\text{local}} + t_{P1} + t_{P2} + t_{P3},
\]

where $t_{\text{local}}$ is the time to execute the account state transition in memory (or the merge computation along the merge path); $t_{P1}$ is the server-side time to compute $\text{MTR}_{\text{acc}}$ over the updated account state, generate the state-transition proof ($\text{update\_payload}$ and $\text{MerkleProofs}$), and issue $\text{STR}_{\text{acc}}^{\text{s}}$; $t_{P2}$ is the user-side time to verify the state-transition proof and $\text{STR}_{\text{acc}}^{\text{s}}$ and to issue $\text{STR}_{\text{acc}}^{\text{u}}$; and $t_{P3}$ is the server-side time to verify $\text{STR}_{\text{acc}}^{\text{u}}$ and for both parties to commit the current account state.

Nested sub-timings within $T_{e2e}$ are reported separately for Merkle- and signature-related overhead:

\begin{itemize}
\item \textbf{Merkle:} nested sub-timing covering server-side account root recomputation and device-side Merkle proof verification.
\item \textbf{Verify:} nested sub-timing covering Ed25519 signature verification of $\text{STR}_{\text{acc}}^{\text{s}}$ and user signing of $\text{STR}_{\text{acc}}^{\text{u}}$.
\end{itemize}

\noindent \textbf{Other Protocol Operation Overhead.}

\begin{itemize}
\item \textbf{Gossip:} the latency of a single fork-detection round is measured, comprising four Ed25519 signature verifications over the account anchor digests of both devices ($\text{STR}_{\text{acc}_A}^{\text{s}}$, $\text{STR}_{\text{acc}_A}^{\text{u}}$, $\text{STR}_{\text{acc}_B}^{\text{s}}$, $\text{STR}_{\text{acc}_B}^{\text{u}}$) and one $\text{MTR}_{\text{acc}}$ comparison.
\item \textbf{Verifiable sharing:} $\textit{share\_generate}$ covers independent share chain construction and dual signing of the sharing snapshot; $\textit{verify\_share}$ covers recipient-side recomputation of $\text{share}_{\text{tail}}$ and verification of $\text{share}_{\text{sig}}^{\text{s}}$ and $\text{share}_{\text{sig}}^{\text{u}}$.
\item \textbf{Merkle proof:} inclusion proof sizes (in bytes) are reported at fixed sampling points on the terminal accumulated account state; $\textit{proof\_sample\_at\_scale}$ additionally measures proof generation latency at the same positions.
\end{itemize}

\noindent \textbf{Reproducibility.} 
Experiment timestamps are simulated using monotonically increasing integers. The $\text{model}_{\text{config}}$ of each Q\&A node is fixed as $\{\texttt{"model\_id"}: \texttt{"eval"},\ \texttt{"temperature"}: 0.0\}$, and $\text{file\_aux\_info}$ is set to an empty dictionary in all synthetic workloads. The conversation creation time $\text{created\_at}$ is taken from the timestamp of the first node; account-level conversation ordering follows $(\text{created\_at},\ \text{session\_id})$, where $\text{session\_id}$ serves as a stable tiebreaker to ensure deterministic Merkle tree construction at all levels during merging.

\subsection{Experimental Workloads and Security Test Coverage}

Three progressively scaled workloads—Basic, Medium, and Large—are defined to evaluate the functional correctness, security-goal coverage, and scalability of VCT across different account sizes. All three workloads employ a consistent protocol feature set and security test matrix, but differ in the number of Q\&A pairs, conversations, and operation executions, serving respectively to assess the intrinsic overhead of individual protocol components and the scalability under accumulated account states.

\noindent \textbf{Isolated-operation baseline.} 
This baseline measures the end-to-end intrinsic overhead of individual protocol operations under a minimal account state. $\texttt{NewSession}$ is measured against an empty account state; all other update and verification operations are executed against an account state containing a single existing conversation. The baseline is designed to isolate the fundamental costs of cryptographic confirmation, Merkle root computation, and signature verification.

\noindent \textbf{Scaled baseline.} 
The same set of protocol operations is measured against the accumulated account state formed by each workload, yielding trends in latency, storage overhead, and Merkle proof size as a function of account scale.

\begin{longtable}{>{\raggedright}p{0.12\textwidth} >{\raggedright}p{0.10\textwidth} >{\raggedright}p{0.10\textwidth} >{\raggedright}p{0.12\textwidth} >{\raggedright\arraybackslash}p{0.42\textwidth}}
\caption{Experimental workload configurations}
\label{tab:workloads}\\
\toprule
\textbf{Workload} & \textbf{Sessions} & \textbf{Total Q\&A} & \textbf{Avg Q\&A / session} & \textbf{Core operations and counts} \\
\midrule
\endfirsthead

\caption*{Table \ref{tab:workloads}: Experimental workload configurations (continued)}\\
\toprule
\textbf{Workload} & \textbf{Sessions} & \textbf{Total Q\&A} & \textbf{Avg Q\&A / session} & \textbf{Core operations and counts} \\
\midrule
\endhead

\midrule
\multicolumn{5}{r}{\textit{Continued on next page}} \\
\endfoot

\bottomrule
\endlastfoot

Basic & 5 & 25 & 5 & Branch commit $\times 2$; session deletion $\times 1$; remote append merge $\times 2$; parallel branch merge $\times 1$; deletion-epoch rejection detection $\times 1$; session sharing $\times 1$; gossip consistency $\times 1$; gossip fork $\times 1$; gossip invalid signature $\times 1$; lagging-device detection $\times 1$; fork evidence size check $\times 1$ \\
Medium & 20 & 100 & 5 & Branch commit $\times 5$; session deletion $\times 4$; remote append merge $\times 5$; parallel branch merge $\times 1$; deletion-epoch rejection detection $\times 1$; session sharing $\times 5$; gossip consistency $\times 10$; gossip fork $\times 5$; gossip invalid signature $\times 5$; lagging-device detection $\times 1$; fork evidence size check $\times 1$ \\
Large & 100 & 500 & 5 & Branch commit $\times 10$; session deletion $\times 20$; remote append merge $\times 10$; parallel branch merge $\times 1$; deletion-epoch rejection detection $\times 1$; session sharing $\times 5$; gossip consistency $\times 25$; gossip fork $\times 10$; gossip invalid signature $\times 10$; lagging-device detection $\times 1$; fork evidence size check $\times 1$ \\
\end{longtable}

Table 5.2 summarizes the correspondence between VCT security goals and specific test cases.

\begin{longtable}{>{\raggedright}p{0.28\textwidth} >{\raggedright\arraybackslash}p{0.64\textwidth}}
\caption{Security goals and test coverage}
\label{tab:security_test_coverage}\\
\toprule
\textbf{Security goal} & \textbf{Test operations} \\
\midrule
\endfirsthead

\caption*{Table \ref{tab:security_test_coverage}: Security goals and test coverage (continued)}\\
\toprule
\textbf{Security goal} & \textbf{Test operations} \\
\midrule
\endhead

\midrule
\multicolumn{2}{r}{\textit{Continued on next page}} \\
\endfoot

\bottomrule
\endlastfoot

Integrity & Message append, branch creation, session deletion, content removal of deleted sessions, Merkle path verification from Q\&A nodes to account root \\
Multi-device consistency and fork detection & Remote append merge, parallel branch merge, deletion strict-freshness check, deletion-epoch check, omitted-interval incremental replay, final view consistency check, fork evidence size verification \\
Share verifiability & Provenance proof verification, independent share chain verification, share chain node tampering and omission, chain-tail signature substitution \\
Non-repudiation & Account root dual-signature verification, share chain tail dual-signature verification, signed payload tampering \\
\end{longtable}

Experimental results demonstrate that across all three workloads, all security functional tests satisfy the expected protocol outcomes. VCT correctly maintains the complete authentication chain from Q\&A nodes through branch tails, conversation roots, and account roots. In concurrent multi-device settings, all valid concurrent increments are deterministically merged, deletion operations are serialized, increments falling within the signed timestamp boundary are rejected, and account-view consistency is detected via gossip. The independent share chain supports independent verification of selectively disclosed nodes and detects modifications, omissions, reorderings, and chain-tail signature substitutions in shared content. These results demonstrate that the hash binding, Merkle authentication structure, dual signatures, deterministic merging, freshness checks, gossip detection, and independent share chain mechanisms of VCT collectively support the core security goals defined in this work across varying account scales.

\subsection{Minimal-State Operation Overhead}

To characterize the intrinsic overhead of individual VCT protocol operations, latency is measured in this section under a minimal-state context. Specifically, $\texttt{NewSession}$ is executed against an empty account state, and all other operations are executed against a minimal account state containing a single existing conversation. The results in this section therefore reflect per-operation protocol costs under a minimal account state and are intended to characterize intrinsic overhead; scalability under accumulated account states is analyzed further in Section 5.5.

\begin{figure}[H]
\centering
\includegraphics[width=0.8\textwidth]{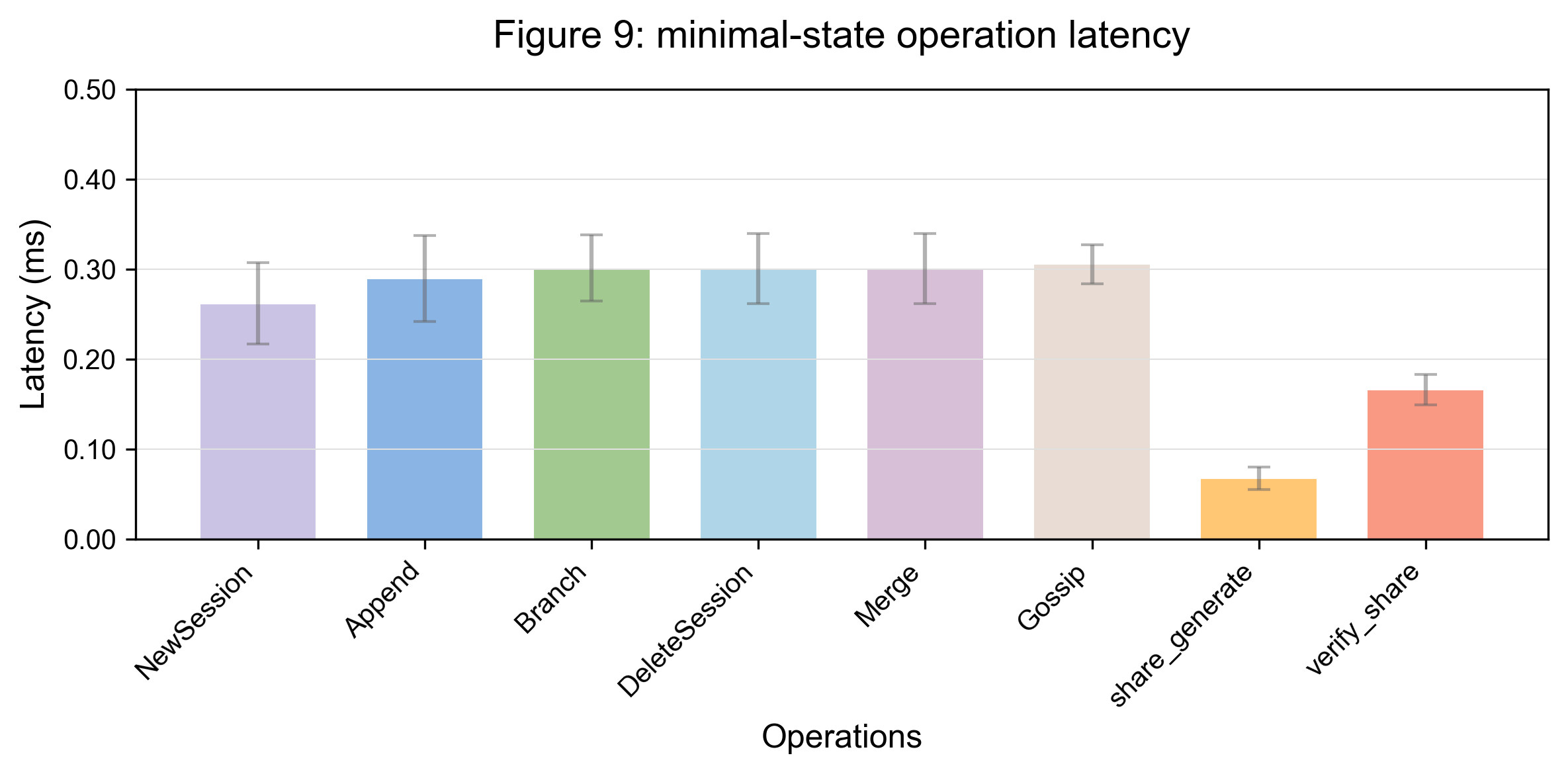}
\caption{Minimal-state operation latency ($N = 50$, synthetic payload). Account updates (including merge) report end-to-end latency $T_{e2e}$. Gossip and sharing are measured as standalone primitives. Share operations use 2 nodes per trial without owner proofs.}
\label{fig:minimal-state-latency}
\end{figure}

As shown in Figure 5.1, all protocol operations under the minimal account state fall within the sub-millisecond to low-millisecond range. The mean latency of $\texttt{NewSession}$, $\texttt{Append}$, $\texttt{Branch}$, $\texttt{DeleteSession}$, $\texttt{Merge}$, and Gossip is concentrated in approximately 0.26–0.31 ms; $\textit{share\_generate}$ (0.068 ms) and $\textit{verify\_share}$ (0.17 ms) exhibit lower absolute latency. These results indicate that the basic operations of the VCT protocol do not introduce significant interaction latency.

To identify the dominant sources of per-operation overhead, Figure 5.2 compares the per-phase latency of the four core update operations, together with the nested Merkle and Verify sub-timings within each phase.

\begin{figure}[H]
\centering
\includegraphics[width=0.8\textwidth]{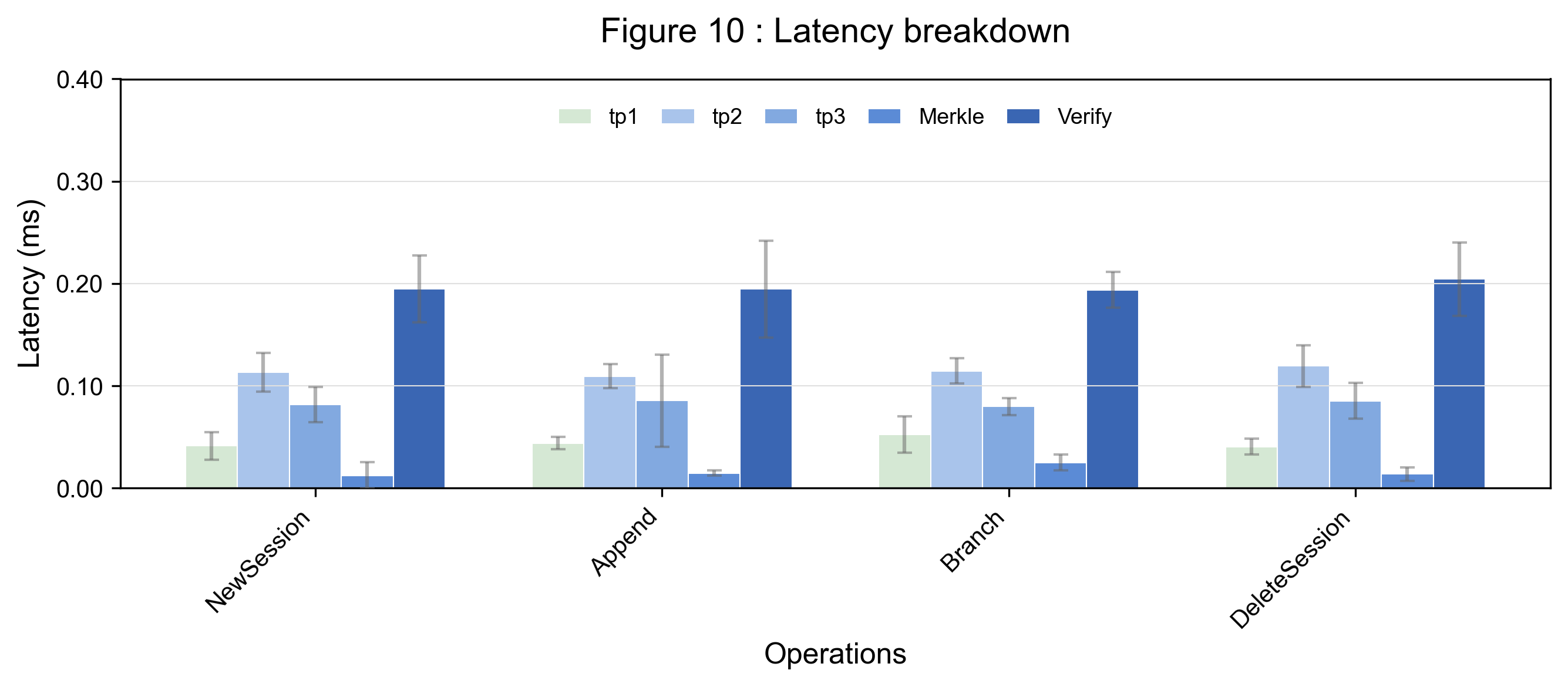}
\caption{Protocol latency breakdown of core update operations. This figure exclusively decomposes protocol-layer cryptographic and synchronization overhead and does not include $t_{\text{local}}$—the latency of in-memory state modifications. Error bars represent standard deviation across 50 repeated trials.}
\label{fig:latency-breakdown}
\end{figure}

Component-level results indicate that Ed25519 signing and verification constitute a stable and relatively dominant source of cryptographic overhead; Merkle-related costs are comparatively low, as the device side executes only logarithmic-depth proof-path verification. Under the minimal account state, the baseline latency of VCT is therefore dominated by dual-signature confirmation rather than by hierarchical state-structure transitions.

\subsection{Cross-Scale Performance Evaluation}

This section evaluates the scalability of the current prototype under accumulated account states. Unlike the isolated-operation baseline of Section 5.4, all operations in this section are executed across the Basic, Medium, and Large account scales to analyze the impact of account-state growth on protocol latency.

\subsubsection{Scalability Trends for Account-State-Transition Operations}

To assess the impact of account-state scale on state-transition operations, Figure 11 compares the latency of $\texttt{NewSession}$, $\texttt{Append}$, $\texttt{Branch}$, $\texttt{DeleteSession}$, $\texttt{Merge}$, and $\textit{proof\_sample}$ across account states of 5, 20, and 100 conversations. All of these operations directly or indirectly depend on account-level Merkle root updates or account-level proof generation, and therefore capture the primary performance bottlenecks of the current prototype as account state scales.

\begin{figure}[H]
\centering
\includegraphics[width=0.8\textwidth]{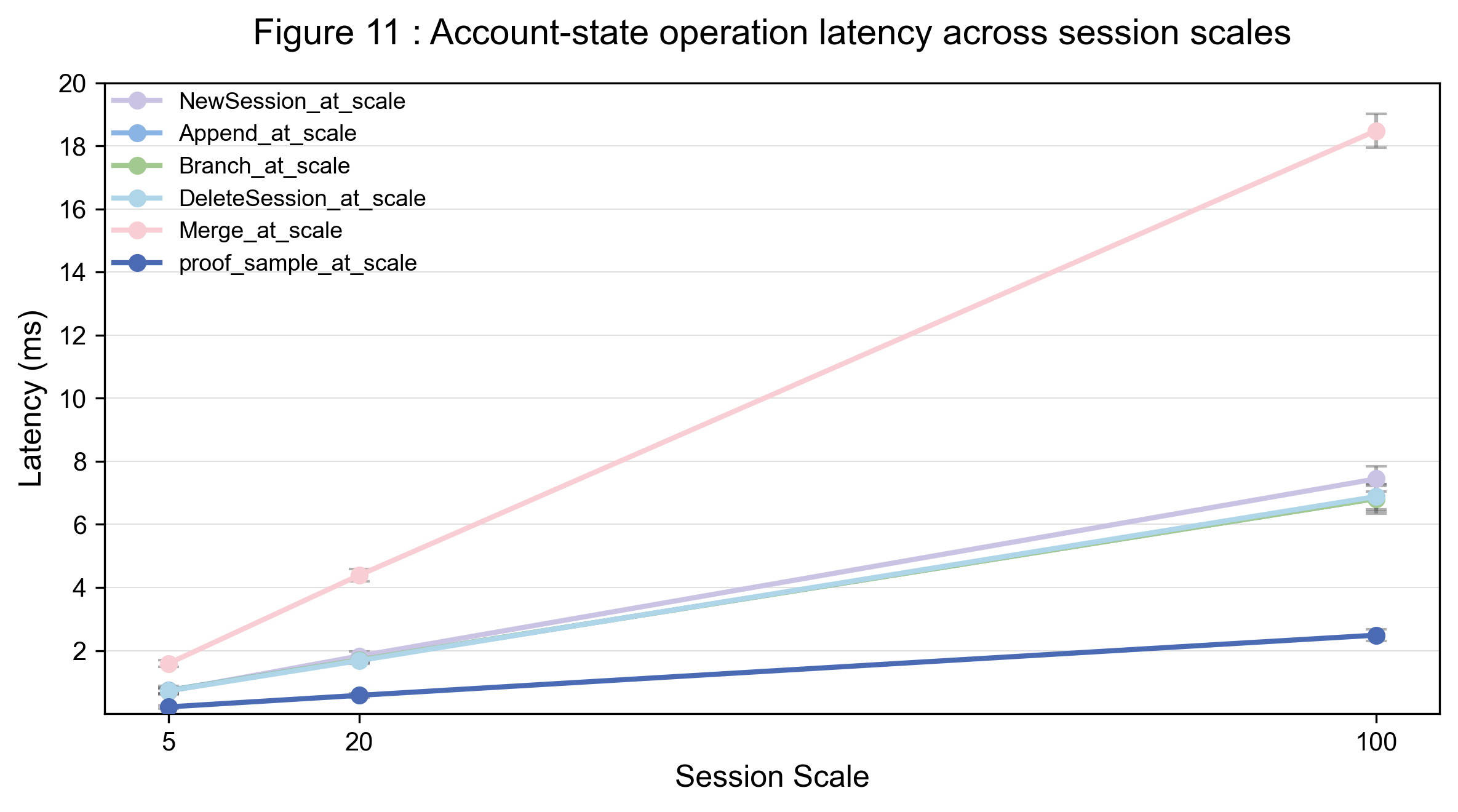}
\caption{Scalability trends for account-root-modifying operations ($N = 50$, synthetic payload). Account sizes range from 5 to 100 conversations.}
\label{fig:scalability-trends}
\end{figure}

As shown in Figure 11, the end-to-end latency of $\texttt{NewSession}$, $\texttt{Append}$, $\texttt{Branch}$, and {\texttt{Delete\-Session}}
increases with account scale, reaching approximately 6.8–7.4 ms at the Large scale. The growth is most pronounced for $\texttt{Merge}$: approximately 4.4 ms at Medium scale and approximately 18.5 ms at Large scale, representing roughly $2.6\times$ and $2.7\times$ the latency of $\texttt{Append}$ at the corresponding scales. This disparity stems from the account-level batch semantics of $\texttt{Merge}$: concurrent modifications from multiple devices are first aligned and consolidated into a unified account state; the account-level and conversation-level Merkle roots are then recomputed, and separate Merkle inclusion proofs are generated for each conversation whose content has changed; the device side must further verify multiple transition proofs and complete dual-signature confirmation. By contrast, $\textit{proof\_sample}$ measures only the cost of generating Merkle inclusion proofs at fixed sampling points—the terminal conversation root and the first active branch tail—on the accumulated account end state, remaining at approximately 2.5 ms at the Large scale. This result indicates that proof-path overhead scales primarily with Merkle tree depth rather than linearly with the full account state; the latency of a single conversation update is dominated by the update to the affected branch or conversation root, account-root reconstruction, transition proof generation, and dual-signature confirmation. $\texttt{Merge}$ additionally incurs the cost of multi-conversation consolidation and multi-path account-level proof generation, making it the primary latency contributor in large-scale settings.

\subsubsection{Lightweight Security Operation Overhead}

To distinguish the performance characteristics of account-state-modifying operations from those of digest-oriented verification operations, Figure 5.4 compares the latency of $\textit{gossip}$, $\textit{share\_generate}$, and $\textit{verify\_share}$ across three account scales. These operations depend primarily on account-root digests, signed states, or independent share chains, and do not require traversal of the full account state.

\begin{figure}[H]
\centering
\includegraphics[width=0.8\textwidth]{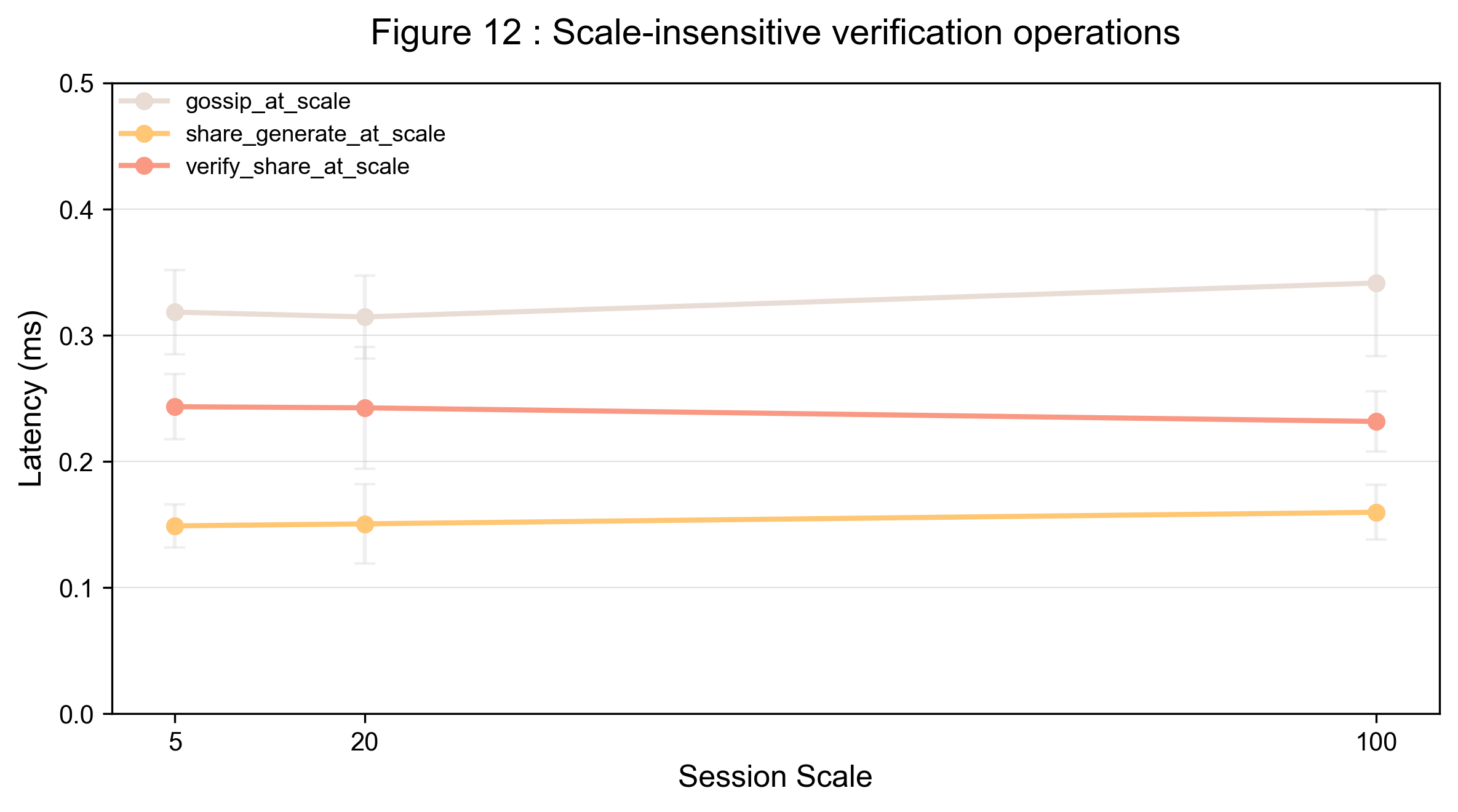}
\caption{Scale experiment for verification-oriented operations ($N = 50$, synthetic payload). The account scale varies from 5 to 100 sessions. $\textit{gossip\_at\_scale}$ verifies only account roots and their dual signatures; sharing operations use public share packages containing 20 selected nodes and omit owner proofs.}
\label{fig:scale-verification}
\end{figure}

As shown in Figure 12, $\textit{gossip}$, $\textit{share\_generate}$, and $\textit{verify\_share}$ remain stable across account scales of 5, 20, and 100 conversations. Gossip latency ranges from 0.31 to 0.34 ms, while $\textit{share\_generate}$ and $\textit{verify\_share}$ remain approximately 0.15 ms and 0.23–0.24 ms, respectively. These results indicate that digest-oriented security operations exhibit weak dependence on account scale: gossip exchanges and verifies only account roots and dual signatures, whereas share generation and verification construct an independent hash chain from the shared nodes without traversing the full account tree.

Taken together, Figures 11 and 12 show that scale-sensitive overhead in the current prototype is concentrated in account-state merging, server-side account-state transitions, and transition-proof generation. In contrast, gossip and public share-package verification maintain low latency with negligible growth as account scale increases. Future engineering optimizations should therefore focus primarily on the $\texttt{Merge}$ path—specifically, conversation-root indexing, $\text{hash}_{\text{tail}}$ caching for unmodified branches, and incremental Merkle-tree maintenance based on intermediate-node caching—so that account-root recomputation is confined to the authentication path from the affected leaf to the root, rather than on gossip or share-verification procedures.

\subsection{Storage Overhead and Merkle Proof Size}

To evaluate the storage overhead introduced by security metadata and Merkle proofs under synthetic workloads, Figure 13 reports the total volume of security metadata together with the sizes of account-level and conversation-level inclusion proofs measured at fixed sampling positions in the final account states of the Basic, Medium, and Large workloads. Security metadata encompasses Q\&A node hashes, conversation roots, account roots, dual signatures, and historical account-root metadata. Proof sizes are sampled at fixed positions, namely the last conversation root in the account tree and the first active branch leaf of the first active conversation, after each workload reaches its final state.

\begin{figure}[H]
\centering
\includegraphics[width=0.8\textwidth]{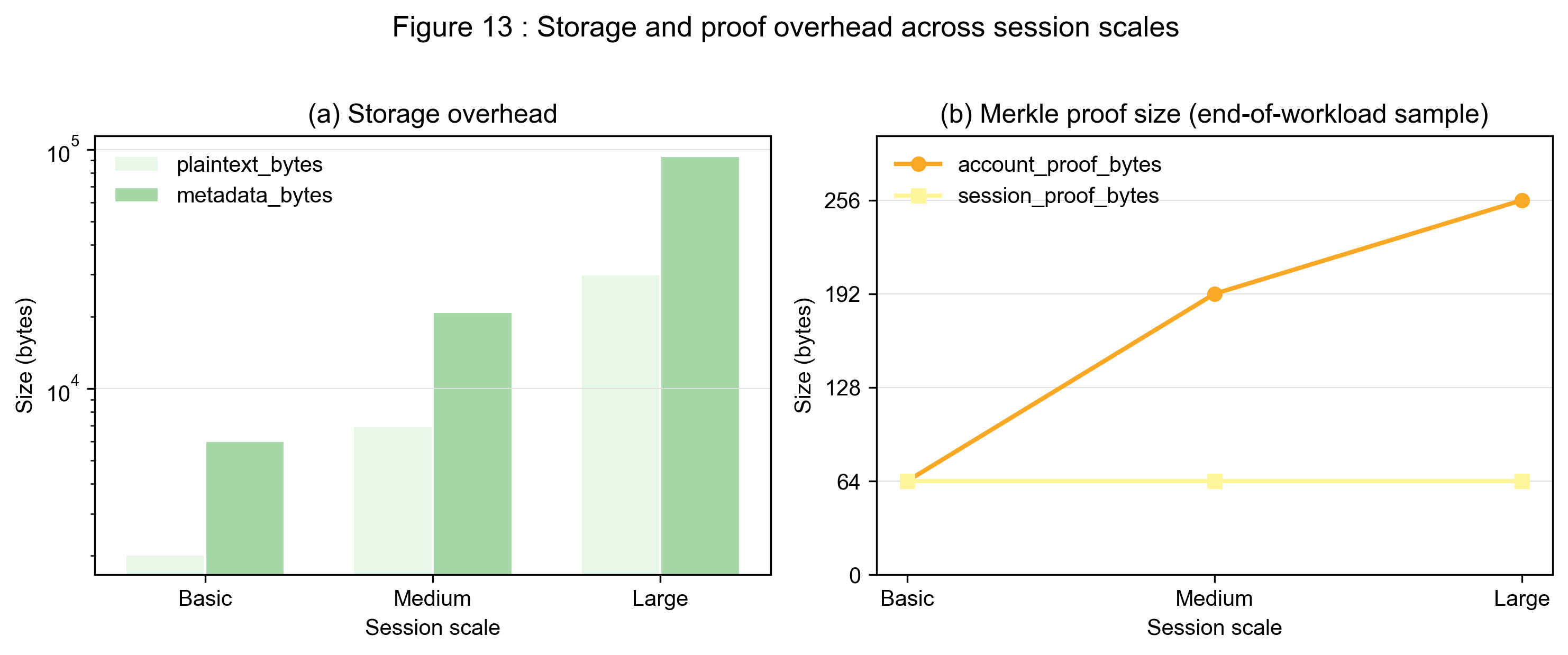}
\caption{Storage overhead and Merkle proof size: (a) security metadata volume; (b) account-level and conversation-level proof sizes.}
\label{fig:storage-proof}
\end{figure}

As shown in Figure 13(a), security metadata grows steadily with the number of nodes and conversations under the synthetic short-text workload. Since the synthetic Q\&A content is extremely short, the ratio of metadata to plaintext is substantially amplified (approximately 3:1 across all three workloads). This ratio should not be extrapolated directly to real-world long-text deployments. The experiment is more appropriately interpreted as characterizing the growth trend of metadata as a function of state scale rather than as an estimate of its absolute storage proportion.

As shown in Figure 13(b), account-level proof size grows sub-linearly with the number of conversations, increasing from 64 B to 192 B and 256 B. Conversation-level proof size remains at 64 B across all three scales, since the fixed sampling position targets the first active conversation whose branch structure is identical across all workloads. This result reflects a key design property of the hierarchical Merkle structure adopted by VCT: account-level proof size scales primarily with the depth of the account tree, while conversation-level proof size is determined by the branch structure within a single conversation and is largely decoupled from the total number of conversations in the account.

A verifier therefore requires only the proof path relevant to the target position rather than the complete account state. Verification can be completed using local authentication information without reconstructing the entire account-level Merkle structure.

\subsection{Real-Payload Evaluation}

To evaluate the impact of conversation content length on performance and storage proportion, two payload configurations are compared under the Medium workload (20-session accumulated account state). The synthetic payload uses minimal placeholder Q\&A content and omits the $\text{file\_aux\_info}$ attachment field, resulting in approximately 41 B of plaintext per node. This configuration is intended to highlight the fixed protocol overhead introduced by hashing, signing, and authenticated-state maintenance. In contrast, the real payload populates $Q$ and $A$ with approximately 4 KB and 16 KB of content, respectively, and attaches approximately 512 B of file metadata per node (including filename, MIME type, SHA-256 digest, and related fields), yielding approximately 21 KB of plaintext per node. This configuration more closely reflects the content volume of practical LLM conversations.

\begin{figure}[H]
\centering
\includegraphics[width=0.8\textwidth]{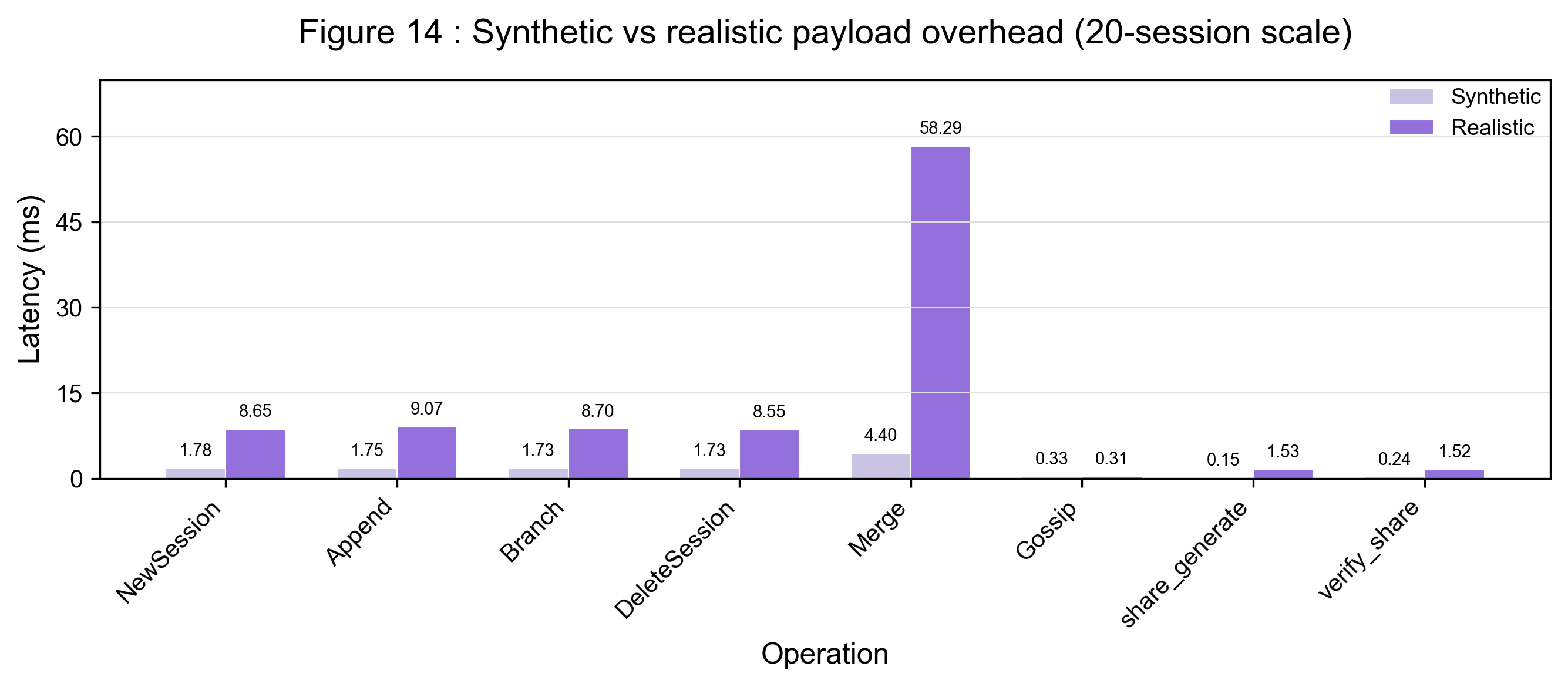}
\caption{Real-payload evaluation: latency comparison between synthetic and real workloads.}
\label{fig:real-payload}
\end{figure}

Storage measurements show that security metadata occupies 20,896 B under both payload configurations. This metadata is determined primarily by fixed structures such as node hashes, conversation roots, account roots, dual signatures, and associated state metadata, and is therefore largely independent of Q\&A content length. In contrast, total plaintext volume increases from approximately 7 KB to approximately 2.30 MB, causing the metadata-to-plaintext ratio to decrease from approximately 299\% to approximately 0.9\%. These results indicate that the elevated metadata ratio observed under the synthetic workload arises from the extremely short placeholder content rather than from any intrinsic inflation of the security metadata itself. In realistic long-text deployments, VCT security metadata grows approximately linearly with conversation scale while remaining a small fraction of the overall conversation content.

Figure 14 further characterizes the impact of payload size on protocol latency. Compared with the synthetic workload, real Q\&A content increases $\texttt{Merge}$ latency from approximately 4.4 ms to approximately 58.3 ms, and increases the latency of $\texttt{NewSession}$, $\texttt{Append}$, $\texttt{Branch}$, and $\texttt{DeleteSession}$ from approximately 1.7–1.8 ms to approximately 8.5–9.1 ms. This increase is attributable primarily to content-dependent node hashing and branch-tail recomputation. Larger Q\&A payloads increase the size of hash inputs and therefore raise the computational cost of updating chain tails along affected branches. In contrast, conversation-level and account-level Merkle trees, together with their authentication paths, operate on fixed-length hash values. Their overhead is determined primarily by the number of branches, the number of conversations, and the scope of state changes, rather than by content length. The amplification is most pronounced for $\texttt{Merge}$, which additionally requires multi-branch tail recomputation, multi-conversation consolidation, account-root reconstruction, and corresponding proof generation.

By comparison, gossip latency remains approximately 0.31 ms under both payload configurations because only account roots and dual signatures are exchanged and verified, with no dependence on conversation content. Among the sharing-related operations, $\textit{share\_generate}$ and $\textit{verify\_share}$ rely on an independent share chain and snapshot dual signatures and likewise do not require traversal of the full account tree. Under the real payload, their latencies are approximately 1.53 ms and 1.52 ms, respectively, compared with approximately 0.15 ms and 0.24 ms under the synthetic workload. The increase arises primarily from recomputation of share-chain hashes over larger shared nodes. Even so, their absolute latencies remain substantially lower than those of account-state update operations under the real workload.

Overall, the primary performance bottleneck of VCT in practical deployments is associated with processing conversation content rather than maintaining authenticated state structures. Gossip and share-verification operations exhibit negligible or only mild sensitivity to payload size and therefore retain favorable deployment characteristics even under realistic workloads.

\subsection{Experimental Discussion}

This chapter presents a systematic evaluation of the functional correctness, security-goal coverage, and performance characteristics of the VCT prototype across three progressively scaled workloads—Basic, Medium, and Large—and two payload configurations—synthetic and real. The experimental results support the following observations.

\noindent \textbf{Functionality and Security.} 
The prototype successfully implements the core mechanisms of three-layer state assertions, multi-device state merging, gossip-based fork detection, and verifiable sharing. Across all experimental configurations, the designed test cases covering integrity, consistency, non-repudiation, and share verifiability are satisfied. These results provide implementation-level evidence that the protocol achieves its intended security objectives and is practically deployable.

\noindent \textbf{Intrinsic Protocol Overhead.} 
Under minimal account-state and short-text workloads, the latency of the core protocol primitives remains predominantly in the sub-millisecond range. This indicates that fixed cryptographic operations—including signature verification, proof validation, digest-oriented gossip, and independent share-chain verification—do not constitute a major source of end-to-end protocol latency. Subsequent performance variation is attributable primarily to account scale, conversation content volume, and merge-path complexity.

\noindent \textbf{Scalability Characteristics.} 
As the number of conversations increases, the latency of single-conversation state updates and $\texttt{Merge}$ grows most noticeably, with $\texttt{Merge}$ emerging as the dominant bottleneck under large-scale workloads. In contrast, gossip and sharing-related operations exhibit only limited sensitivity to account scale and maintain relatively stable latency across all tested configurations.

\noindent \textbf{Storage and Proof Overhead.} 
Security metadata grows approximately linearly with the number of conversations. Account-level and conversation-level proof sizes are bounded primarily by account-tree depth and single-conversation branch structure, respectively, and therefore do not scale with the size of the complete account state. Under realistic payloads, the ratio of metadata to conversation content is substantially lower than under synthetic short-text workloads, indicating that the cryptographic assertions introduced by VCT impose only modest storage overhead in practice.

\noindent \textbf{Real-Payload Behavior and Optimization Directions.} 
Conversation content volume is the primary performance-sensitive factor in realistic deployments. State-update paths are affected most significantly, with $\texttt{Merge}$ exhibiting the largest amplification effect. By contrast, anchor verification and gossip are essentially unaffected by payload size, while sharing operations exhibit only mild sensitivity. Future optimizations should therefore focus on branch-tail caching, incremental merge processing, and server-side incremental Merkle-tree maintenance rather than on the dual-signature mechanism or digest-exchange procedures themselves.

Overall, the experimental results provide evidence that VCT achieves its functional and security objectives while maintaining manageable overhead under both scaled workloads and realistic payload conditions. Further reductions in latency and storage overhead remain achievable through engineering optimizations such as branch-tail caching, incremental merge processing, and incremental Merkle maintenance.

\section{Discussion and Future Work}
\subsection{Scope and Limitations}

The primary objective of VCT is to provide verifiable authenticity, integrity, and accountability for large language model conversation records. The protocol enables users to verify whether a given conversation state has been correctly generated, stored, synchronized, and shared, and to detect unauthorized tampering with conversation history.

VCT does not, however, attempt to verify the factual correctness, reasoning validity, or semantic quality of model-generated responses. The protocol guarantees that a response has been faithfully recorded and cryptographically anchored to a signed account state, but provides no evidence regarding the correctness of the response content itself. VCT should therefore be understood as a conversation-record verification framework rather than a verifiable reasoning framework.

In multi-device settings, VCT adopts eventual-consistency semantics: account consistency is established primarily through verifiable synchronization between devices and the server and through $\texttt{Merge}$ commits, while gossip is employed for fork detection and reconciliation across multi-device views. Prior to synchronization completion, transient inconsistencies may exist between the local views of different devices; under the assumption that honest devices eventually complete synchronization with the server and participate in gossip, the system achieves eventual convergence while preserving an accountable operation chain.

Security metadata accumulates over prolonged use. Account state assertions, dual signatures, and deletion-related records must be retained to support auditability and verifiability; the total volume grows near-linearly with the number of conversations and the frequency of state commits. The evaluation in Section 5 demonstrates that, under real long-text workloads, the ratio of security metadata to conversation content is low; however, this result is based on the Medium workload end state and does not cover long-retention or high-commit-frequency scenarios representative of extended operation. Long-term deployments may require checkpointing, archiving, or pruning mechanisms, which must be designed without compromising the verifiability of historical states.

Several limitations exist at the experimental and prototype levels. The current implementation does not incorporate branch-tail caching, incremental merge indexing, or server-side incremental Merkle maintenance; the latency results in Section 5 should therefore be interpreted as a characterization of prototype-level performance rather than as a lower bound on what optimized implementations can achieve. Performance evaluation is conducted primarily through local prototype execution, excluding network round-trip latency, model inference delay, and production-level concurrency pressure; synthetic short-text workloads are used to highlight fixed protocol overhead, and the real-payload comparison is conducted only under the 20-session accumulated account state. Security analysis is carried out under standard assumptions—collision-resistant hashing, unforgeable signatures, and honest-but-curious participants—and does not cover malicious collusion, side-channel attacks, or availability threats that fall outside the scope of the formal proof.

\subsection{Design Trade-offs}

Several design choices in VCT deliberately prioritize auditability and verifiability over stronger consistency semantics or higher concurrent throughput.

First, $\texttt{DeleteSession}$ is modeled as an account-level serialized state transition rather than an operation eligible for concurrent merging alongside non-deletion updates. Before initiating a deletion, a device must synchronize to the latest account root; non-deletion updates must be submitted against an account anchor within the current deletion epoch to prevent ambiguous merge semantics between deletions and concurrent updates. This design reduces concurrency in deletion-intensive scenarios but eliminates semantic conflicts between deletions and concurrent updates, and guarantees determinism in account-state evolution.

Second, account state transitions in VCT are confirmed by dual signatures from the user and the server. Compared with a server-only logging approach, this introduces additional communication and storage overhead; the resulting account states are, however, non-repudiable from both parties' perspectives and provide stronger evidentiary value for subsequent verification, synchronization, and sharing. Under real payloads, the ratio of fixed-structure security metadata to conversation content is substantially lower than under the synthetic short-text workload, indicating that the relative storage burden introduced by dual signing is acceptable under deployment conditions.

Third, the protocol adopts a hierarchical Merkle tree structure at the conversation and account levels. This structure incurs additional metadata and proof-generation costs, but enables a verifier to perform local verification without access to the full account state. Evaluation results indicate that proof size scales primarily with Merkle tree depth rather than linearly with conversation content volume; under a fixed sampling strategy, conversation-level proof length is determined by the branch structure within a single conversation and grows logarithmically with the number of branches, largely decoupled from the total number of conversations in the account.

Overall, VCT trades moderate metadata and synchronization overhead for auditability, verifiable synchronization, and evidentiary integrity. The experiments in Section 5 support the deployability of this trade-off under scaled and real-payload conditions.

\subsection{Future Directions}

The following directions are identified as avenues for further improving the practicality and scalability of VCT.

\noindent \textbf{Verifiable History Compression and Checkpointing.} 
Cryptographic checkpointing and metadata pruning of distant historical states warrant investigation, subject to the constraint that deletion barriers and checkpoint root verifiability are preserved. Such mechanisms would bound storage growth for long-lived accounts without compromising auditability or accountability.

\noindent \textbf{Incremental Maintenance of State Updates.} 
As demonstrated in Section 5, repeated recomputation during account state transitions constitutes the primary scalability bottleneck, most notably for $\texttt{Merge}$ commits. Under the existing merge semantics and deletion-epoch constraints, branch-tail caching, changed-conversation indexing, and incremental Merkle tree maintenance are promising directions, with the goal of limiting state updates to recomputation along affected paths only.

\noindent \textbf{Integration with Verifiable Reasoning.} 
VCT guarantees the provenance and state evolution of conversation records but does not cover the correctness of model reasoning. Combining VCT with verifiable reasoning, model execution environment attestation, and verifiable proofs of inference processes could yield end-to-end assurance spanning both whether a conversation has been faithfully recorded and whether the model generation process is auditable, thereby extending verifiability from conversation history to model behavior itself.

\noindent \textbf{Production-Scale Evaluation and Deployment Validation.} 
Future work should reproduce the benchmarks of this chapter under larger account sizes, real network conditions, and full long-text workloads, and assess the impact of operational requirements—including archiving strategies, key rotation, and compliance-driven retention—on protocol overhead and security boundaries.

\section{Conclusion}

This paper addresses the verifiability of large language model interaction records and proposes the Verifiable Conversation Transcript system, VCT. Unlike conventional logging systems, transparency logs, or secure messaging protocols, VCT does not aim to record a linear event sequence; instead, an independently verifiable account-level authenticated state is established over dynamically evolving LLM conversation states. The design targets the distinctive characteristics of LLM conversations—including Q\&A atomic structure, history-node-based editing, response regeneration, legitimate branching, session deletion, multi-device concurrency, and selective sharing—and provides unified mechanisms for integrity, consistency, share verifiability, and non-repudiation.

VCT adopts a three-layer verifiable structure. At the interaction layer, each Q\&A pair is modeled as a hash-chain node. At the conversation layer, branch tails are used to construct a conversation-level Merkle root. At the account layer, all conversation roots under an account are further aggregated into an account-level Merkle root, jointly confirmed by dual signatures from the user and the server. This structure preserves legitimate intra-conversation branches without collapsing all interactions into a single linear log. Building on this foundation, VCT defines a state-update protocol, deletion serialization rules, a multi-device state-merge protocol, a gossip-based fork detection mechanism, and a verifiable sharing protocol. Deletion serialization prevents semantic conflicts between deletion states and ordinary concurrent increments. The gossip protocol detects inconsistent views served by the server through cross-device exchange of dual-signed account roots. The sharing protocol reorganizes the user-confirmed set of shared nodes into an independent hash chain, over whose tail dual signatures are produced, enabling recipients to verify that nodes in the public share package have not been tampered with, omitted, reordered, or substituted.

Security analysis demonstrates that, under standard assumptions of collision-resistant hashing, unforgeable digital signatures, and secure key derivation, VCT achieves integrity, consistency, share verifiability, and operation non-repudiation. Experimental evaluation further validates the practical achievability of the design: core operations on the prototype under minimal account states remain in the sub-millisecond to low-millisecond range; all cross-scale security tests pass; gossip and share verification exhibit negligible growth with account scale; and storage and proof overhead remain manageable. Under real long-text workloads, the ratio of security metadata to conversation content decreases substantially; however, state-update and $\texttt{Merge}$ latency increase noticeably as conversation content volume and account scale grow. This increase is attributable primarily to branch-tail recomputation, Merkle root updates, and multi-conversation consolidation in $\texttt{Merge}$, rather than to fixed cryptographic primitives such as signature verification and proof checking.

Overall, VCT provides a verifiable infrastructure for LLM interaction records oriented toward forensic, audit, and accountability-tracing scenarios, enabling third parties to verify the authenticity and integrity of conversation records without placing full trust in the platform server.

\bibliographystyle{unsrt}
\bibliography{VCT_full_arxiv_en}

@article{schneier1999secure,
  author    = {Bruce Schneier and John Kelsey},
  title     = {Secure Audit Logs to Support Computer Forensics},
  journal   = {ACM Transactions on Information and System Security},
  volume    = {2},
  number    = {2},
  pages     = {159--176},
  year      = {1999}
}

@techreport{bellare1997forward,
  author    = {Mihir Bellare and Bennet S. Yee},
  title     = {Forward Integrity for Secure Audit Logs},
  institution = {University of California at San Diego},
  type      = {Technical Report},
  number    = {CS98-580},
  year      = {1997},
  month     = nov
}

@inproceedings{holt2006logcrypt,
  author    = {Jason E. Holt},
  title     = {Logcrypt: Forward Security and Public Verification for Secure Audit Logs},
  booktitle = {Proceedings of the 15th USENIX Security Symposium},
  pages     = {53--62},
  year      = {2006}
}

@inproceedings{ma2009new,
  author    = {Di Ma and Gene Tsudik},
  title     = {A New Approach to Secure Logging},
  booktitle = {Proceedings of the 24th IFIP International Information Security Conference (SEC 2009)},
  pages     = {48--62},
  year      = {2009}
}

@inproceedings{merkle1987digital,
  author    = {Ralph C. Merkle},
  title     = {A Digital Signature Based on a Conventional Encryption Function},
  booktitle = {Advances in Cryptology --- {CRYPTO} '87},
  series    = {Lecture Notes in Computer Science},
  volume    = {293},
  pages     = {369--378},
  publisher = {Springer},
  year      = {1987},
  doi       = {10.1007/3-540-48184-2_32}
}

@article{papamanthou2011authenticated,
  author    = {Charalampos Papamanthou and Roberto Tamassia and Nikos Triandopoulos},
  title     = {Authenticated Data Structures for Outsourced Databases},
  journal   = {Information Systems},
  volume    = {36},
  number    = {2},
  pages     = {195--214},
  year      = {2011}
}

@inproceedings{miller2014authenticated,
  author    = {Andrew Miller and Michael Hicks and Jonathan Katz and Elaine Shi},
  title     = {Authenticated Data Structures, Generically},
  booktitle = {Proceedings of the 41st ACM SIGPLAN-SIGACT Symposium on Principles of Programming Languages (POPL)},
  pages     = {411--424},
  year      = {2014}
}

@misc{laurie2013certificate,
  author       = {Ben Laurie and Adam Langley and Emilia Kasper},
  title        = {Certificate Transparency},
  howpublished = {RFC 6962},
  year         = {2013}
}

@inproceedings{melara2015coniks,
  author    = {Marcela S. Melara and Aaron Blankstein and Joseph Bonneau and Edward W. Felten and Michael J. Freedman},
  title     = {{CONIKS}: Bringing Key Transparency to End Users},
  booktitle = {Proceedings of the 24th USENIX Security Symposium},
  pages     = {383--398},
  year      = {2015}
}

@inproceedings{cohn2017signal,
  author    = {Katriel Cohn-Gordon and Cas Cremers and Benjamin Dowling and Luke Garratt and Douglas Stebila},
  title     = {A Formal Security Analysis of the Signal Messaging Protocol},
  booktitle = {Proceedings of the 2017 IEEE Symposium on Security and Privacy (S\&P)},
  pages     = {840--858},
  year      = {2017}
}

@misc{barnes2023mls,
  author       = {Richard Barnes and Benjamin Beurdouche and Jon Millican and Emad Omara and Konrad Kohbrok and Raphael Robert},
  title        = {The Messaging Layer Security (MLS) Protocol},
  howpublished = {RFC 9420},
  year         = {2023}
}

@article{lightman2023verify,
  author       = {Hunter Lightman and Vineet Kosaraju and Yura Burda and Harri Edwards and Bowen Baker and Teddy Lee and Jan Leike and John Schulman and Ilya Sutskever and Karl Cobbe},
  title        = {Let's Verify Step by Step},
  journal      = {arXiv preprint},
  volume       = {arXiv:2305.20050},
  year         = {2023}
}

@inproceedings{ling2023deductive,
  author    = {Ziyi Ling and Yunfang Wu and Yuxiang Wang and Xinyu Zhang and Wei Li},
  title     = {Deductive Verification of Chain-of-Thought Reasoning},
  booktitle = {Advances in Neural Information Processing Systems (NeurIPS)},
  year      = {2023}
}

@inproceedings{jacovi2024weakest,
  author    = {Alon Jacovi and Yonatan Bitton and Bernd Bohnet and Jonathan Herzig and Orith Toledo-Ronen and Alon Halfon and Matan Vaxman and Ilya Shnayderman and Yonatan Katz and Yoav Levine and Noam Slonim and Chulaka Gunasekara and Benjamin Sznajder},
  title     = {A Chain of Thought Is as Strong as Its Weakest Link},
  booktitle = {Proceedings of the 62nd Annual Meeting of the Association for Computational Linguistics (ACL)},
  year      = {2024}
}

@misc{rinderknecht2025,
  title        = {United States v. Rinderknecht, Criminal Complaint and Affidavit in Support of Criminal Complaint},
  howpublished = {Case No. 2:25-mj-01161, C.D. California},
  year         = {2025},
  month        = jan
}

@misc{garcia2025character,
  title        = {Garcia v. Character Technologies, Inc.},
  howpublished = {No. 6:24-cv-01903-ACC-DJK, M.D. Florida},
  year         = {2025},
  month        = may
}

@misc{raine2025openai,
  title        = {Raine v. OpenAI, Complaint},
  howpublished = {No. 3:25-cv-04827, N.D. California},
  year         = {2025},
  month        = aug
}

@article{li2025fundamental,
  author    = {Jiawei Li and Yu Gao and Yifan Yang and Zhongyi Zhang and Xiaojie Yuan and Jie Tang and Juanzi Li},
  title     = {Fundamental Capabilities and Applications of Large Language Models: A Survey},
  journal   = {ACM Computing Surveys},
  volume    = {58},
  number    = {2},
  pages     = {38:1--38:45},
  year      = {2025}
}

@article{ye2025llms4all,
  author    = {Yifan Ye and Zhen Zhang and Tianyi Ma and Fangrui Liu and Haotian Zhang and Shiyu Zhao and Yang Liu},
  title     = {{LLMs4All}: A Systematic Review of Large Language Models Across Academic Disciplines},
  journal   = {arXiv preprint},
  volume    = {arXiv:2509.19580},
  year      = {2025}
}

@misc{embry2026rule707,
  author       = {Stephen Embry},
  title        = {Proposed Evidentiary Rule 707: Addressing a Nonexistent Problem Instead of Real Ones},
  howpublished = {Above the Law},
  year         = {2026},
  month        = jan
}

@misc{unodc2017digital,
  author       = {{United Nations Office on Drugs and Crime}},
  title        = {Cybercrime Module 6: Key Issues --- Digital Evidence Admissibility},
  year         = {2017},
  note         = {Education for Justice Initiative}
}

@misc{anvar2014basheer,
  title        = {Anvar P.V. v. P.K. Basheer \& Ors.},
  howpublished = {(2014) 10 SCC 473, Supreme Court of India},
  year         = {2014}
}

@article{goldwasser1988digital,
  author    = {Shafi Goldwasser and Silvio Micali and Ronald L. Rivest},
  title     = {A Digital Signature Scheme Secure Against Adaptive Chosen-Message Attacks},
  journal   = {SIAM Journal on Computing},
  volume    = {17},
  number    = {2},
  pages     = {281--308},
  year      = {1988},
  doi       = {10.1137/0217017}
}

@article{bernstein2012high,
  author    = {Daniel J. Bernstein and Niels Duif and Tanja Lange and Peter Schwabe and Bo-Yin Yang},
  title     = {High-speed high-security signatures},
  journal   = {Journal of Cryptographic Engineering},
  volume    = {2},
  number    = {2},
  pages     = {77--89},
  year      = {2012},
  doi       = {10.1007/s13389-012-0027-1}
}

@techreport{kaliski2000pkcs5,
  author      = {Burt Kaliski},
  title       = {{PKCS} \#5: Password-Based Cryptography Specification Version 2.0},
  institution = {RSA Laboratories},
  type        = {RFC},
  number      = {2898},
  year        = {2000},
  month       = sep
}

@techreport{nist2015fips180,
  author      = {{National Institute of Standards and Technology}},
  title       = {Secure Hash Standard ({SHS})},
  institution = {National Institute of Standards and Technology},
  type        = {{FIPS} Publication},
  number      = {180-4},
  year        = {2015},
  doi         = {10.6028/NIST.FIPS.180-4}
}

@article{huang2023survey,
  author    = {Lei Huang and Weijiang Yu and Weitao Ma and Weihong Zhong and Zhangyin Feng and Haotian Wang and Qianglong Chen and Weihua Peng and Xiaocheng Feng and Bing Qin and Ting Liu},
  title     = {A Survey on Hallucination in Large Language Models: Principles, Taxonomy, Challenges, and Open Questions},
  journal   = {ACM Transactions on Information Systems},
  year      = {2025},
  doi       = {10.1145/3703155}
}

@article{scanlon2023chatgpt,
  author    = {Mark Scanlon and Frank Breitinger and Christopher Hargreaves and Jan-Niclas Hilgert and John Sheppard},
  title     = {{ChatGPT} for Digital Forensic Investigation: The Good, the Bad, and the Unknown},
  journal   = {Forensic Science International: Digital Investigation},
  volume    = {46},
  pages     = {301609},
  year      = {2023},
  doi       = {10.1016/j.fsidi.2023.301609}
}

@inproceedings{tomescu2019transparency,
  author    = {Alin Tomescu and Vivek Bhupatiraju and Dimitrios Papadopoulos and Charalampos Papamanthou and Nikos Triandopoulos and Srinivas Devadas},
  title     = {Transparency Logs via Append-Only Authenticated Dictionaries},
  booktitle = {Proceedings of the 2019 ACM SIGSAC Conference on Computer and Communications Security (CCS)},
  pages     = {1299--1316},
  year      = {2019},
  doi       = {10.1145/3319535.3345652}
}

@inproceedings{shapiro2011conflict,
  author    = {Marc Shapiro and Nuno Pregui{\c{c}}a and Carlos Baquero and Marek Zawirski},
  title     = {Conflict-Free Replicated Data Types},
  booktitle = {Proceedings of the 13th International Symposium on Stabilization, Safety, and Security of Distributed Systems (SSS)},
  series    = {Lecture Notes in Computer Science},
  volume    = {6976},
  pages     = {386--400},
  year      = {2011},
  doi       = {10.1007/978-3-642-24550-3_29}
}

@inproceedings{ryan2014enhanced,
  author    = {Mark D. Ryan},
  title     = {Enhanced Certificate Transparency and End-to-End Encrypted Mail},
  booktitle = {Proceedings of the Network and Distributed System Security Symposium (NDSS)},
  year      = {2014}
}

@inproceedings{kremer2002formal,
  author    = {Steve Kremer and Olivier Markowitch},
  title     = {Optimistic Non-Repudiable Information Exchange},
  booktitle = {Proceedings of the 21st IFIP WG 6.1 International Conference on Formal Techniques for Networked and Distributed Systems (FORTE)},
  series    = {Lecture Notes in Computer Science},
  volume    = {2529},
  pages     = {258--271},
  year      = {2002},
  doi       = {10.1007/3-540-36135-9_17}
}

@article{walfish2015verifying,
  author    = {Michael Walfish and Andrew J. Blumberg},
  title     = {Verifying Computations without Reexecuting Them},
  journal   = {Communications of the ACM},
  volume    = {58},
  number    = {2},
  pages     = {74--84},
  year      = {2015},
  doi       = {10.1145/2641562}
}

@misc{iso27037,
  author    = {{International Organization for Standardization}},
  title     = {{ISO/IEC} 27037:2012 --- Information Technology --- Security Techniques --- Guidelines for Identification, Collection, Acquisition and Preservation of Digital Evidence},
  year      = {2012}
}

@article{haber1991timestamp,
  author  = {Stuart Haber and W. Scott Stornetta},
  title   = {How to Time-Stamp a Digital Document},
  journal = {Journal of Cryptology},
  volume  = {3},
  number  = {2},
  pages   = {99--111},
  year    = {1991}
}

@inproceedings{crosby2009tamper,
  author    = {Scott A. Crosby and Dan S. Wallach},
  title     = {Efficient Data Structures for Tamper-Evident Logging},
  booktitle = {Proceedings of the 18th USENIX Security Symposium},
  pages     = {317--334},
  year      = {2009}
}

@article{mahajan2011depot,
  author  = {Prince Mahajan and Srinath Setty and Sangmin Lee and Allen Clement and Lorenzo Alvisi and Mike Dahlin and Michael Walfish},
  title   = {Depot: Cloud Storage with Minimal Trust},
  journal = {ACM Transactions on Computer Systems},
  volume  = {29},
  number  = {4},
  pages   = {12:1--12:38},
  year    = {2011}
}

@inproceedings{mazieres2002byzantine,
  author    = {David Mazi{\`e}res and Dennis Shasha},
  title     = {Building Secure File Systems out of Byzantine Storage},
  booktitle = {Proceedings of the 21st Annual ACM SIGACT-SIGOPS Symposium on Principles of Distributed Computing (PODC)},
  pages     = {108--117},
  year      = {2002}
}

@inproceedings{li2004sundr,
  author    = {Jinyuan Li and Maxwell N. Krohn and David Mazi{\`e}res and Dennis Shasha},
  title     = {Secure Untrusted Data Repository ({SUNDR})},
  booktitle = {Proceedings of the 6th USENIX Symposium on Operating Systems Design and Implementation (OSDI)},
  pages     = {121--136},
  year      = {2004}
}

@article{kleppmann2017json,
  author  = {Martin Kleppmann and Alastair R. Beresford},
  title   = {A Conflict-Free Replicated {JSON} Datatype},
  journal = {IEEE Transactions on Parallel and Distributed Systems},
  volume  = {28},
  number  = {10},
  pages   = {2733--2746},
  year    = {2017}
}

@techreport{rfc8032,
  author      = {Simon Josefsson and Ilari Liusvaara},
  title       = {Edwards-Curve Digital Signature Algorithm ({EdDSA})},
  institution = {Internet Research Task Force},
  type        = {RFC},
  number      = {8032},
  year        = {2017}
}

@techreport{nist800132,
  author      = {Meltem S{\"o}nmez Turan and Elaine Barker and William Burr and Lily Chen},
  title       = {Recommendation for Password-Based Key Derivation: Part 1: Storage Applications},
  institution = {National Institute of Standards and Technology},
  type        = {{NIST} Special Publication},
  number      = {800-132},
  year        = {2010}
}

@inproceedings{zhou1996fair,
  author    = {Jianying Zhou and Dieter Gollmann},
  title     = {A Fair Non-Repudiation Protocol},
  booktitle = {Proceedings of the 1996 IEEE Symposium on Security and Privacy},
  pages     = {55--61},
  year      = {1996}
}

\end{document}